\documentclass[3p,authoryear,times]{elsewier/elsarticle}

\makeatletter
\def\ps@pprintTitle{%
 }
\makeatother

\journal{}

\usepackage{longtable}
\usepackage{amstext,amsfonts,amsmath,amsxtra,amssymb,amsthm}
\usepackage{algorithm}
\usepackage{algpseudocode}
\usepackage{tabularx}
\usepackage{array}
\usepackage{minitoc}
\usepackage{multicol, multirow}
\usepackage[toc, nonumberlist,acronym,nomain]{glossaries}
\usepackage{graphicx}
\usepackage{caption}
\usepackage{subcaption}

\graphicspath{{images/}, {images/plot/}}
\DeclareGraphicsExtensions{.pdf,.jpg,.png}

\newtheorem{defi}{Definition} 
\newtheorem{prop}{Proposition}

\newcommand*{\QEDA}{\hfill\ensuremath{\blacksquare}}%

\usepackage{color}
\usepackage[dvipsnames]{xcolor}

\usepackage{tabularx,ragged2e,booktabs,caption}
\newcolumntype{C}[1]{>{\Centering}m{#1}}

\makeglossaries

\begin{document}
\newacronym{TEM}{TEM}{scanning electron microscope}

\newacronym{BOPI}{BOPI}{Bounded Oscillation Prediction Intervals}
\newacronym{LLR}{LLR}{Local Linear Regression}
\newacronym{MIS}{MIS}{Mean Interval Size}
\newacronym{EGSD}{EGSD}{Equivalent Gaussian Standard Deviation}
\newacronym{SVM}{SVM}{Support Vector Machines}
\newacronym{QR}{QR}{Quantile Regression}
\newacronym{LQR}{LQR}{Linear Quantile Regression}
\newacronym{KNN}{KNN}{K-Nearest Neighbors}
\newacronym{LOO}{LOO}{Leave-One-Out}
\newacronym{LPR}{LPR}{Local Polynomial Regression}
\newacronym{LHNPE}{LHNPE}{Local Homoscedastic Normal Prediction Error}

%
%
%

%
%
%
%
%
%
%
%
%
%
%
%
%
%
%
%
%
%
%
%
%
%
%
%
%

\begin{frontmatter}



\title{Reliable Prediction Intervals for Local Linear Regression}

%
\author[ifsttar]{Mohammad Ghasemi Hamed\corref{cor1}}
\ead{mohammad.ghasemi-hamed@ifsttar.fr}
\author[enac]{Masoud Ebadi Kivaj}

\cortext[cor1]{Corresponding author}

\address[ifsttar]{IFSTTAR-COSYS-LIVIC, 25 All\'ee des Marronniers 78000 Versailles, France\\ }
\address[enac]{Independent Researcher, 3 Alborz st. Koohsar, 35731 Tehran, Iran}

\begin{abstract}

This paper introduces two methods for estimating reliable prediction intervals for local linear least-squares regressions, named Bounded Oscillation Prediction Intervals (BOPI). It also proposes a new measure for comparing interval prediction models named Equivalent Gaussian Standard Deviation (EGSD). The experimental results compare BOPI to other methods using coverage probability, Mean Interval Size and the introduced EGSD measure. The results were generally in favor of the BOPI on considered benchmark regression datasets. It also, reports simulation studies validating the BOPI method's reliability.

\end{abstract}


\end{frontmatter}
\section{Introduction}
Almost all methods aiming at learning a continuous response variable predict a conditional distribution for the response variable. Having an estimated regression model built upon a finite sample, one may be interested in providing inferences more than what point-wise model's predictions would provide. In particular, dealing with high dimensional datasets or considering a complex model naturally demands a more comprehensive study of the predicted value's dispersion. This demand becomes fundamental in applications requiring a high level of confidence, like aircraft trajectory prediction, health informatics, security and safety systems, etc. For this purpose, one may use a high confidence prediction interval: a prediction interval with a high probability $\beta$ of containing the next observation of the regression output.

\subsection{Motivation}
This work considers prediction intervals for least-squares \gls{LLR}. A common practice in the interval prediction is to take $\hat{f}(x) \pm Z_{\frac{1-\beta}{2}}MSE^{\frac{1}{2}}$ as prediction intervals, where $Z_{\frac{1-\beta}{2}}$ and $MSE$ are respectively the $\frac{1-\beta}{2}$-quantile of the standard normal distribution and the mean squared error of the regression method given by a leave-one-out or a cross validation scheme. This method, described as the ``conventional method'' in Section \ref{ref_tol_convetional}, have some drawbacks which cause their $\beta$-content prediction intervals to be less reliable when the desired content is high $\beta \geq 0.9$. However high confidence prediction intervals are very commonly used in machine learning and statistical hypothesis-testing.

\begin{figure}[htbp!]
        \centering
        \begin{subfigure}[b]{0.5\textwidth}
                                \includegraphics[height=5cm,width=6cm]{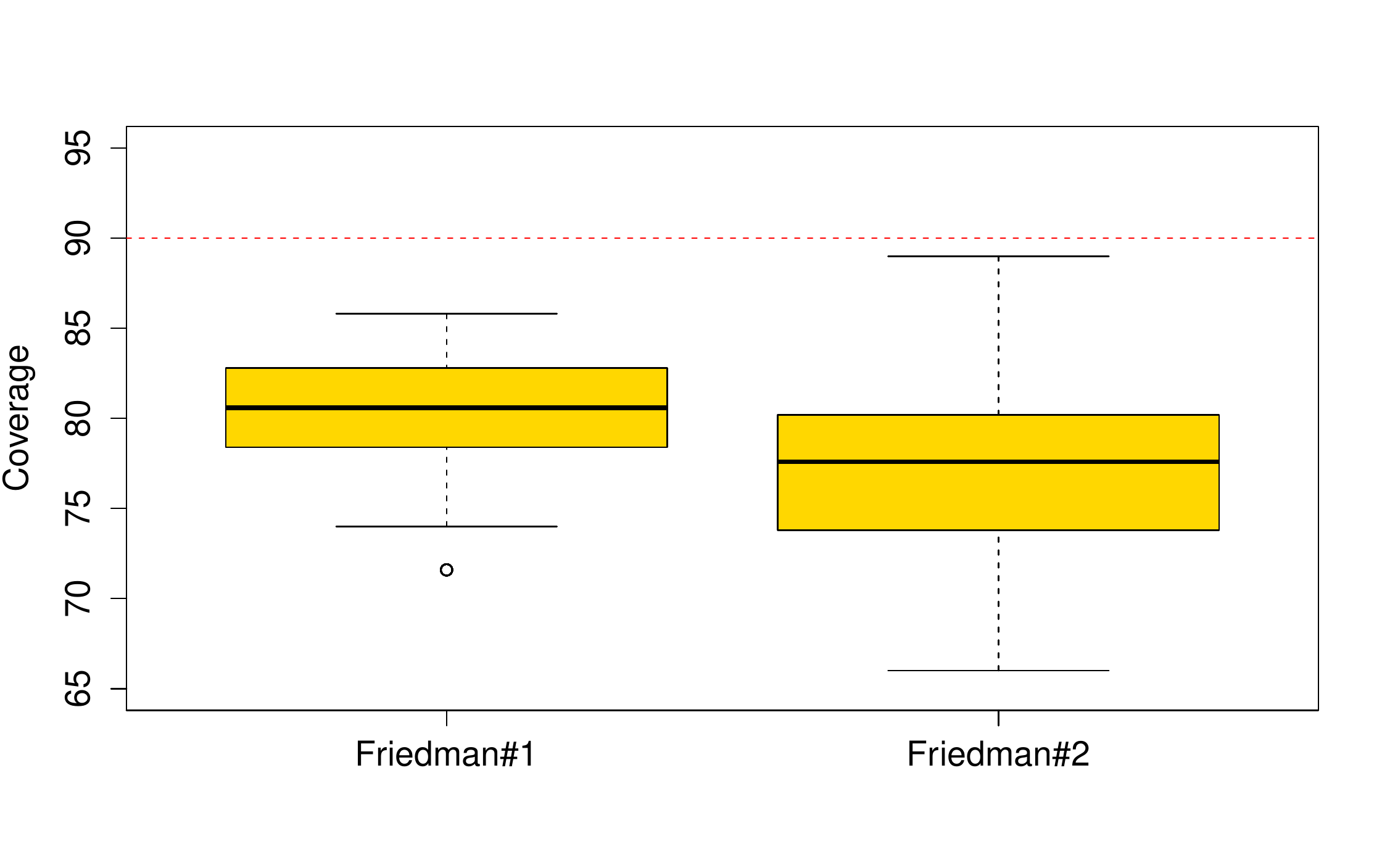}
                \caption{$\beta=0.9$}
                \label{fig_freids_a90}
        \end{subfigure}%
        ~ 
        \begin{subfigure}[b]{0.5\textwidth}
                               \includegraphics[height=5cm,width=6cm]{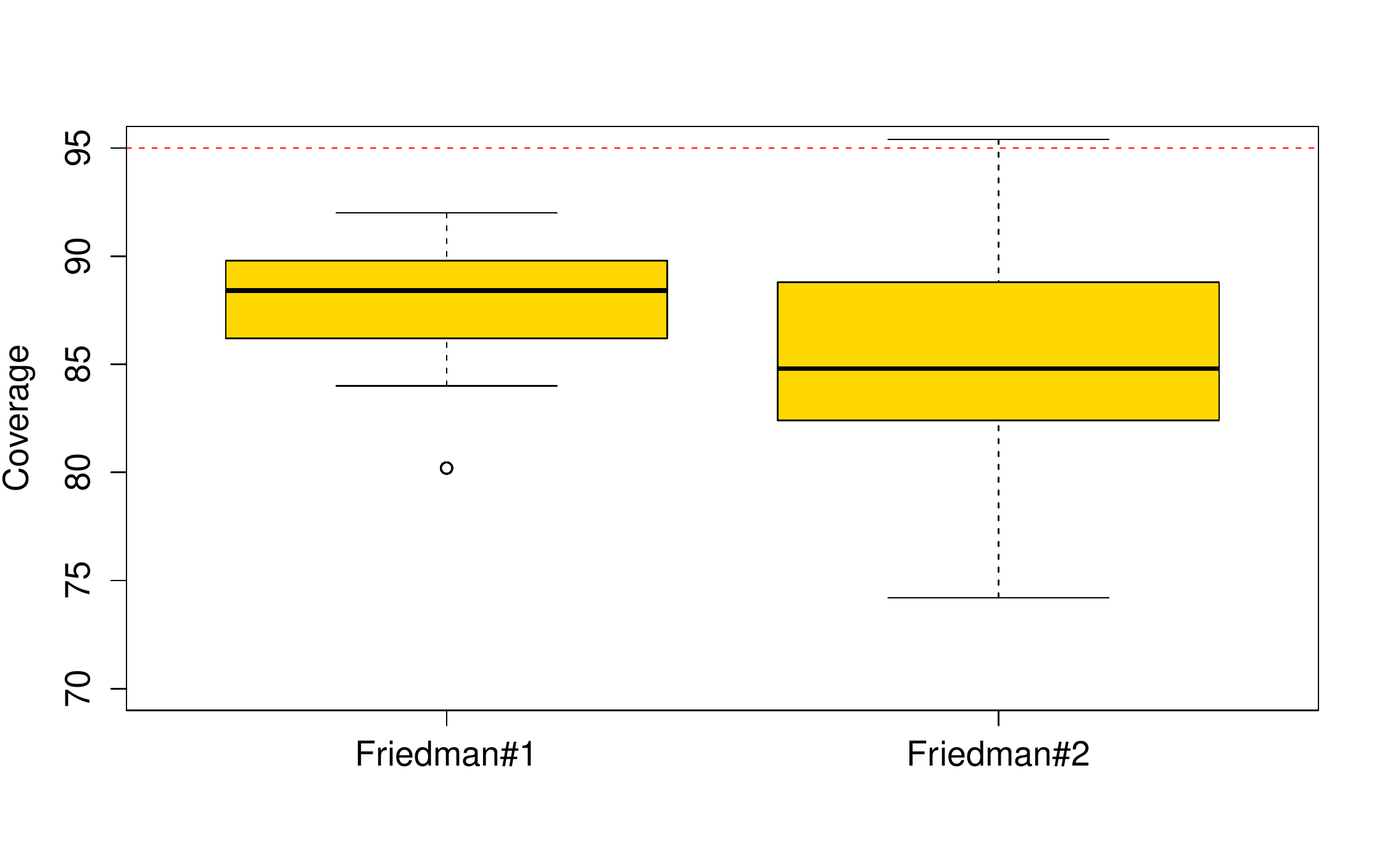}
                \caption{$\beta=0.95$}
                \label{fig_freids_a95}
        \end{subfigure}

        ~ 
        \caption{The empirical distribution of coverage probability (see Equation\eqref{eq_coverage_def}) of a simulation study with $1000$ loess models, each built on a training set of $1000$ instances. All $\beta$-content intervals, $\beta= {0.9,0.95}$, are obtained using the conventional method on separated tests set of $500$ instances. The dotted (red) lines display the desired coverage. The datasets were generated with regression functions Friedman\#1 and Friedman\#2 \citep{Breiman_1996,friedman_MARS}.}\label{fig_freids}
\end{figure}

Figure \ref{fig_freids} displays empirical distribution of coverage probability (see Equation\eqref{eq_coverage_def} of a {simulation study with loess models (see Section \ref{ref_def_loess}) estimated on a training set of $1000$ instances. All $\beta$-content prediction intervals, $\beta= {0.9,0.95}$, are obtained using the conventional method on separated test sets of $500$ instances. The whole process is iterated $1000$ times. The datasets were generated with data generating processes Friedman\#1 and Friedman\#2 \citep{Breiman_1996,friedman_MARS}}. One could observe that the conventional intervals are not reliable, only $20$ of the $4000$ tested models obtained a coverage greater than or equal to their nominal content which is far below the desired $\beta$ rate. This problem is the motivation of the current work. 

\subsection{Related works}
Non-parametric regression has been widely studied since 1975. Several monographs like \citep{eubank_spline_1988}, \citep{hasti_tibshirani_1990}, \citep{Applied_non_Hardle}, \cite{Wahba_1990} and \citep{fan_gijbels_1996} have discussed this topic. The idea of Local Polynomial Regression (LPR) appeared in \citep{local_stone_1977} and \citep{local_clev_1979}. \citep{local_clev_1979} introduced Locally Weighted Regression (LWR) and a robust version of locally weighted regression known as Robust Locally Weighted regression Scatter plot Smoothing (LOWESS). \citep{local_clev_1988} shown that locally weighted linear regression could be very useful in real data modeling applications. They introduced "loess" which is a multivariate version of locally weighted regression. Their work includes the application of loess with multivariate predictor dataset and an introduction of some statistical procedures analogous to those usually used in parametric regression. They also proposed an ANOVA test for loess. \citep{kernel_fan_1992,local_linear_fan_1993} studied some theoretical aspects of local polynomial regression. He showed that Locally Weighted Linear Regression (LWLR) (or weighted local linear regression) is design-adaptive, it adapts to random and fixed design. LWLR can be used as well in highly clustered than nearly uniform design. He also showed that the best local linear smoother has $100\%$ efficiency among all possible linear smoothers, including kernel regression, orthogonal series and splines in minimax sense. Another important property of LWLR is their adaptation to boundary points. As shown in \citep{fan_gijbels_1992}, the LWLR estimator does not have boundary effects and therefore it does not require any modifications at the boundary points. This is an attractive property of these estimators, because in practice, a large proportion of the data can be included in the boundary regions. Then \citep{ruppert_wand_1994} extended Fan's results on asymptotic bias and variance to the case of multivariate predictor variables. \\

Prediction intervals along with other statistical intervals have been rigorously studied for the linear model in \citep{book_lm_rao, Mathew_Tolerance_book,paulson_1943,stat_intervals_guide_1991}. There are currently some prediction intervals for the regression problems with a non-linear model, however their applications in literature remain limited for non-parametric regression models. \citep{ghasemi_reg_interval_sim} proposed a \gls{KNN} based interval prediction method, called simultaneous interval regression for \gls{KNN}. Unlikely to that work, here the authors are not looking after models that guarantee the simultaneous condition or the reliability conditions of tolerance intervals. Furthermore, the prediction intervals introduced in this work are based on \gls{LLR} instead of \gls{KNN}.

\subsection{The contribution}

We introduce two methods for obtaining Bounded Oscillation Prediction Intervals (BOPI) for local linear regression. It is assumed that the mean regression function is locally linear and the prediction error is locally homoscedastic and normal. The BOPI methods consider regression bias and find variable size intervals that work properly with biased regression models. The proposed prediction intervals are constructed using prediction errors of the estimated local linear regression model. These errors are obtained by a cross validation schema, for instance a leave-one-out or a $10$-fold cross validation.\\

In order to estimate prediction intervals, the current work introduces a bandwidth called LHNPE bandwidth (Local Homoscedastic Normal Prediction Error bandwidth) which is different from the regression bandwidth, as explained in Section \ref{ref_theoritical}. One of the introduced prediction intervals method, ``Farness BOPI'', has a bandwidth with a fixed number of neighbors and the other one, ``Adaptable BOPI'', uses a LHNPE bandwidth with varying number of neighbors. Both methods obtain variable size intervals which will be discussed in Section \ref{ref_pred_loess}. The idea behind the variable LHNPE bandwidth selection method is to find the ``best'' LHNPE bandwidth for each input vector $x^{*}$. This iterative procedure, described in Section \ref{ref_pred_loess_varK}, leads one to choose the prediction interval that has the best trade-off between the precision (in term of interval size) and the uncertainty to contain the response value. It is achieved by finding a balance between the faithfulness of the local assumptions (LHNPE conditions) and the required sample size to contain the desired $\beta$ proportion of the response value. In the same context, the Equivalent Gaussian Standard Deviation (EGSD) measure is used for ranking interval prediction models. This measure rates the efficiency of an interval prediction method.

In order to validate the introduced methods, several artificial and real datasets are used to compare the introduced prediction interval methods for local linear regression (Section \ref{ref_algorithm}) with commonly used interval predictions method. These methods are the linear prediction intervals, \gls{SVM} quantile regression and a common interval prediction technique, that we call the conventional prediction intervals. The conventional prediction intervals, described in Section \ref{ref_tol_convetional}, are similar to the Wald method for obtaining confidence intervals. They use Gaussian confidence intervals with mean and variance, respectively equal to the prediction value and the mean squared error of the regression given by a \gls{LOO} or (the same measure in) a 10-fold cross validation scheme. Selected methods are tested upon their capacity to provide two-sided $\beta$-content prediction intervals. The models are compared for the reliability and efficiency of their obtained envelope as described in Section \ref{ref_comp_section}. This comparison is performed with simulation studies on two artificial data generating process (DGP) and a $10$-fold cross validation schema on $11$ benchmark regression datasets with sample sizes and number of independent variables varying respectively from $N$ = 103 to $N$ = 8192 and from $p$ = 1 to $p$ = 21. Some of the real datasets contain numerical variables and some datasets have numerical and categorical variables. \\

This work is organized as follows: Section 2 is a background on regression and prediction intervals. Section 3 describes the local linear regression and particularly the loess method which is used in the experimental section. Section is a discussion on the selection criterion over different prediction interval methods. Section 5 explains the idea and hypothesis behind BOPI for LLR. Section 6 introduces the BOPI algorithms while Section 7 provides a detailed explanation for their application using linear loess. Section 8 uses experiments to compare our methods with other least squares and quantile regression methods on artificial and real benchmark dataset. The final section is a discussion with concluding remarks.

\section{Background}
\subsection{Context and Notation} 

This work considers prediction intervals for local linear regression in fixed design. Fixed design  assumes that the regression dataset $\mathcal{S}$ is a random sample composed of $N$ pairs $(x_i,Y(x_i))$, where $x_i$ is a deterministic (non-random) vector composed of $p-1$ variables (non-random observation) and the $Y(x_i)$ observations are Independent Identically Distributed (iid) random variables. The function $f(\cdot)$ denotes the mean of $Y(x)$'s distribution with a zero mean error and an unknown variance $\sigma^2$. Hereafter the following notations are used:

\begin{itemize}
\item $\mathcal{S}=(x_1,Y(x_1)), \ldots, (x_n,Y(x_N))$: the random sample of regression;
\item $N$: the number of observations in the regression dataset $\mathcal{S}$;
\item $p$: the number of independent variables plus one;
\item $f(x)$: the conditional mean of the response variable for a specified combination of the predictors;
\item $\hat{f}(x)$: the estimated regression function given $x$;
\item $\hat{f}^{-i}(x)$ the estimated regression function given $x$ without using the $i^{th}$ observation $(x_i,y_i)$;
\item $\varepsilon$: the random  error term;
\item $\varepsilon^{pred}_{x}$: the prediction error at $x$, $\varepsilon^{pred}_{x}=  Y(x)-\hat{f}(x)$;
\item $\hat{\sigma}^2$: the estimated variance of the error term;

\item $x^{*}$: a new observation in the predictor space that may not exist in the training set;
\item $Y(x)$: the conditional response variable for a given combination of the predictors, $Y(x)=f(x)+\varepsilon$;
\item $Y_i$: the $i^{th}$ random response variable, $Y_i=Y(x_i)$;
\item $y_i$: an observation of the random variable $Y_i$;

\item $I(\varepsilon^{pred}_{x})^{Tol}_{\gamma,\beta}$: $\beta$-content $\gamma$-coverage tolerance interval for  the distribution of the prediction error at $x$;
\item $I(x)^{Pred}_{\beta}$: $\beta$-content prediction interval for the response variable at point $x$;
\item $I(\varepsilon^{pred}_{x})^{Pred}_{\beta}$: $\beta$-content prediction interval of the prediction error at point $x$;
\item $Z_{\beta}$: the $\beta$-quantile of a standard normal distribution;
\item $\chi^2_{\beta,n}$: the $\beta$-quantile of a chi-square distribution with $n$ degrees of freedom.

\end{itemize}
 Note that, in this work, we suppose that the prediction error at any $x_i$ is obtained with $\varepsilon^{pred}_{x_i}= Y(x_i)-\hat{f}^{-i}(x_i)$, where the mean estimate $\hat{f}^{-i}(x_i)$ is obtained without using the observation $(x_i,y_i)$. 

\subsection{Least-squares Regression}
Regression analysis is a statistical technique for estimating the value of one variable as a function of independent variables. As mentioned in fixed-design regression, the random variable $Y_i$ or $Y(x_i)$ follows a mean function $f(x_i)$ with a random error term $\varepsilon_i$ defined as:

\begin{equation}
\label{eq_regerssion_fixed_def}
\displaystyle Y_i = f(x_i) + \varepsilon_i \textit{, where } E(\varepsilon_i)=0.
\end{equation}

The model supposes that the $\varepsilon_i$ are Independent and Identically Distributed (iid) random variables. The objective is to estimate the mean function $f(\cdot)$ by $\hat{f}(\cdot)$. The usual assumption is to suppose that the variance of the error is the same everywhere (homoscedasticity). Least-squares regression takes an estimator $\hat{f}(\cdot)$ that minimizes the Mean of Squared Errors (MSE):
 \begin{equation}
 \label{eq_reg_MSE}
MSE(\hat{f})=\frac{1}{N}\sum_{i=1}^N(y_i-\hat{f}(x_i))^2
 \end{equation}


\subsection{Conventional Interval prediction for least-squares Regression}
\label{ref_tol_convetional}
One of the common interval prediction techniques used in practice is to take $\hat{f}(x) \pm Z_{\frac{1-\beta}{2}}RMSE$ as the interval which contains a $\beta$ proportion of $Y(x)$'s population, where $RMSE$ is the root mean squared error of the regression method given by a \gls{LOO} or a 10-fold cross validation scheme. 

\begin{equation}
\label{eq_tol_convetional_interval2} 
P\bigg( Y(x) \in \bigg[\hat{f}(x) - Z_{\frac{1-\beta}{2}}RMSE,\hat{f}(x) + Z_{1-\frac{1-\beta}{2}}RMSE \bigg] \bigg) = \beta. 
\end{equation}

While the conventional interval prediction method is simple, it has some drawbacks:
\begin{itemize}
\item The estimation does not take into account the regression sample size, unlike prediction interval or tolerance intervals;
\item It estimates global inter-quantile for the conditional response variable;
\item It supposes that the estimated regression function is non-biased, but we know that the regression bias term in non-parametric regression methods does not disappear when the sample size $N$ goes to infinity \citep{fan_gijbels_1996,locally_weighted_learning}. 
\end{itemize}


\subsection{Prediction interval for normal distribution}
The prediction interval for the future observation from a normal distribution is given by \citep{prediction_normal_hahn_1969}:
\begin{equation*}
\frac{X_{n+1}-\overline{X}}{\hat{\sigma} \sqrt{1+1/n}} \sim t_{n-1},\\
\end{equation*}
where $X_{n+1}$, $\overline{X}$ and $\hat{\sigma}$ respectively represent the $(n+1)^{th}$ observation, the sample mean and sample standard error on the $n$ past observations. A two-sided $\beta$-content prediction interval for the future observation $X_{n+1}$ is obtained as:
\begin{equation}
\label{eq_prediction_normal_interval}
P(X_{n+1} \in I^{Pred}_{\beta})=\beta, \quad I^{Pred}_{\beta}= \left[ \overline{X} \pm t_{(\frac{1-\beta}{2},n-1)} \hat{\sigma} \sqrt{1+ \frac{1}{n}} \right]. 
\end{equation}
where $\overline{X}$ is the estimated mean from the $n$ past observations, $t_{(\frac{1-\beta}{2},n-1)}$ is the $(\frac{1-\beta}{2})$-quantile of Student's t-distribution with $n-1$ degrees of freedom.

\subsection{Tolerance interval for normal distribution}
\label{ref_tolerance}
Let $\mathbb{X}=(X_1, \ldots, X_n)$ denote a random sample from a continuous probability distribution. A tolerance interval is an interval that is guaranteed, with a specified confidence level $\gamma$, to contain a specified proportion $\beta$ of the population. A $\beta$-content $\gamma$-coverage tolerance interval, denoted by $I^{Tol}_{\gamma,\beta}$, is defined as: \citep{Mathew_Tolerance_book}:

 \begin{equation}
 \label{prob_tol}
 P_{\mathbb{X}} \bigg( P( X \in I^{Tol}_{\gamma,\beta}|\mathbb{X} ) \geq \beta \bigg) = \gamma.
 \end{equation}
When the sample set (of size $n$) follows a univariate normal distribution, the lower and upper tolerance bounds ($X_{l}$ and $X_{u}$, respectively) are obtained as follows:
\begin{eqnarray}
\label{eq_normal_tolrance}
    X_{l}= \hat{\theta} - \mathbf{c} \hat{\sigma} , X_{u}= \hat{\theta} + \mathbf{c} \hat{\sigma} \\
\label{eq_normal_tolrance_formula_def}
    \mathbf{c} = \sqrt{ \frac{(n-1) (1 + \frac{1}{n}) Z^2_{1- \frac{1-\beta}{2} } } {\chi^{2}_{1-\gamma ,n-1} } } 
\end{eqnarray}
where $\hat{\theta}$ is the sample mean of the distribution, $\hat{\sigma}$ is the sample standard deviation, $\chi^{2}_{1-\gamma ,n-1}$ represents the $1-\gamma$ quantile of the chi-square distribution with $n-1$ degrees of freedom and $Z^2_{1- \frac{1-\beta}{2}}$ is the square of $(1- \frac{1-\beta}{2})$ quantile of the standard normal distribution \citep{normal_tolrance_howe_1969}.

\subsection{Prediction intervals in Regression} 

\begin{defi}
A $\beta$-content prediction interval for $x$ , denoted here by $I(x)^{Pred}_{\beta}$, has a probability of $\beta$ to contain the next observation of $Y(x)$. It is defined by the equation below \citep{book_lm_rao}:

Equation (\ref{eq_def_reg_pred})
 \begin{equation}
 \label{eq_def_reg_pred}
 P_{\mathcal{S},Y(x)}\left( Y\left(x\right) \in I\left(x\right)^{Pred}_{\beta} \right) = \beta,
 \end{equation}
 where the interval $I(x)^{Pred}_{\beta}=[L(x)^{Pred}_{\beta},U(x)^{Pred}_{\beta}]$ is an estimation of the true interval (population interval) $I(x)^{Pop}_{\beta}$ such that: 
 $${P_{Y(x)}\left( Y\left(x\right) \in I\left(x\right)^{Pop}_{\beta} \right)=\beta}.$$

\end{defi}

A prediction interval $I(x)^{Pred}_{\beta}$ is obtained with a random sample $\mathcal{S}$. It is an estimator for the unknown population interval $I(x)^{Pop}_{\beta}$, which in contrary to prediction intervals, is not random. The population interval $I(x)^{Pop}_{\beta}$ has fixed interval limits that are obtained by the true distribution of $Y(x)$. 
Since for a given value of $x$, the distribution of the bounds of $I(x)^{Pred}_{\beta}$ depends on the random sample $\mathcal{S}$, prediction intervals have a joint probability distribution $P\left(s,y\left(x\right)\right)$ for the random variables $\mathcal{S}$ and $Y(x)$. For a detailed discussion about the differences between prediction and other statistical intervals, see \citep{paulson_1943,stat_intervals_guide_1991,Mathew_Tolerance_book}.

\section{Local regression methods}
\label{ref_local_methods}

Local Polynomial Regression (LPR) assumes that the unknown function $f(\cdot)$ can be locally approximated by a low degree polynomial. \gls{LPR} fits a low degree polynomial model in the neighborhood ($x_i$) of $x$. The estimated vector of parameters used in the fitted \gls{LPR} is the vector that minimizes a locally weighted sum of squares. Thus for each $x$ a new polynomial is fitted to its neighborhood and the response value is estimated by evaluating the fitted local polynomial with the vector $x$ as covariate. In general the polynomial degree $d$ is $1$ or $2$; for $d=0$, \gls{LPR} becomes a kernel regression and when $d=1$ it changes to \gls{LLR}.\\

This LPR estimator is computed as follows \citep{fan_gijbels_1996}: 

\begin{equation}
 \label{eq_loess_w_pol_reg}
  \hat{f}(x)= \sum_{i=1}^{N}a_i(x)Y_i,
\end{equation}
$\text{ where } a(x) =\mathbf{1}^{T} L_x \text{, } \mathbf{1}^{T}= (1,0, \cdots ,0)$ and $L_x$ is computed as follow:

\begin{equation}
 \label{eq_loess_local_matrix}
 L_x= (\mathbf{X_x}^{T} \mathbf{W_x}^{T} \mathbf{X_x})^{-1} \mathbf{X_x}^{T} \mathbf{W_x},
\end{equation} 
where $\mathbf{Y} = (Y_1, \cdots ,Y_N )^T$ is the vector of response variables and for each $x$, $\mathbf{X_x} \text{ and } \mathbf{W_x}$ are respectively its predictor matrix and weight matrix as described below: 

\begin{equation}
\label{eq_loess_xvector}
 \mathbf{X_x}=\left( \begin{matrix} 
1 & (x_{1}-x) & \cdots & \frac{(x_{1}-x)^d}{d!}\\
\vdots & \vdots & \vdots & \vdots \\
1 & (x_{n}-x) & \cdots & \frac{(x_{n}-x)^d}{d!}\\
\end{matrix} \right), \mathbf{W_x}= \mathrm{diag} (\mathcal{K}(\frac{x_i-x}{b}) )_{N\times N}.
 \end{equation}

where, the Kernel function $\mathcal{K}(\cdot)$ is used to weight the observations. Kernel functions are chosen so that observations closer to the fitting point $x$ have larger weights and those far from $x$ have smaller weights. If $\mathcal{K}(\cdot)$ is a kernel, then $\mathcal{K}_b(\cdot)$ is also a kernel function:
 
 $$\mathcal{K}_b(u) = \frac{1}{b} \mathcal{K}(\frac{u}{b}) , \text{ where } b > 0.$$ 

 The term $b$, known as the bandwidth, is a constant scalar value used to select an appropriate scale for the data. In this work, we use the following kernel:

\begin{equation}
\label{eq_def_loess_multivariare_simple_kernel}
 \mathcal{K}_{b}(u)=\frac{1}{b} \mathcal{K} \bigg( \frac{D(u)}{b}\bigg),
\end{equation}

where $D(\cdot)$ is a distance function like the $L_2$-norm\footnote{Discussing more about local regression methods and their computational and practical aspects are not among the scope of this work. For a review on local regression methods see \citep{locally_weighted_learning} and for a discussion about the computational and practical aspects of nonparametric smoothing see \citep{Bowman_comp_nonparam,Hart_nonparam_test}.}. 

\subsection{Bandwidth Selection}
\label{ref_loess_bandwidth}
A commonly used bandwidth selection method is \gls{LOO} technique suggested in \citep{local_stone_1977} which chooses the following bandwidth $b$:
\begin{equation}
 \label{eq_loess_bandwidth_loo}
 b= \underset{}{\operatorname{Arg min} } \sum_{i=1}^{N} (y_i- \hat{f}^{-i}(x_i))^2,
\end{equation}
where $\hat{f}^{-i}(x_i)$ is the estimation without using the $i^{th}$ observation. Estimating the bandwidth by \gls{LOO} is a time-consuming task, so it is common to minimize the $k$-fold\footnote{Note that the $k$ used in $k$-fold cross-validation is different from the $k$ denoting the number of neighbors in the forthcoming sections.} cross-validation score with $k=5$ or $k=10$ instead of \gls{LOO}. This leads to an approximation of \gls{LOO}. In this work, we use $10$-fold cross validation to estimate the bandwidth of our dataset. For more details about the use of a local version of PRESS statistics (which is also called leave-one-out MSE in the literature) to speed up the cross-validation procedure see \citep{locally_weighted_learning}. For more details on other bandwidth selection strategies, see \citep{locally_weighted_learning,fan_gijbels_1996,Applied_non_Hardle,Bowman_comp_nonparam}.

\subsection{Loess}
\label{ref_def_loess}
Loess was introduced by \citep{local_clev_1988}, and is a multivariate version of Locally Weighted Scatterplot Smoothing (LOWESS) \citep{local_clev_1979}. It is another version of LPR. Loess is described by injecting Equations (\ref{eq_loess_local_matrix} and \ref{eq_loess_xvector}) in (\ref{eq_loess_w_pol_reg}) and taking the degree of the polynomial term $d=1$ or $d=2$ in Equation (\ref{eq_loess_xvector}). 
For the bandwidth selection and weight calculation, loess applies similar bandwidths to \gls{KNN}. Its weights are calculated with (\ref{eq_def_loess_multivariare_simple_kernel}) where $u= (x_i-x)$, $D(\cdot)$ is $u$'s $L_2$-norm in the predictor space and $b$ is the Euclidean distance between the input vector $x$ and its $K^{th}$ nearest neighbor. The weight function chosen by \citep{local_clev_1988} was the Tricube kernel, however it is not mandatory.\\
 In this work, we used loess of degree one as the non-parametric smoother function. For each input vector $x$, we use Equation (\ref{eq_loess_xvector}), with $d=1$, to estimate the vector of parameter $\hat{\theta}_x$ by using the training set.

\section{Comparing Interval Prediction Methods}
\label{ref_comp_section}

In this section we discuss the selection criterion over different prediction interval methods. For a given dataset, we may use several prediction intervals methods but we need some quality measure to compare them. For this purpose, we define the dataset measures listed below. 

The $\beta$-content prediction intervals $I(x_i)^{Pred}_{\beta}=[L(x_i)^{Pred}_{\beta},U(x_i)^{Pred}_{\beta}]$ must be obtained for observations not contained in the training set $\mathcal{S}$. Therefore, for small to large datasets, these measures are obtained by  a cross-validation or a \gls{LOO} schema.

\subsection{Coverage Probability}
coverage probability is the fraction of the response values that are contained in the $\beta$-content prediction interval $I(x_i)^{Pred}_{\beta}$. 
\begin{equation}
\label{eq_coverage_def}
coverage_{\beta}= N^{-1} \sum_{i= 1}^{N} V(x_i)
\end{equation}

where is $V(x_i)$ is defined as:
$$ V(x_i)= \begin{cases} 1 & \text{if } Y(x_i) \in I(x_i)^{Pred}_{\beta}, \\ 0 & \text{otherwise}. \end{cases} $$

\subsection{Mean of Interval Size (MIS)}
Mean of Interval Size (MIS) is the average size of prediction intervals estimated on the training set: 

 $$ MIS_{\beta}= N^{-1} \sum^{N}_{i=1} size(I(x_i)^{Pred}_{\beta}).$$
Another criterion is to report the sample standard deviation of interval sizes $\sigma_{is}$.


 \subsection{Equivalent Gaussian Standard Deviation (EGSD)}
 \label{ref_egsd_def}
 

If we have different interval prediction models estimated on the same dataset giving different coverage values but approximately equal MIS values, one generally select the estimated model with the higher coverage. However, this model selection criteria would not be make sense when confronted to models (estimated on the same dataset) giving different values for both MIS and coverage. Let $m$ be a $\beta$-content interval prediction model estimated on the dataset $\mathcal{S}$, yielding $MIS^{m} \text{ and } coverage^{m}_{\beta}$. The Equivalent Gaussian Distribution (EGD) for $m$ is the normal distribution of the length of intervals obtained by method $m$ that contains their response variable. Therefore, the EGD with the smallest standard deviation (EGSD) corresponds to the ``best'' model. So, for a model $m$ giving  $coverage^{m}_{\beta}$, its EGSD is the standard deviation of the normal distribution, $\theta$-content inter-quantile size of which is be equal to $MIS^{m}$ and it is calculated as:  

\begin{equation}
\label{eq_egsd_def}
EGSD^m_{\mathcal{S}}= \frac{MIS^{m}_{\mathcal{S}}}{2 Z_{1-\frac{1-\theta}{2}}} \text{, where } \theta= Cover^{m}_{\beta}
\end{equation}
  
EGSD measures the trade-off between average interval size and the fraction of successful predictions. \textit{Smaller EGSD values denote more efficient interval prediction models.} Finally, for the sake of readability, all computed EGSD are normalized on each dataset. This normalized value is the ratio of the method's $EGSD^{m}$ to the maximum $EGSD$ value on the underlying dataset:
$$ normalized EGSD^m= \frac{EGSD^m}{ \underset{i \in (1, \ldots, c)}{max} {(EGSD^i)}}. $$
Note that if the method $m_1$ has a smaller EGSD than the model $m_2$, it does not mean that the $m_2$'s envelope is wider than the $m_1$'s envelope. As seen above, smaller normalized MIS values means tighter envelopes and smaller EGSD values means more efficient methods.

\section{Bounded Oscillation Prediction Intervals (BOPI) for Local Linear Regression}
	\label{ref_pred_loess}
In this section two methods for obtaining Bounded Oscillation Prediction Intervals (BOPI) for local linear regression are introduced. It is assumed that the mean regression function is locally linear and the prediction error is locally homoscedastic and normal. The introduced methods consider regression bias and find variable size intervals that work properly with biased regression models. The BOPI are constructed using prediction errors of the local linear regression which are obtained by a cross validation schema, for instance a \gls{LOO} or a $10$-fold cross validation. In order to estimate local linear regression, one should consider a regression bandwidth; the authors consider the \gls{KNN} bandwidth. However, the choice of the regression bandwidth is independent of the BOPI methods (for more on bandwidth selection in regression see Section \ref{ref_def_loess}). In order to estimate bounded oscillation prediction intervals, the current work introduces a bandwidth called LHNPE bandwidth. This work suggests two different LHNPE bandwidths: a bandwidth having a fixed number of neighbors and a bandwidth having a variable number of neighbors. Both of them result variable size intervals which will be discussed in more details on sections bellow. 


The idea behind BOPI methods is to exploit the local density of the prediction errors ($Y_i-\hat{f}(x_i)$) inside the LHNPE neighborhood (explained further in the next section) of a new observation $x^{*}$ and then, to find the most appropriate interval which should contain a desired proportion $\beta$ of the $Y(x^{*})$'s distribution. The introduced prediction intervals are estimated by adding the mean regression estimates $ \hat{f}(x^{*})$ to the tolerance intervals for the prediction error $I(\varepsilon^{pred}_{x})^{Tol}_{\gamma,\beta}$. This technique should be efficient, since as we will see later, these tolerance intervals are centered on the negative of estimated bias and when added to the regression estimates, the bias term (which is always present) is treated properly. The presence of bias is due to the fact that, the optimal smoothing in non-parametric regression consists of a balance between the variance and the squared bias of the regression estimator. Therefore, the regression bias in non-parametric regression is a non-vanishing term, even asymptotically(\cite{Applied_non_Hardle}). 


\subsection{Definition}
\label{ref_theoritical}
This part describes the context and idea behind of bounded oscillation prediction intervals for local linear regression. We first define the concept of a Local Homoscedastic Normal Prediction Error (LHNPE) regression estimator. Then we define the LHNPE neighborhood at $x^{*}$ to obtain the bounded oscillation prediction interval at $x^{*}$. In fact, if a regression method satisfies the LHNPE conditions, then for every $x^{*}$ we can use its LHNPE neighborhood to estimate the bounded oscillation prediction interval of $x^{*}$. Finally we obtain the equation of the estimator of bounded oscillation prediction intervals for local linear regression. Let us begin with definitions of the assumptions that will be used in this work:

\begin{defi}
The oscillation of the function $f:X\to\mathbb{R}$ on an open set $U$ is defined as: 
$$ \omega_{f}(U) = \underset{x\in U} {\sup} f(x) - \underset{x\in U} {\inf} f(x). $$
\end{defi}

\begin{defi}
\label{def_lhnpe}
A regression estimator $\hat{f}(x)$ is a \textit{Local Homoscedastic Normal Prediction Error (LHNPE)} regression estimator if its prediction errors satisfy the following conditions:
\begin{itemize}
 \item[$\bullet$] \textit{Normal prediction error:} the prediction error $\varepsilon^{pred}_{x}= Y(x)-\hat{f}(x)$ follows a normal distribution.
 \item[$\bullet$] \textit{Almost constant distribution of the prediction error:} the mean $\mu(\varepsilon^{pred}_x)$ and the standard deviation $\sigma(\varepsilon^{pred}_x)$ of the distribution for the prediction error have small local oscillations. This is defined formally as:

For all x, there exists an open set $U \ni x $, such that:
 $$\omega_{\mu(\varepsilon^{pred}_x)} (U) \leq \upsilon_{1} \text{ and } \omega_{\sigma(\varepsilon^{pred}_x)} (U) \leq \upsilon_{2}, $$ 
 where $\upsilon_{1}$ and $\upsilon_{2}$ are small fixed positive values.
\end{itemize}
\end{defi}

\begin{defi}
Let $\hat{f}(x^{*})$ be a LHNPE regression estimator for $x^{*}$ defined in Definition \ref{def_lhnpe}. The \textit{LHNPE neighborhood for $x^{*}$} is defined as instances for which the prediction error satisfies the LHNPE conditions. This neighborhood is described as below:
\begin{equation}
\label{eq_kset}
Kset_{x^{*}}=\{(x_i,Y_i)|\  d(x^{*},x_i)\leq b\},
\end{equation}
where $d(x^{*},x_i)$ is a distance function in the feature space and $b$ denote the LHNPE bandwidth.
\end{defi}

Note that the LHNPE neighborhood $Kset_{x^{*}}$ is different from the regression neighborhood $Reg_{x^{*}}$ in local linear regression. The regression neighborhood is described as below: 
\begin{equation}
\label{eq_reg_set}
Reg_{x^{*}}=\{(x_i,Y_i) | \  d(x^{*},x_i)\leq b_{reg}\}.
\end{equation}

It should be noted that while the regression bandwidth ($b_{reg}$) finds a trade-off between regression's variance and squared bias, the LHNPE bandwidth is used to find the neighborhood where the oscillation of the mean and variance of the distribution of the prediction error is bounded by a small positive value.

The LLR assumptions constraint the regression neighbors to be the set of observations for which the mean regression function is almost linear. This is less restrictive than the LHNPE conditions. So, the LHNPE neighborhood of an input vector $x^{*}$, is more likely to be included in its regression neighborhood:
\begin{equation}
\label{eq_reg_kset}
Kset_{x^{*}} \subseteq Reg_{x^{*}}.
\end{equation}

However, it is possible to have a dataset with two or more instances having different regression neighborhoods and approximately the same LHNPE neighborhood. 
%

\begin{prop}
\label{ref_prop_tol}
Let $Y(x)= f(x)+\varepsilon_x$ and let $\hat{f}(x)$ denote its local linear regression estimator. If this regression estimator satisfies the conditions below:
\begin{itemize}
 \item[$\bullet$] \textit{Normal error distribution:} $\varepsilon_{x} \sim \mathcal{N}(0, \sigma^2_{x})$;
 \item[$\bullet$] \textit{$\hat{f}(x)$ has an almost constant distribution as} defined as in Definition \ref{def_lhnpe}.
\end{itemize}
Then we have these following statements:

\begin{enumerate}[(a)]
\item $\hat{f}(x)$ is an LHNPE regression estimator;
\item The interval $I(x^{*})^{Pred}_{\beta}$ for the input $x^{*}$ obtained by Equation \eqref{eq_tol_def_local_linear_1} is a $\beta$-content prediction interval for $Y(x^{*})$; 
\begin{equation}
\label{eq_tol_def_local_linear_1}
I(x^{*})^{Pred}_{\beta} = \hat{f}(x^{*}) + I(\varepsilon^{pred}_{x^{*}})^{Pred}_{\beta}, 
\end{equation}

$$\text{ } \varepsilon^{pred}_{x^{*}} = Y(x^{*})-\hat{f}(x^{*}),$$
where $I(x^{*})^{Pred}_{\beta}$ and $I(\varepsilon^{pred}_{x^{*}})^{Pred}_{\beta}$ respectively denote the response prediction interval and the prediction interval for the normal distribution (computed using Equation~\eqref{eq_prediction_normal_interval}) on the prediction errors $\varepsilon^{pred}_{x^{*}}$ within the LHNPE neighborhood.

\item The sample bias of the prediction error in the LHNPE neighborhood is a consistent estimator of the regression bias:
\begin{equation*}
\underset{ K \rightarrow \infty}{\operatorname{ plim}} \left( \hat{f}(x^{*})- \widehat{bias}_{\hat{f}(x^{*})} \right) = f(x^{*}),
\end{equation*}
where $K/N \rightarrow 0$ as $N \rightarrow \infty$, $K=card(Kset_{x^{*}})$  and $\widehat{bias}_{\hat{f}(x^{*})}$ are respectively the cardinal of $Kset_{x^{*}}$ and the sample bias of $\hat{f}(x^{*})$.

\end{enumerate}
\end{prop}

\textit{Proof:} See Appendix B.\\

LHNPE conditions assume that the prediction error has an unknown normal distribution with an unknown mean and an unknown variance being respectively the negative bias and the variance of the prediction error. By adding $I(\varepsilon^{pred}_{x^{*}})^{Pred}_{\beta}$ to the biased regression estimator, the bias is reduced by the estimated bias and results to prediction intervals that work better with local linear regression estimators. Proposition (\ref{ref_prop_tol}) shows that the knowledge of $x^*$'s LHNPE neighborhood enables us to calculate its prediction interval by Equation (\ref{eq_tol_def_local_linear_1}). However the LHNPE neighborhood of $x^*$ is \textit{not known}, so it should be estimated on the training set. Having a finite sample, the estimation of the LHNPE neighborhood may lead to prediction intervals smaller than the true model prediction intervals (less reliable because its actual content is less than the desired $\beta$). Therefore, we approximate the prediction interval for the prediction errors obtained in $x^{*}$'s LHNPE neighborhood ($I(\varepsilon^{pred}_{x^{*}})^{Pred}_{\beta}$ in Equation~\eqref{eq_tol_def_local_linear_1}) by an upper bounds described below.

Proposition \ref{eq_interv_compare} shows that for any $\beta$-prediction intervals of the standard normal distribution, we always have a tolerance interval that is wider than or equal to it. Having in mind Propositions \ref{ref_prop_tol} and \ref{eq_interv_compare} and in order to avoid smaller prediction intervals, one can approximate the unknown prediction intervals on prediction error $I(\varepsilon^{pred}_{x^{*}})^{Pred}_{\beta}$ in Equation (\ref{eq_tol_def_local_linear_1}) by the tolerance interval on the prediction errors within the estimated LHNPE neighborhood. Proposition \ref{eq_interv_compare} is verified numerically, so it impose the LHNPE neighborhood to be such as $20 \leq K\leq 10000$. The lower limit of $20$ is chosen because methods are not intended to be used for very small values of LHNPE neighborhood and the upper limit is justified by the fact that a LHNPE neighborhood of size greater than $10000$ may occur in very limited cases.

The LHNPE neighborhood can be estimated on the training set. In other words, the $\beta$-content prediction interval on the response variable for $x^{*}$ is obtained as:
\begin{equation}
\label{eq_tol_def_local_linear}
I(x^{*})^{Pred}_{\beta} \simeq \hat{f}(x^{*}) + \hat{I}(\varepsilon^{pred}_{x^{*}})^{Tol}_{\gamma,\beta} \text{, } \gamma \geq 0.7
\end{equation}
where $\hat{I}(\varepsilon^{pred}_{x^{*}})^{Tol}_{\gamma,\beta}$ denote a $\gamma$-coverage $\beta$-content tolerance interval for the normal distribution (obtained by Equations~\eqref{eq_normal_tolrance} and \eqref{eq_normal_tolrance_formula_def}) on the prediction errors within the \textit{estimated} LHNPE neighborhood. The optimal value of $\gamma$ will vary depending on the underlying dataset, the desired content $\beta$ and the required reliability of the final prediction intervals obtained using Equation~\eqref{eq_tol_def_local_linear}. 

\subsection{Tolerance intervals as upper limits of prediction interval}

By properties of tolerance intervals for a normal distribution (see Section \ref{ref_tolerance}), if one fixes $n$ and $\beta$ such that $20 \leq n\leq 10000$ and $ 0 <\beta < 1$, and let the confidence $\gamma \geq 0.7$, then $\gamma$-coverage $\beta$-content tolerance intervals of the normal distribution are greater than or equal to its $\beta$-content prediction intervals.

\begin{prop}
\label{eq_interv_compare}
For any random sample larger than $20 \leq n\leq 10000$, if we set $\gamma$ and $\beta$, then the $\gamma$-coverage $\beta$-content tolerance interval of the standard normal distribution is greater than or equal to its $\beta$-prediction intervals. This is stated formally below:

\begin{equation}
\forall 20 \leq n\leq 10000 , \gamma \geq 0.7 , \beta \in [0.01,0.99] , \ size(I^{Tol}_{\gamma,\beta}) \geq size(I^{Prev}_{\beta}) .
\end{equation}

where $size(I)= U -L$, $I=[L,U]$ and the terms $I^{Tol}_{\gamma,\beta}$ and $I^{Prev}_{\beta}$ refer to $\gamma$-coverage $\beta$-content tolerance interval and $\beta$-prediction interval of the standard normal distribution.
\end{prop} 

\textit{Proof:} See Appendix B.\\

Figure (\ref{fig_tol_factors}) compares the size of tolerance intervals and prediction intervals of the standard normal distribution for $n\geq 20, \gamma = 0.7 $ and $ 0.01 \leq \beta \leq 0.99$. The mentioned tolerance interval and prediction intervals are obtained respectively by Equations (\ref{eq_normal_tolrance}) and (\ref{eq_prediction_normal_interval}). We can see that in this case, tolerance intervals are always greater than or equal to prediction intervals. 

 \begin{figure}[htbp!]
 \centering
	\includegraphics[scale=0.5]{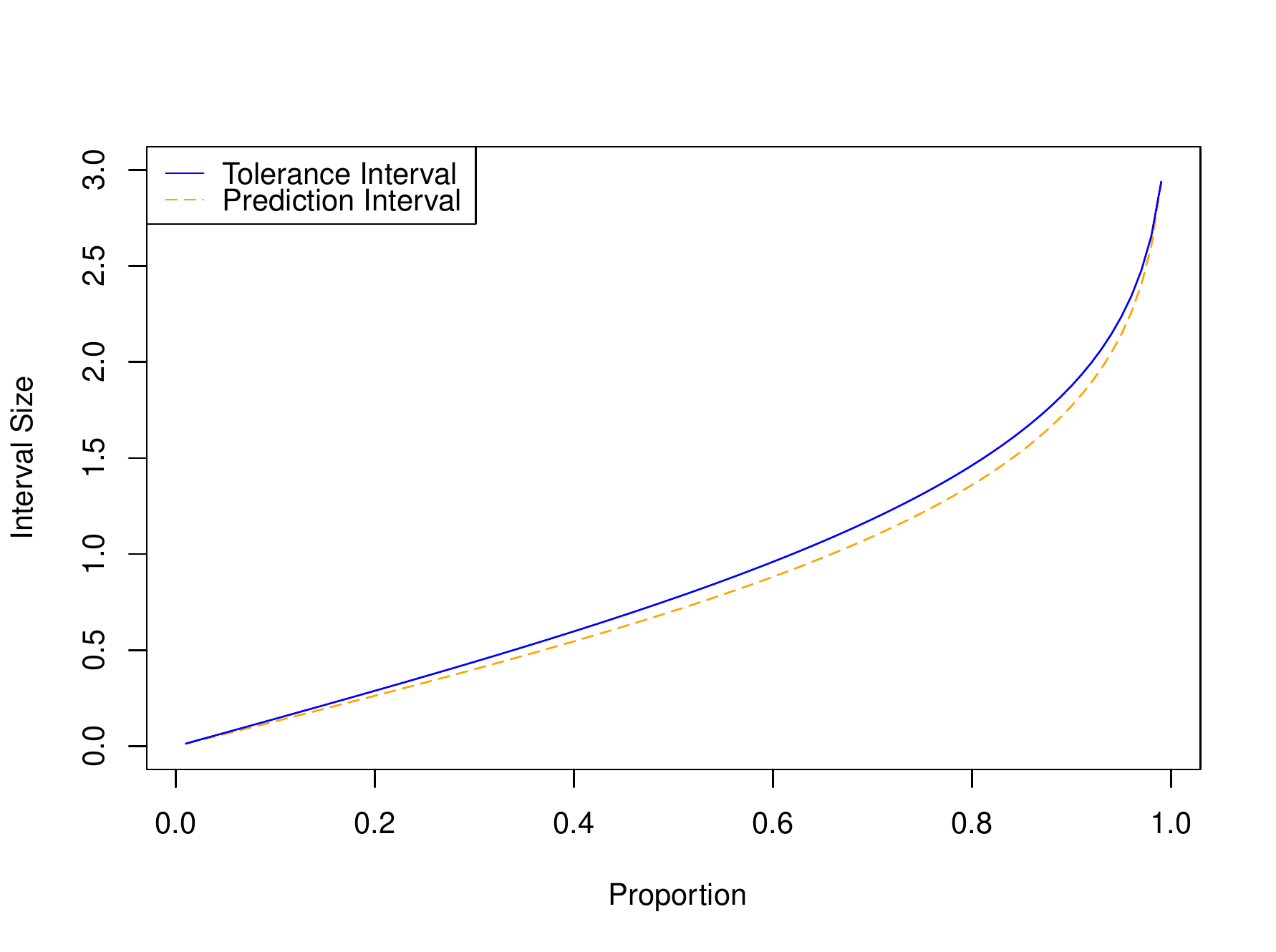} 		
 	\caption{This Figure compares the size of tolerance intervals (solid blue line) and prediction intervals (dashed orange line) of the standard normal distribution for $n\geq 20, \gamma = 0.7 $ and $ 0.01 \leq \beta \leq 0.99$. The mentioned tolerance interval (solid blue) and prediction intervals (dashed orange) are obtained respectively by Equations (\ref{eq_normal_tolrance} and \ref{eq_prediction_normal_interval}). }   
  \label{fig_tol_factors}
\end{figure}

Table \ref{tb_tol_gamma} represents the smallest sample size such that a two-sided $\gamma$-coverage $\beta$-content tolerance interval contains its corresponding two-sided $\beta$-content prediction interval. The tolerance intervals and prediction intervals are computed for the standard normal distribution and they are respectively obtained by Equations (\ref{eq_normal_tolrance}) and (\ref{eq_prediction_normal_interval}). As one see in the table, the required sample size is decreasing with the desired proportion $\beta$. For example, consider the comparison between $\beta=0.8$ and $\beta=0.95$. By looking at Table \ref{tb_tol_gamma}, one can see that the two-sided $0.55$-coverage $0.8$-content tolerance interval for the standard normal distribution contains its $0.8$-content prediction interval. However, for $\beta=0.95$, we need to have a sample of $n\ge 100$ to guarantee that the two-sided $0.55$-coverage $0.95$-content tolerance interval will contain its $0.95$-content prediction interval. Since these methods are not intended to be used for very small datasets, this table does not show $n < 20$.\\

\begin{table}[!htbp]
 \center
 \begin{tabular}{|l|l|l|l|l|}
 \hline
 $\gamma$ & \multicolumn{4}{c|}{Desired Proportion} \\ 
 \cline{2-5}
	& $0.8$ & $0.9$ & $0.95$ & $0.99$ \\ \hline
 \hline 
 0.55 & 20 & 50 & 100 & 350 \\ \hline
 0.6 & $\leq 20 $ & 20 & 50 & 80 \\ \hline
 0.65 & $\leq 20 $ & $\leq 20 $ & 20 & 40 \\ \hline
 0.7 & $\leq 20 $ & $\leq 20 $ & $\leq 20 $ & 20		\\ \hline
 \end{tabular}
 
 \vspace{0.3cm}
\caption{\label{tb_tol_gamma} Smallest sample size $n$ for which a two-sided $\gamma$-coverage $\beta$-content tolerance interval contains its corresponding two-sided $\beta$-content prediction interval. The mentioned tolerance and prediction intervals are computed for the standard normal distribution and they are respectively obtained by Equations (\ref{eq_normal_tolrance}) and (\ref{eq_prediction_normal_interval}). Note that by properties of tolerance intervals, when $\gamma$ increases and $\beta$ is fixed, the required sample size $n$ decreases.}
\end{table}


\section{The BOPI Algorithms}
\label{ref_algorithm}
As described before, having a local linear model satisfying the LHNPE conditions, one can take advantage of the LHNPE conditions for the local linear estimator and, as described by (\ref{eq_tol_def_local_linear}), use the tolerance interval of the normal distribution on the prediction errors within the estimated LHNPE neighborhood to approximate the BOPI on the response value at $x^{*}$. Let $Eset_{x^{*}}$ denote the prediction error inside the estimated LHNPE neighborhood of $x^{*}$ and it is defined as:
\begin{equation}
\label{eq_eset}
Eset_{x^{*}}= \{ \varepsilon^{pred}_{x_i} \ | \   (x_i,Y_i) \in Kset_{x^{*}} \} , \text{ where } \varepsilon^{pred}_{x_i} = Y_i-\hat{f}^{-i}(x_i).
\end{equation}
 where $\hat{f}^{-i}(x_i)$ is the local linear estimation without using the $i^{th}$ observation, obtained by (\ref{eq_loess_w_pol_reg}). Note that when $(x_i,Y_i)$ belongs to the training set, $Y_i-\hat{f}(x_i)$ becomes a residual and it depends on the random variable $Y_i$; however, $Y_i-\hat{f}^{-i}(x_i)$ and $Y_i$ are independent. \\

Given an input vector $x^{*}$, $\widehat{K}_{lhnpe}$ the number of neighbors in $Eset_{x^{*}}$, $\beta$ the desired content and $\gamma$ the confidence level, the tolerance interval for the prediction error variable $\varepsilon^{pred}_{x^{*}}$ is computed by replacing $\hat{\theta}, \hat{\sigma}$ and $n$ in Equations (\ref{eq_normal_tolrance}) and (\ref{eq_normal_tolrance_formula_def}):
\begin{eqnarray}
\label{eq_reg_tol_loess_fixed_1}
 \hat{I}(\varepsilon^{pred}_{x^{*}})^{Tol}_{\gamma,\beta}  = \hat{\theta} \pm \mathbf{c}\hat{\sigma} , \text{ where } \mathbf{c} = \sqrt{ \frac{(\widehat{K}_{lhnpe}-1) (1 + \frac{1}{\widehat{K}_{lhnpe}}) Z^2_{1- \frac{1-\beta}{2} } } {\chi^{2}_{1-\gamma ,\widehat{K}_{lhnpe}-1} } }, \\
\label{eq_reg_tol_loess_fixed_2} 
\hat{\theta}=\overline{\varepsilon}^{pred}_{x_i}  = \widehat{K}_{lhnpe}^{-1} \underset{ \varepsilon^{pred}_{x_i} \in Eset_{x^{*}}}{\operatorname{ \sum} \varepsilon^{pred}_{x_i} }
\text{ and } \hat{\sigma}^2 = \frac{\sum_{\varepsilon^{pred}_{x_i} \in Eset_{x^{*}}} (\varepsilon^{pred}_{x_i} - \overline{\varepsilon}^{pred}_{x_i} )^2} {(\widehat{K}_{lhnpe}-1) }.
\end{eqnarray}


In this work the authors suggest two methods for computing the $\widehat{K}_{lhnpe}$-nearest neighbors of $x^{*}$. One of them deals with estimated LHNPE neighborhood as fixed and the other as variable number of neighbors and both of them are tuned on the training set, so this results in two methods for obtaining BOPI. The tolerance interval $\hat{I}(\varepsilon^{pred}_{x^{*}})^{Tol}_{\gamma,\beta}$ in Equation (\ref{eq_tol_def_local_linear}) uses $\widehat{K}_{lhnpe}$ of prediction errors (obtained by a cross validation schema) in the training set inside the estimated LHNPE neighborhood of $x^{*}$. Prediction error of the whole training set is denoted by $error\_set$:
\begin{equation}
\label{eq_error_set_def}
error\_set= \{ \varepsilon^{pred}_{x_i} | \  (x_i,Y_i), i \in (1, \cdots, N) \} , \text{ where } \varepsilon^{pred}_{x_i}= Y_i- \hat{f}^{-i}(x_i).
\end{equation}


%

For the relationship between the minimum coverage level $\gamma$ in tolerance intervals and $\widehat{K}_{lhnpe}$, see Table \ref{tb_tol_gamma}.


\subsection{LHNPE bandwidth with Farness BOPI (F-BOPI)}
\label{ref_pred_loess_fixK}
This method considers a fixed number of the nearest neighbors of $x^{*}$ as its LHNPE neighborhood. We denote this interval prediction method for LLR by Farness BOPI (F-BOPI) and  the fixed number of returned neighbors is denoted by ${K}^{f}_{lhnpe}$. ${K}^{f}_{lhnpe}$ is a hyper-parameter to be tuned on the training set such that the LHNPE conditions are respected for the majority of instances in training set. This neighborhood is generally selected in such a way to keep the most of training instances' LHNPE neighborhood inside their corresponding regression neighborhood. Once the local linear model is built and the $error\_set$ is computed, the computational complexity of F-BOPI for a new instance is the same as under the conventional prediction intervals.


\subsection{LHNPE bandwidth with Adaptable BOPI (A-BOPI)}
\label{ref_pred_loess_varK}
The idea behind this LHNPE bandwidth selection, denoted by Adaptable BOPI (A-BOPI), method is to find the ``best'' LHNPE bandwidth for each input vector $x^{*}$. Here, the best number of LHNPE neighbors is denoted by ${K}^{a}_{lhnpe}$. 
 For a fixed value of $\beta$, and for each input vector $x^{*}$, the $\beta$-content $\gamma$-coverage normal tolerance interval of errors in $Eset_{x^{*}}$ defined in (\ref{eq_eset}) is calculated and this process is repeated for the same input vector $x^{*}$ but with different values of ${K}^{a}_{lhnpe}, {K}^{min}_{lhnpe} \leq {K}^{a}_{lhnpe} \leq {K}^{max}_{lhnpe}$. Finally, the $\hat{I}(\varepsilon^{pred}_{x^{*}})^{Tol}_{\gamma,\beta}$ having the smallest size is chosen and is added to $\hat{f}(x^{*})$. This iterative procedure leads us to choose the interval that has the best trade-off between the precision (size of intervals) and the uncertainty (number of observations used to obtain the interval) to contain the response value. The more ${K}^{a}_{lhnpe}$ increases, the less the local homoscedasticity assumption (bounded oscillation of the prediction error) match the reality and this yields a prediction error variance different from the true one. If by increasing ${K}^{a}_{lhnpe}$, the local estimation of the prediction error variance exceeds its true value, the fact that the tolerance interval size decreases when ${K}^{a}_{lhnpe}$ increases could partially compensates the interval size growth caused by this over estimation. However, an increase in ${K}^{a}_{lhnpe}$ may also reduce prediction variance; this issue is controlled by ${K}^{max}_{lhnpe}$. On the contrary, when ${K}^{a}_{lhnpe}$ is too small, the LHNPE conditions are more likely to be respected but the tolerance intervals size get larger (due to the small ${K}^{a}_{lhnpe}$). Thus choosing the value of ${K}^{a}_{lhnpe}$ that minimizes a fixed $\beta$-content $\gamma$-coverage tolerance interval ensures that we will have the best\footnote{Assuming $\gamma$ as fixed and ${K}^{min}_{lhnpe} \leq {K}^{a}_{lhnpe} \leq {K}^{max}_{lhnpe}$.} trade-off between the faithfulness of the local assumptions (LHNPE conditions) and the required neighborhood size to contain the desired $\beta$ proportion of the response value. The optimal value of ${K}^{a}_{lhnpe}$ may vary much more on heterogeneous datasets.


%

In order to find to keep the ${K}^{a}_{lhnpe}$-nearest neighbors of $x^{*}$ in its LHNPE neighborhood, we put two global limits for the search process: the variables ${K}^{min}_{lhnpe}$ and ${K}^{max}_{lhnpe}$. ${K}^{min}_{lhnpe}$ is the smallest number of neighbors which is assumed here to be greater than or equal to $20$. The upper bound ${K}^{max}_{lhnpe}$, is used to stop the search process if by growing the number of neighbors we constantly decrease the interval size. This break may be necessary when an increase in the number of neighbors result in adding new neighbors all having smaller prediction errors than the current neighbors. In practice, these smaller prediction errors usually belong to a different neighborhood in the feature space with different error variances and/or prediction error distributions. Therefore these two bounds serve to restrict the search process in a region where it is most likely to contain the LHNPE neighborhood of $x^{*}$. ${K}^{max}_{lhnpe}$ should almost always be included in the regression neighborhood. However one can take it greater than the regression bandwidth and let the search process 
find the neighborhood which gives the smallest tolerance interval.\\

Once the local linear model is built and $error\_set$ is found on the training set, the computational complexity of interval prediction for a new instance is $({K}^{max}_{lhnpe}- {K}^{min}_{lhnpe}+1)$ times higher than the complexity of an evaluation under the local linear regression. This is because an interval prediction for $x^*$ with A-BOPI has a $Kset_{x^*}$-finding step in which $({K}^{max}_{lhnpe}- {K}^{min}_{lhnpe}+1)$ different intervals are evaluated. 
In this step, A-BOPI finds the tightest interval among the computed ones and shifts its center to the LLR' estimation. More explanation on the LLR complexity can be found in \citep{locally_weighted_learning,fan_marron_1994,gasser_kneip_1989}

\begin{figure}[htbp!]
        \centering
                                \includegraphics[height=6cm,width=9cm]{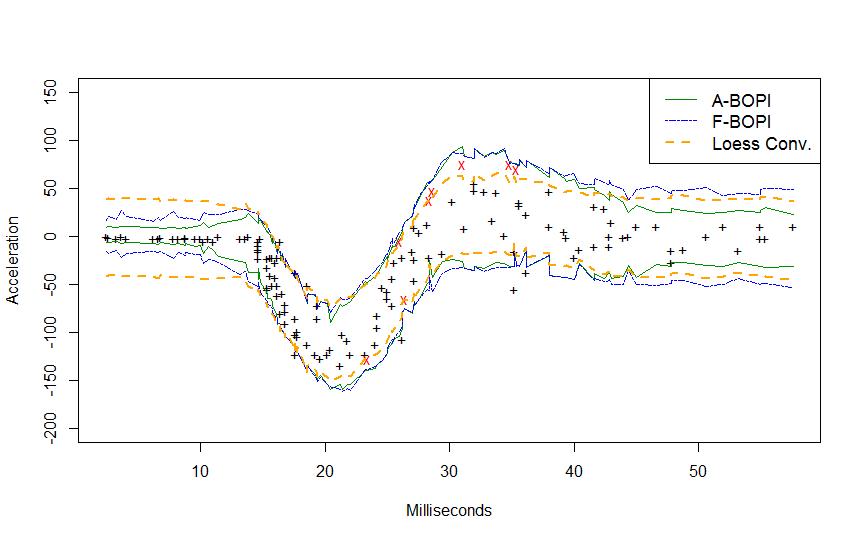}
        ~ 
        \caption{The comparison of A-BOPI (in green) and F-BOPI (in blue) prediction intervals to the conventional (F-BOPI in orange) prediction intervals on the Motorcycle dataset (described in Section \ref{ref_datasets}) when $\beta=0.9$. The three envelopes around the mean are obtained on the same loess regression model with a 10-fold cross validation schema. The red crosses in the plot show the points where the conventional method fail to cover wheras both BOPI methods cover successfuly.}\label{fig_moto_interval_09_3}
\end{figure}

Figure \ref{fig_moto_interval_09_3} illustrates an example of the comparison of the BOPI methods with the conventional prediction intervals when $\beta=0.9$. The results are explained in detail in Table \ref{tab_loess_tol_bechmark_80_2}. In this example, our introduced methods are more reliable than the conventional method; they provide variable size intervals with a good trade-off between the interval size and the model coverage.\\

\begin{figure}[htbp!]
        \centering
        \begin{subfigure}[b]{0.5\textwidth}
                                \includegraphics[height=5cm,width=6cm]{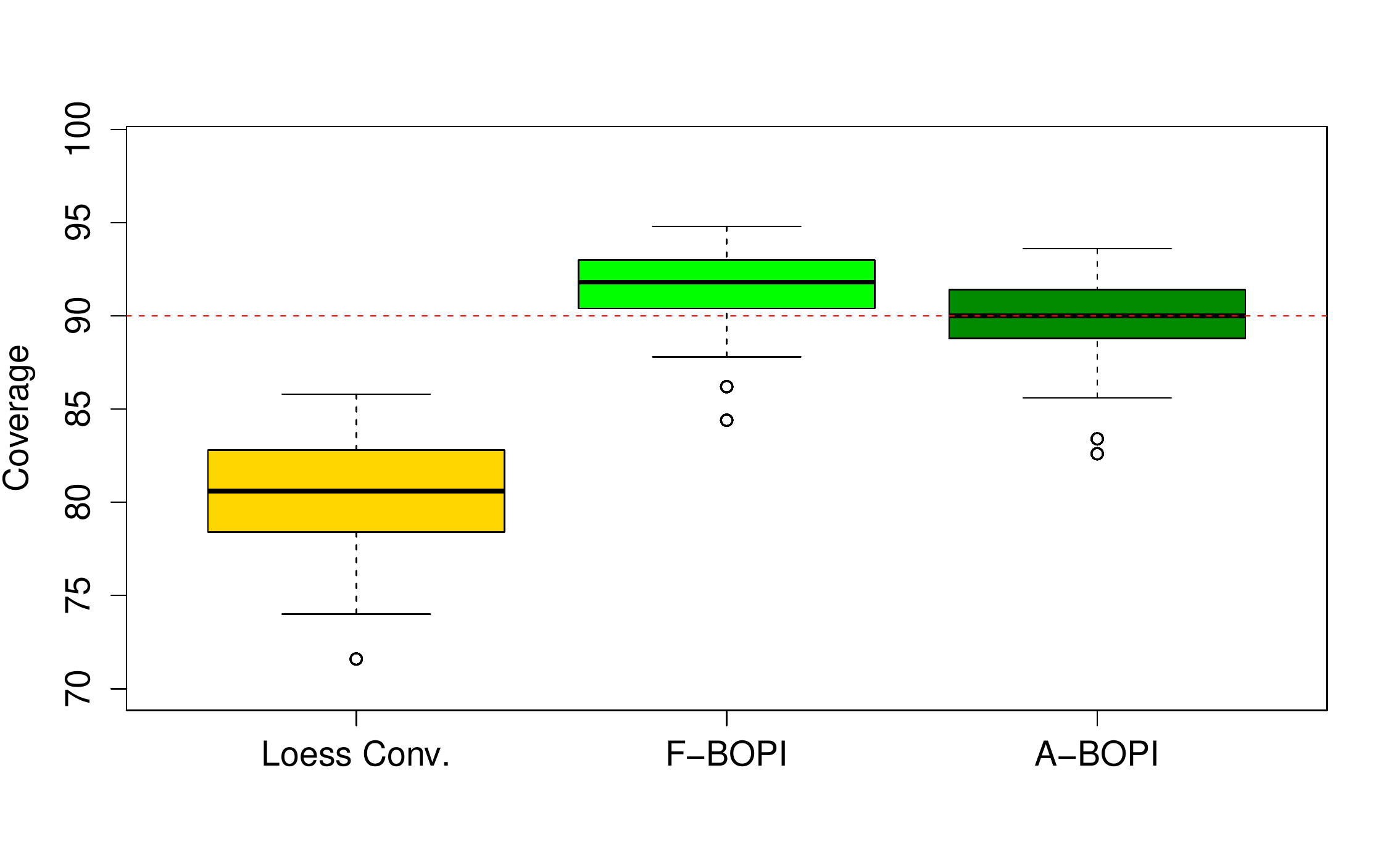}
                \caption{Friedman\#1, $\beta=0.9$}
                \label{fig_f1_g95_a9_MIP}
        \end{subfigure}%
        ~ 
        \begin{subfigure}[b]{0.5\textwidth}
                               \includegraphics[height=5cm,width=6cm]{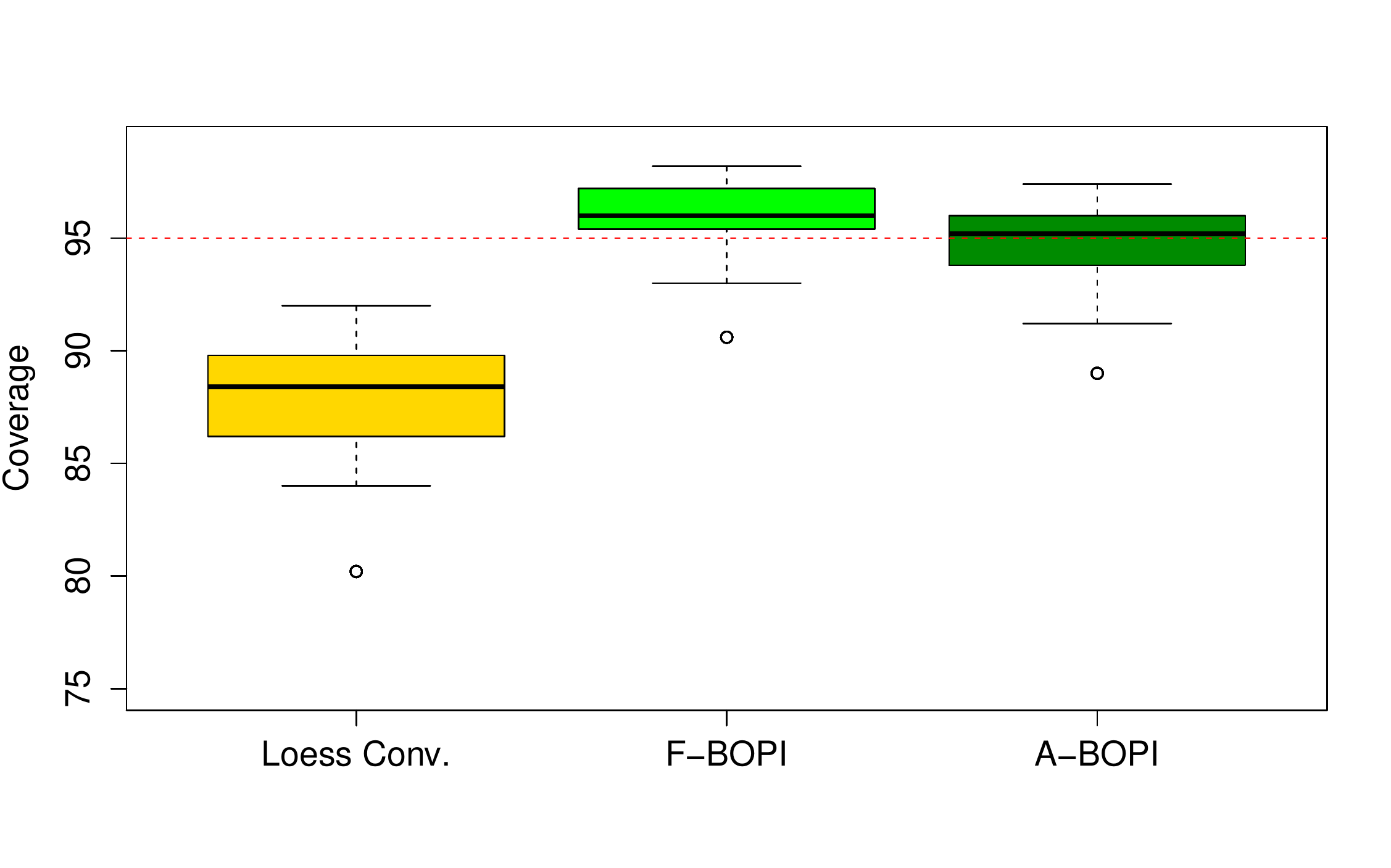}
                \caption{Friedman\#1, $\beta=0.95$}
                \label{fig_f1_g99_a95_MIP}
        \end{subfigure}
        
                \begin{subfigure}[b]{0.5\textwidth}
                                \includegraphics[height=5cm,width=6cm]{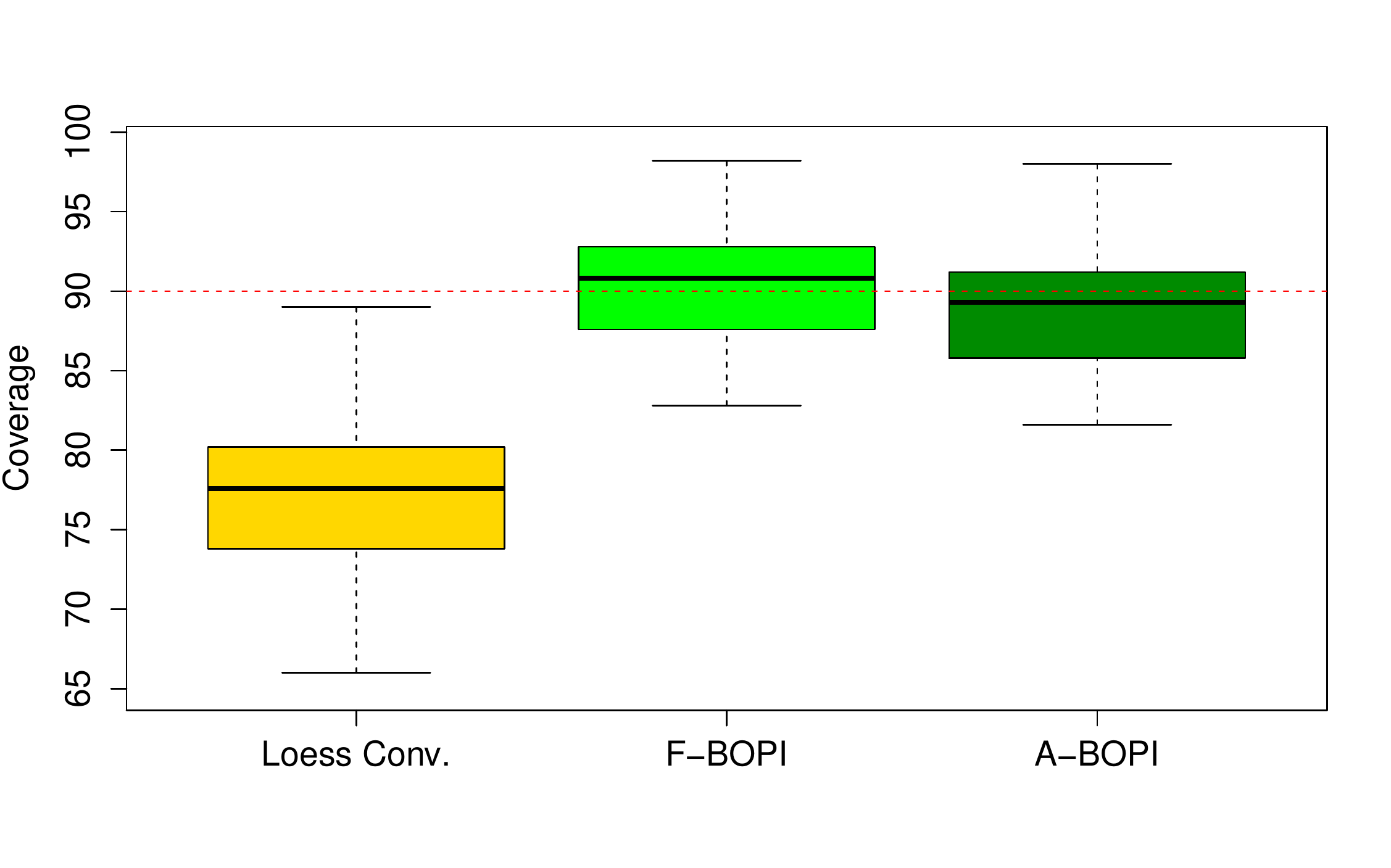}
                \caption{Friedman\#2 $\beta=0.9$}
                \label{fig_f2_g95_a9_MIP}
        \end{subfigure}%
        ~ 
        \begin{subfigure}[b]{0.5\textwidth}
                               \includegraphics[height=5cm,width=6cm]{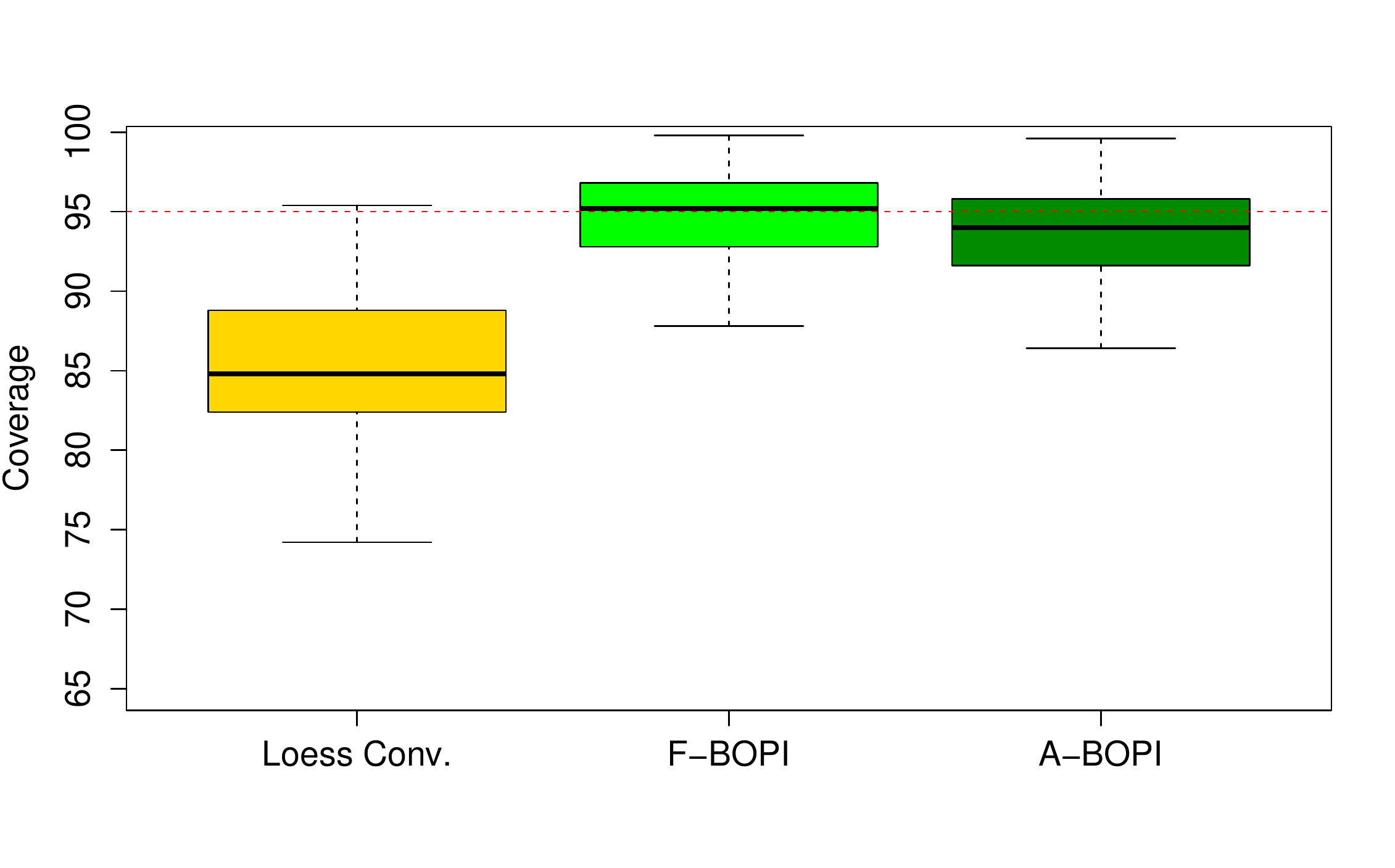}
                \caption{Friedman\#2 $\beta=0.95$}
                \label{fig_f2_g99_a95_MIP}
        \end{subfigure}

        ~ 
        \caption{coverage of a simulation study with $1000$ loess models, each built on a training set of $1000$ instances. The $\beta$-content intervals, $\beta= {0.9,0.95}$, are obtained using the conventional method (Loess Conv.), F-BOPI and A-BOPI methods on separated test sets of $500$ instances. The dotted (red) lines display the desired coverage. The datasets (same as Figure \ref{fig_freids}) were generated using data generatipn processes Friedman\#1 and Friedman\#2 (described in \ref{ref_simu}) \citep{friedman_MARS,Breiman_1996}. $K_{loess}=100$ as the regression bandwidth; it is constant for the three methods, ${K}^{f}_{lhnpe}=40$ for F-BOPI and $({K}^{min}_{lhnpe}=30,{K}^{max}_{lhnpe}=50)$ for A-BOPI. For more details see Table \ref{tab_simu_f1Table} and Table \ref{tab_simu_f2Table}.}\label{fig_freids_rc}
\end{figure}

Figure \ref{fig_freids_rc} displays the coverage, of a simulation study with $1000$ loess models, each built on a training set of $1000$ instances. The $\beta$-content intervals, $\beta= {0.95,0.95}$, are obtained using the conventional method (Loess Conv.) and F-BOPI and A-BOPI methods on separated test sets of $500$ instances. The datasets are the same (generated) datasets used in Figure \ref{fig_freids}, they were generated using data generating processes Friedman\#1 and Friedman\#2 (described in \ref{ref_simu}) \citep{friedman_MARS,Breiman_1996}. $K_{loess}=100$ as the regression bandwidth is constant for the three methods, ${K}^{f}_{lhnpe}=40$ for F-BOPI and $({K}^{min}_{lhnpe}=30,{K}^{max}_{lhnpe}=50)$ for A-BOPI. We can observe that F-BOPI and A-BOPI methods are much more reliable than the conventional method. These simulations are reported in Section \ref{ref_simu} (Tables \ref{tab_simu_f1Table} and \ref{tab_simu_f2Table}).

\subsection{Hyper-parameter Tuning}
\label{ref_predictive_hyper}
Assuming that the regression model is already built (the regression bandwidth $b_{reg}$ is already estimated for the dataset), one needs to find the optimal vector of hyper-parameters for the prediction interval methods proposed above. The hyper-parameter tuning problem is first converted into an optimization problem and then an optimization algorithm is proposed. The tuning process uses prediction errors obtained by LLR on the training set to find optimal solutions. 

Let $\lambda$ denote the vector of hyper-parameters for A-BOPI or F-BOPI with $\beta$ as their desired proportion of content. The optimization problem is the following:

\begin{equation}
\label{eq_pred_tuning_lambda}
\lambda_0 =\underset{ }{\operatorname{Arg min} } (MIS^{\lambda}_{\beta}) \text{, where } MIS^{\lambda}_{\beta} = \frac{1}{N} \sum_{i=1}^{N} size(I(x_{i})^{Tol}_{\lambda,\beta}) 
\end{equation}
\text{\ \ \ \ \ \ \ \ \ \ \ \ \ \ \ \ \ \ \ \ \ \ \ \ Subject to:} 
\begin{eqnarray}
\label{eq_pred_tuning_constraints_2}
\begin{aligned}
&\textit{ coverage Tuning Constraint: }coverage^{\lambda_0}_{\beta} = \beta \\
&\textit{$\lambda$-Specific Constraints: } \text{ depends on the prediction intervals.} 
\end{aligned}
\end{eqnarray}

\textit{Note that the coverage Tuning Constraint is a hard constraint and there is no trade-off between satisfying this constraint and minimizing the MIS.} Once $\lambda_0$ which satisfies the constraint defined above is found it would, results in intervals having the smallest MIS measure where coverage and MIS are computed based on a leave-one-out or 10-fold cross validation scheme on the training set.

\section{Application to loess}
\label{ref_pred_loess_app}
This subsection briefly reviews an application with the loess of degree one regression method. As described in Section (\ref{ref_def_loess}), loess is a version of linear polynomial regression that, for each observation, takes its $K$ nearest instances in the feature space as its neighborhood. Let us denote loess's regression bandwidth with $K_{loess}$. Loess could use among others a first or second degree polynomial. \\

Prediction intervals with loess of degree one have three or four hyper-parameters: $K_{loess}$ and the prediction hyper-parameters which are the confidence level $\gamma$ and the estimated LHNPE bandwidth. As seen above, $({K}^{f}_{lhnpe})$ and $({K}^{min}_{lhnpe},{K}^{max}_{lhnpe})$ are respectively the LHNPE bandwidth for prediction intervals obtained with fixed and variable number of instances. Based on (\ref{eq_reg_kset}), for A-BOPI we usually have:
\begin{equation*}
 {K}^{max}_{lhnpe} \leq K_{loess} 
\end{equation*}

and for F-BOPI, we have:
\begin{equation*}
 {K}^{f}_{lhnpe} \leq K_{loess}. 
\end{equation*}


\subsection{Optimization problem for loess of degree one}
\label{ref_tol_hyper-params}

As described in (\ref{ref_predictive_hyper}), it is assumed that at this stage the loess bandwidth $K_{loess}$ has been found. The difference between A-BOPI and F-BOPI is in their LHNPE bandwidth hyper-parameters, so we have $\lambda=(K_{loess},(\gamma,{K}^{min}_{lhnpe},{K}^{max}_{lhnpe}))$ for A-BOPI and $\lambda=(K_{loess},(\gamma,{K}^{f}_{lhnpe}))$ for F-BOPI\textit{ Once the loess regression model is estimated, the prediction interval hyper-parameter tuning reduces to the constraint optimization problem listed below where {all the constraints are hard constraints}}.\\

Optimization problem for fixed $\widehat{K}_{lhnpe}$:
\begin{equation*}
(\gamma,{K}^{f}_{lhnpe}) =\underset{ }{\operatorname{Arg min} } (MIS^{\lambda}_{\beta}) \text{, where } MIS^{\lambda}_{\beta} = \frac{1}{N} \sum_{i=1}^{N} \hat{I}(\varepsilon_{x_{i}})^{Tol}_{\gamma,\beta}\\
\end{equation*}
\vspace{-0.4cm}
\begin{eqnarray*}
\begin{aligned}
\text{\small{With }\textit{Tuning Constraints:}} 
\left\{
 \begin{array}{ll}
   \textit{\small{coverage Tuning Constraint: }} coverage^{\lambda_0}_{\beta} =\beta \\
   \textit{$\lambda$-specific Constraints:} 
   		\left\{
			 	\begin{array}{ll} 
		 	 	 t\leq \gamma < 1, \text{see Table \ref{tb_tol_gamma}} \\
						 0< {K}^{f}_{lhnpe} \leq N \\ 
				 \end{array}
				\right. 
 \end{array}
\right. 
\end{aligned}
\end{eqnarray*}\\

Optimization problem for variable $\widehat{K}_{lhnpe}$:
\begin{equation*}
(\gamma,{K}^{min}_{lhnpe},{K}^{max}_{lhnpe}) =\underset{ }{\operatorname{Arg min} } (MIS^{\lambda}_{\beta}) \text{, where } MIS^{\lambda}_{\beta} = \frac{1}{N} \sum_{i=1}^{N} I(\varepsilon_{x_{i}})^{Tol}_{\gamma,\beta}\\
\end{equation*}
\vspace{-0.4cm}
\begin{eqnarray*}
\begin{aligned}
\text{\small{With }\textit{Tuning Constraints:}} 
\left\{
 \begin{array}{ll}
   \textit{\small{coverage Tuning Constraint: }} coverage^{\lambda_0}_{\beta} =\beta \\
   \textit{\small{$\lambda$-specific Constraints:}}
   		\left\{
			 	\begin{array}{ll} 
		 	 	 t\leq \gamma < 1, \text {see Table \ref{tb_tol_gamma}} \\
						     0< {K}^{min}_{lhnpe} \leq {K}^{max}_{lhnpe} \leq N \\ 
				 \end{array}
				\right. 
 \end{array}
\right. 
\end{aligned}
\end{eqnarray*}

Note that for F-BOPI, the smallest value of $\gamma$ (which is denoted by $t$ in the optimization problem above), depends on ${K}^{f}_{lhnpe}$ and this relationship is shown in Table \ref{tb_tol_gamma}. For A-BOPI the same dependency exists between the smallest value of $\gamma$ and ${K}^{min}_{lhnpe}$.

\subsection{Hyper-parameter tuning for loess of degree one}
\label{ref_lhne_hyper}
Algorithm \ref{alg_loess_tol_vark_hyperparam} describes how to tune the prediction interval hyper-parameters for variable $\widehat{K}_{lhnpe}$. The algorithm used for the fixed $\widehat{K}_{lhnpe}$ is almost the same, except that it computes the hyper-parameter ${K}^{f}_{lhnpe}$ instead of the pair $({K}^{min}_{lhnpe},{K}^{max}_{lhnpe})$, so we omit its description. In a first attempt, $\gamma$ is considered as a fixed high value like $\gamma=0.9$ or $\gamma=0.99$ and the focus is on finding the LHNPE neighborhood hyper-parameter: the hyper-parameter ${K}^{f}_{lhnpe}$ or the pair $({K}^{min}_{lhnpe},{K}^{max}_{lhnpe})$. as described before the variable $coverage^{\lambda}_{\beta}$ defined by Equation (\ref{eq_pred_tuning_constraints_2}) must be greater than or equal to $\beta$. Thus the LHNPE neighborhood hyper-parameter(s) which find(s) intervals that, based on a \gls{LOO} or 10-fold cross validation scheme on the training set, satisfies the coverage tuning constraint defined in (\ref{eq_pred_tuning_constraints_2}) and also have the smallest Mean Interval Size (MIS) is selected. Once  ${K}^{f}_{lhnpe}$ or $({K}^{min}_{lhnpe},{K}^{max}_{lhnpe})$ is found, one searches for the smallest value of $\gamma$ that satisfies the coverage tuning constraint.

\begin{figure}[htbp!]
        \centering
        \begin{subfigure}[b]{0.5\textwidth}
                 \includegraphics[height=5cm,width=6cm]{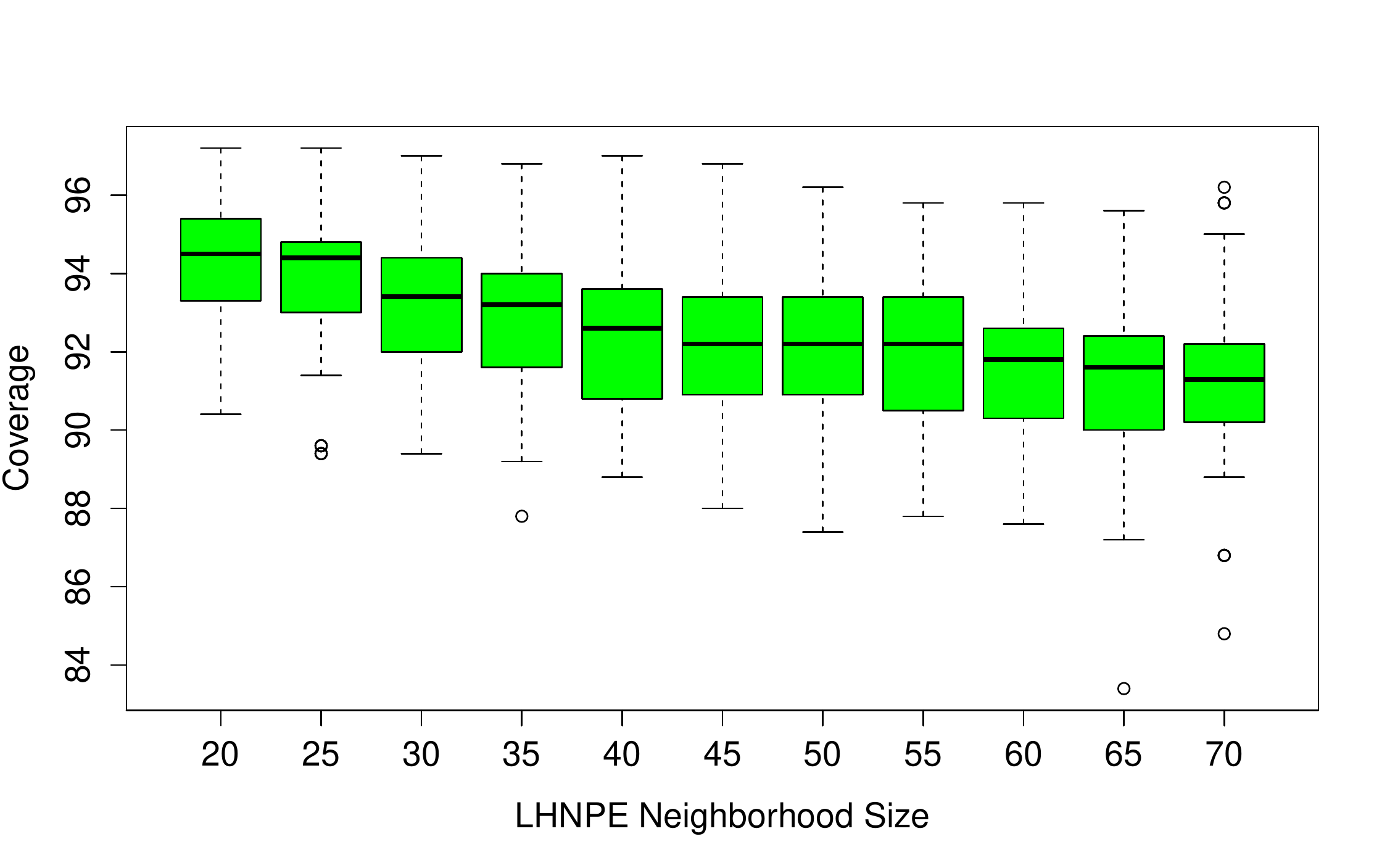}
                \caption{$Friedman\#1$}
                \label{fig_freids_a90}
        \end{subfigure}%
        ~ 
        \begin{subfigure}[b]{0.5\textwidth}            
                \includegraphics[height=5cm,width=6cm]{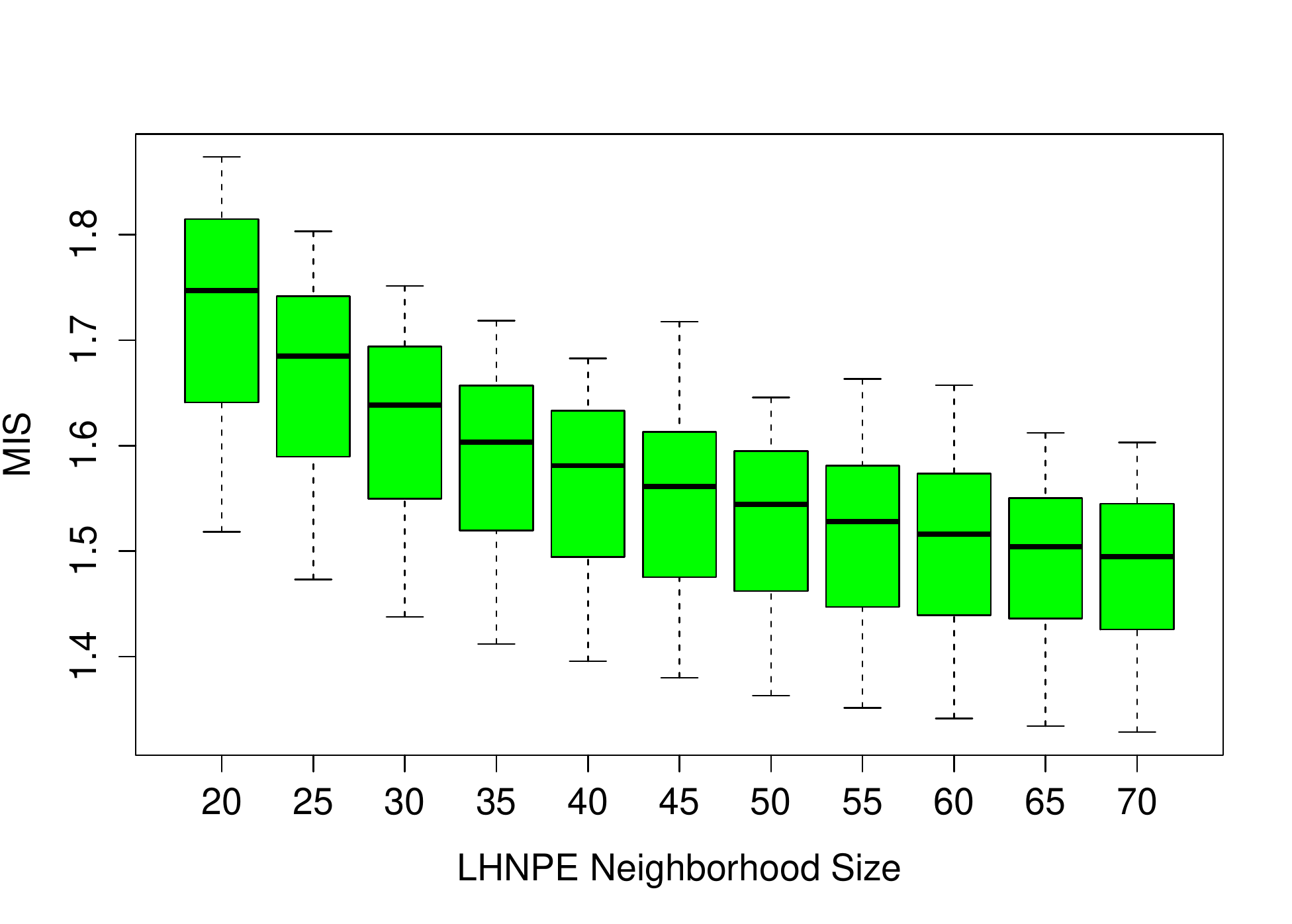}
                \caption{$Friedman\#1$}
                \label{fig_freids_a95}
        \end{subfigure}      
         \begin{subfigure}[b]{0.5\textwidth}
                 \includegraphics[height=5cm,width=6cm]{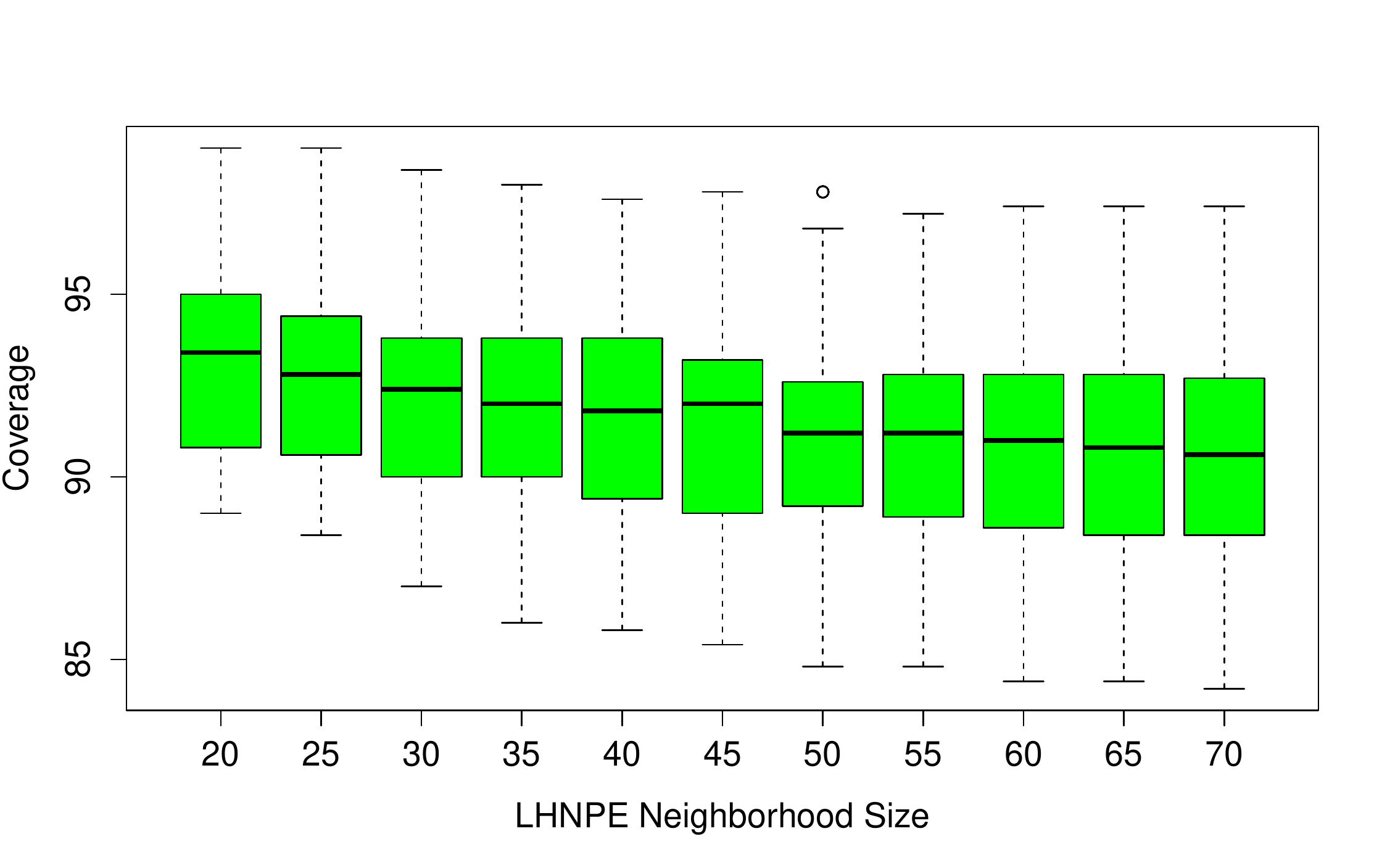}
                \caption{$Friedman\#2$}
                \label{fig_freids_a90}
        \end{subfigure}%
        ~ 
        \begin{subfigure}[b]{0.5\textwidth}            
                \includegraphics[height=5cm,width=6cm]{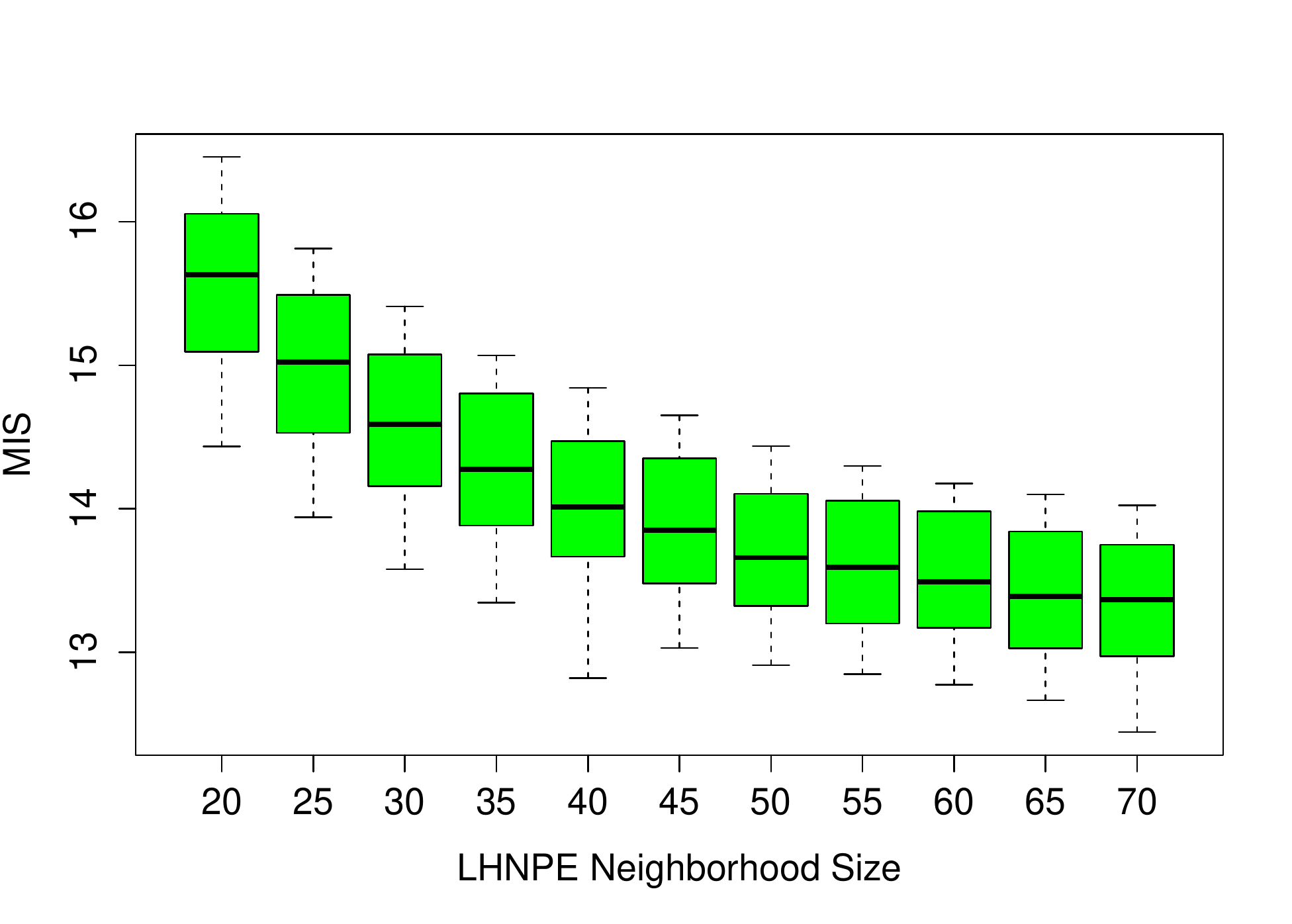}
                \caption{$Friedman\#2$}
                \label{fig_freids_a95}
        \end{subfigure}    
                        
        ~ 
        \caption{ The distribution of coverage and MIS of F-BOPI for $\gamma=0.9, \beta=0.95$ and ${K}^{f}_{lhnpe}$ varying from $20$ to $70$ by steps of $5$. For each value of ${K}^{f}_{lhnpe}$, $100$ loess models are estimated, each trained and tested on separated datasets generated by the Friedman\#1 Friedman\#2 data generating processes (described in \ref{ref_simu}) \citep{friedman_MARS,Breiman_1996}.}\label{fig_lhnpe_coverage}
\end{figure}

Figure \ref{fig_lhnpe_coverage} displays the variation of the distribution of coverage and MIS of F-BOPI, with constant values for $\gamma=0.9$ and $\beta=0.95$, by changing ${K}^{f}_{lhnpe}$ from $20$ to $70$ by steps of $5$. For each value of ${K}^{f}_{lhnpe}$, $100$ loess models are estimated, each built on a training set of $1000$ instances. Plot of distribution of coverage and MIS values obtained on separated test sets of $500$ instances where the datasets were generated with the Friedman\#1 and Friedman\#2 data generating processes (described in \ref{ref_simu}) \citep{friedman_MARS,Breiman_1996}. As the results in this example show, coverage and MIS decrease by increasing ${K}^{f}_{lhnpe}$, the MIS average does not decrease much more for ${K}^{f}_{lhnpe} \geq 45$ and the coverage distribution begins to have smaller minimum values for ${K}^{f}_{lhnpe}\geq 50$ or ${K}^{f}_{lhnpe} \geq 55$. The results suggest that the function LHNPE neighborhood may be between $30$ and $50$, so setting $({K}^{min}_{lhnpe}=30,{K}^{max}_{lhnpe}=50)$ and ${K}^{f}_{lhnpe}=40$ may be an acceptable solution.

 As seen before, when the neighborhood size ${K}^{f}_{lhnpe}$ decreases the tolerance interval size increases, as a result small LHNPE neighborhoods lead in larger tolerance interval sizes and thus a higher coverage. 
 
{By taking ${K}^{f}_{lhnpe} \geq 20$ the estimated neighborhood is enlarged with instances that generally satisfy the LHNPE conditions, so the MIS decreases rapidly (tolerance intervals decrease faster than prediction intervals) while in the same time, the coverage decreases and converges to the desired $\beta$. In fact, as long as $\gamma$ and ${K}^{f}_{lhnpe}$ are chosen as recommended by Table~\ref{tb_tol_gamma}, and the ${K}^{f}_{lhnpe}$ nearest instances to $x^{*}$ are in its LHNPE neighborhood, tolerance intervals of the estimated normal distribution are wider than its prediction intervals. 
On the other hand due to the limited sample size, the number of instances in the training set that belong to the LHNPE neighborhood of $x^{*}$ usually changes based on the location of $x^{*}$ in the feature space. 
Indeed, using a too large ${K}^{f}_{lhnpe}$ leads to considering far neighbors of $x^{*}$, those not belonging to its LHNPE neighborhood, as if they were in so. If this happens for a significant number of tested instances during the hyper parameter tuning, we end up with many intervals being wider than necessary, causing a large MIS. This difficulty in estimating the $K^{f}_{lhnpe}$ are caused by the heterogeneity and heteroscedasticity of the underlying dataset, and directly influence the mentioned MIS and coverage variations. 
Therefore, we can state that: although increasing ${K}^{f}_{lhnpe} \geq 20$ generally decreases MIS without having a significant impact on coverage, this situation usually changes after a threshold and the variation of coverage and MIS after the ${K}^{f}_{lhnpe}$'s threshold depends on the testing model. 
}

  
 In practice, evaluating the efficiency of both methods on datasets, and incorporating obtained a priori knowledge in the hyper-parameter tuning phase is suggested. One could find ${K}^{f}_{lhnpe}$ for F-BOPI method and when it comes to the finding $({K}^{min}_{lhnpe},{K}^{max}_{lhnpe})$, she can try to choose the $[{K}^{min}_{lhnpe}, {K}^{max}_{lhnpe}]$ interval such that it contains the fixed ${K}^{f}_{lhnpe}$ value found before.

$$ {K}^{min}_{lhnpe} \leq {K}^{f}_{lhnpe} \leq {K}^{max}_{lhnpe}. $$ 

Once ${K}^{f}_{lhnpe}$ or the pair $({K}^{min}_{lhnpe},{K}^{max}_{lhnpe})$ is found, next step is decreasing value of $\gamma$, which decreases the mean interval size. The goal is to have the smallest mean tolerance interval size that satisfies coverage tuning constraint. The idea is to set the value of the neighborhood parameters with those found in the previous process and decrease $\gamma$. This procedure is repeated as long as the inclusion constraint is satisfied and $\gamma$ is larger than its minimum value shown in Table \ref{tb_tol_gamma}. High values of $\gamma$ will guarantee the satisfaction of the coverage Tuning constraint but the computed intervals can be very large. Note that, with this approach, the value of $\gamma$ can be less than $\beta$ and this may happen when the local density of the response variable is quite high. Based on the new value of $\gamma$, one can go to the first step and recalculate new values for the neighborhood hyper-parameter (${K}^{f}_{lhnpe}$ or the pair $({K}^{min}_{lhnpe},{K}^{max}_{lhnpe})$) and this can be repeated for one or two iterations until the 
coverage Tuning constraint is satisfied and the obtained MIS change is negligible.

\begin{algorithm}[htbp!]
\caption{Hyper-parameter tuning for prediction intervals with variable $\widehat{K}_{lhnpe}$. }
\label{alg_loess_tol_vark_hyperparam}
\begin{algorithmic}[1]
\Function{TuneHyper-Params}{$error\_set,\beta$} 
\State $\gamma \leftarrow 0.99$ \Comment{ or $\gamma \leftarrow 0.9$ depending on the dataset}
\State $({K}^{min}_{lhnpe},{K}^{max}_{lhnpe}) \leftarrow (MIN_{K_0},MAX_{K_0})$ initial values 
\State $ \lambda \leftarrow (\gamma,{K}^{min}_{lhnpe},{K}^{max}_{lhnpe})$
\For {$iteration =1..3$}
\State $(Coverage,MIS) \leftarrow$ \Call{ComputeOnTrainigSet}{$\beta,\lambda$} \label{marker}
\State $MIS_{min} \leftarrow MIS$.
\While{ $Coverage \geq \beta$	and $MIS \leq MIS_{min}$ } 
 \State $({K}^{min}_{lhnpe},{K}^{max}_{lhnpe}) \leftarrow ({K}^{min}_{lhnpe},{K}^{max}_{lhnpe}) + somestep$
 \State $ \lambda \leftarrow (\gamma,{K}^{min}_{lhnpe},{K}^{max}_{lhnpe})$
 \State $MIS_{min} \leftarrow MIS$
\State $(Coverage,MIS)\leftarrow$\Call{ComputeOnTrainigSet}{$\beta,\lambda$} 
 \EndWhile

\While{ $Coverage \geq \beta$	and $MIS \leq MIS_{min}$ } 
 \State $\gamma \leftarrow \gamma - step$
\If {$\gamma < t$} \Comment{ $t$ value can be found in Table \ref{tb_tol_gamma}.} 
\State{Break;} \Comment{ Goes outside of the loop} 
 \EndIf
 
 \State $ \lambda \leftarrow (\gamma,{K}^{min}_{lhnpe},{K}^{max}_{lhnpe})$
 \State $MIS_{min} \leftarrow MIS$
\State $(Coverage,MIS)\leftarrow$\Call{ComputeOnTrainigSet}{$\beta,\lambda$} 
 \EndWhile
\EndFor
 \State \Return $({K}^{min}_{lhnpe},{K}^{max}_{lhnpe},\gamma)$
 \EndFunction
 \\
 \Function{ComputeOnTrainigSet}{$\beta,\lambda$} 
 \State Use Equations \eqref{eq_reg_tol_loess_fixed_1} and \eqref{eq_reg_tol_loess_fixed_2} to obtain BOPI intervals on the training set with a \gls{LOO} or 10-fold cross-validation schema.
 
 \State $Coverage \leftarrow$ use Equation (\ref{eq_coverage_def}) on the intervals calculated in the previous step.
\State $MIS \leftarrow$ compute the mean size of these intervals found above.
 \State \Return $(Coverage,MIS)$
\EndFunction
\end{algorithmic}
\end{algorithm}


\section{Experiments}
In this section, several artificial and real datasets are used to compare the introduced prediction intervals methods for local linear regression described in Section \ref{ref_algorithm}) with the conventional prediction intervals described by Equation~\eqref{eq_tol_convetional_interval2}, the linear prediction intervals and SVM quantile regression. The selected methods will be tested upon their capacity to provide two-sided $\beta$-content prediction intervals. The estimated prediction intervals are compared for their reliability and efficiency of their obtained envelope as described in Section \ref{ref_comp_section}. Note that we are interested in comparing the aforementioned methods, regardless of any variable selection or outliers detection pre-processing.

\subsection{Prediction Intervals Methods}
This part involves the description of the tested prediction intervals. The numerical study in \ref{ref_simu} uses three of these methods (F-BOPI, A-BOPI and Loess Conv.), whereas Section \ref{ref_exp_bech} reports the application of  all of them on real datasets.

\subsubsection{Method's Implementation}
\label{ref_method_def}
The tested methods are the followings:

\begin{itemize}
\item F-BOPI: two-sided prediction interval for linear loess as explained in Section \ref{ref_pred_loess} with the fixed $K$ LHNPE neighborhood (${K}^{a}_{lhnpe}$). The prediction intervals are obtained on the same estimated linear loess model as A-BOPI and Loess Conv. F-BOPI hyper-parameters values for the real datasets can be found in Tables \ref{tab_method_80} and \ref{tab_method_95}.
	
\item  A-BOPI: two-sided prediction interval for linear loess as explained in Section \ref{ref_pred_loess} with the variable $K$ LHNPE neighborhood (${K}^{f}_{lhnpe}$). The prediction intervals are obtained on the same estimated linear loess model as A-BOPI and Loess Conv. The A-BOPI hyper-parameters values for the real datasets can be found in Tables \ref{tab_method_80} and \ref{tab_method_95}.
	
\item Loess Conv. the conventional interval prediction method explained by Equation \ref{eq_tol_convetional_interval2} obtained with the same estimated linear loess model as F-BOPI and values of the estimated $K_{loess}$ for the real datasets can be found in Table \ref{tab_dataset_hyper}.

\item OLS prediction intervals for classical linear regression (Ordinary Least-Squares) obtained by:
\begin{equation}
\label{eq_lr_prevision_interval}
  \hat{f}(x) \pm \mathbf{c} t_{({1-\frac{1-\beta}{2}},N-p)},  \mathbf{c}= \sqrt{ \frac{N\hat{\sigma}^{2}}{N-p} ( 1+ x^{*T} (X^{T}X)^{-1} x^{*}) }
\end{equation}
where $p$ and $\hat{\sigma}^{2}$ are respectively the number of independent variables plus one, and the estimated variance of the error term \citep{book_lm_rao}.

\item  LS-SVM Conv.: the conventional interval prediction method explained in Equation \ref{eq_tol_convetional_interval2} obtained with a least-square SVM regression. We used the {\tt ksvm} function in {\tt R}'s {\tt kernlab} package. This function is used with the following arguments: kernel=``rbfdot'' (for a radial basis kernel function), kpar= ``automatic'' (default value for radial basis functions), tau = 0.01, cross=10, reduced =	TRUE, tol = 0.0001.

	\item SVM Quantile: two-sided interval prediction by two SVM quantile regression models \citep{nonparam_quantile_estimation}. For this purpose, one must build two distinct quantile regression models: a lower $(\frac{1-\beta}{2})$-quantile regression model and an upper $(1-(\frac{1-\beta}{2}))$-quantile regression model. This method's hyper-parameter minimizes the Pin-ball loss function with a $10$-fold CV on the training set. This method is implemented by the {\tt kqr} function in {\tt R}'s {\tt kernlab} package. $kqr$ is used with the following arguments: kernel=``rbfdot'' (radial basis kernel function), kpar= ``automatic'' (default value for radial basis functions), the cost regularization parameter is set between 3.8 and 5, depending on the dataset; its values for the real datasets can be found in Table \ref{tab_dataset_hyper}.

\item  SVM Quantile CV : two-sided interval prediction by two SVM quantile regression models \citep{nonparam_quantile_estimation}. This method is similar to SVM Quantile mentioned above. It also requires a lower $(\frac{1-\beta}{2})$-quantile regression model and an upper $(1-(\frac{1-\beta}{2}))$-quantile regression model. The ``NPQR CV'' hyper-parameters are tuned in a way to find intervals that, in a $10$-fold CV on the training set, have the smallest MIS and satisfy the tuning coverage constraint. We use the {\tt kqr} function in {\tt R}'s {\tt kernlab} package with the following arguments: kernel=``rbfdot'' (radial basis kernel function), kpar= ``automatic'' (default value for radial basis functions), the cost regularization parameter is chosen to lie 0.05 and 0.2, depending on the dataset; its values for the real datasets can be found in Table \ref{tab_dataset_hyper}. Satisfying the tuning coverage constraint on the training set requires us to select small values of cost regularization parameters.

\end{itemize}

Tricube kernel, as in \citep{local_clev_1988}, is the kernel function used in all local linear models above.

\subsubsection{Hyper-parameter tuning for real datasets}
\label{ref_hyper_strategy}
 In a first attempt, datasets are divided into two sub-samples of size $\frac{2}{3}N$ and $\frac{1}{3}N$, where $N$ represents the dataset size. The part containing $\frac{2}{3}$ of observations is used to tune the estimated model's hyper-parameters. Then, all of the instances serve to validate the results using a $10$-cross validation scheme.
 
Once the optimal value of $K_{loess}$ has been found for each dataset, the aforementioned tuning strategy is used to find the prediction intervals hyper-parameters. Linear loess regression uses the $K_{loess}$-nearest neighbors as the bandwidth. This $K_{loess}$ is found by minimizing the $10$-fold cross validation error on the training set; its values for the real datasets can be found in Table \ref{tab_dataset_hyper}. For more details about linear loess see Section \ref{ref_def_loess}. Tables \ref{tab_dataset_hyper}, \ref{tab_method_80} and \ref{tab_method_95} show the hyper-parameters values for the methods described in Section \ref{ref_method_def}.

\subsection{Simulations}
\label{ref_simu}
This part, compares the BOPI methods for local linear regression in Section \ref{ref_algorithm} with the conventional prediction intervals described by Equation~ \eqref{eq_tol_convetional_interval2} on two artificial benchmark data generating process (DGP) Friedman\#1 DGP and Friedman\#2 DGP \citep{friedman_MARS,Breiman_1996}, also available in {\tt mlbench} package of {\tt R}.\\

The results are based on a 3 fold cross-validation schema where $\frac{2}{3}$ of the generated sample is taken as training set and $\frac{1}{3}$ as validation set. The method is applied to these simulated samples and computed results, that is coverage and MIS are reported. For simplicity and based on some experience (see Section \ref{ref_lhne_hyper}) the methods' hyper-parameters are selected as follows: $K_{loess}=100$ as the regression bandwidth; it is constant for the three methods, ${K}^{f}_{lhnpe}=40$ for F-BOPI and $({K}^{min}_{lhnpe}=30,{K}^{max}_{lhnpe}=50)$ for A-BOPI.\\

Friedman\#1 DGP is consisted of $10$ independent predictors, $x=\left(x_1,\ldots,x_{10} \right)$, uniformly distributed over $[0, 1]$ and the response variable is given by:
\begin{equation*}
  Y(x) = 10 sin (\pi x_{1}x_2)+ 20(x_{3}- 0.5)^2 + 10 x_{4}+ 5x_{5} +\varepsilon,~ \varepsilon \sim \mathcal{N}(0 , 1).
\end{equation*}

Friedman\#2 DGP response is is given by:
\begin{eqnarray*}
  Y(x) = \left( x_1^2+ \left( x_2x_3 -  \frac{1}{x_2x_4}  \right) ^2\right)^{\frac{1}{2}} +\varepsilon, \varepsilon \sim \mathcal{N}(0 , 125).
\end{eqnarray*}

It is consisted of $4$ independent predictors, $x=\left(x_1,\ldots,x_4\right)$, uniformly distributed over:
\begin{eqnarray*}
  0 \leq x_1 \leq 100\\
40\pi \leq  x_2 \leq 560\pi \\
0 \leq x_3 \leq 1 \\
1 \leq x_4 \leq 11 
\end{eqnarray*}

Table~\ref{tab_simu_f1covTable} reports coverage and MIS of the tested methods on $500$ samples of $1500$ observations each of them generated by Freidman\#1 with $\beta=0.95$. The whole process is iterated for $\gamma= 0.8,0.9,0.95$ and $0.99$ and the average results and their standard errors are reported. As one could observe, while the conventional method's coverage is always significantly less than the desired content ($\beta=0.95$), the coverage rates of the BOPI methods are not significantly different from the desired content, except when $\gamma=0.8$. These results show that by increasing $\gamma$, coverage increases which by its turn increases MIS substantially. This latter is in the same line of our method's description in Section~\ref{ref_algorithm}. Note that F-BOPI is always more reliable than A-BOPI. Table~\ref{tab_simu_f2covTable} reports the same simulation experiment for Friedman\#2 DGP. The results are generally the same as above except that coverage does not exceed the desired content $\beta$ and coverages' standard error are higher. 

Table \ref{tab_simu_f1Table} and Table \ref{tab_simu_f2Table} report coverage of the tested methods on samples of different size ($1500$ and $3000$) generated respectively by Friedman\#1 and Friedman\#2 with $\beta= 0.8,0.9,0.95$ and $0.99$ and $\gamma=0.99$. While considering Table \ref{tab_simu_f1Table}, one could note that: the coverage of the conventional method is always significantly below the desired content $\beta$, the coverage of F-BOPI is always a bit higher than $\beta$ and the coverage of A-BOPI is always a bit lower than $\beta$. Doubling the sample size from $1500$ observations to $3000$ observations does not alter the results significantly, and as expected, doubling the iteration steps lessens standard errors of coverage rates. Reported coverages in Table \ref{tab_simu_f2Table} are generally lower than the desired content $\beta$. Results shows that all methods yield coverages lower than the desired $\beta$, although the BOPI's coverages are much more closer to $\beta$ than the conventional method's coverage. Doubling the sample size does not alter the results, but doubling the iterations from $500$ to $1000$ steps worsens the results and yield higher standard errors. Since it happens for all the three methods, it could be due to the fact that the local linear regression with chosen hyper-parameters is not a suitable method for capturing Friedman\#2 DGP characteristics.

\subsection{Real Datasets}
\label{ref_datasets}
Eleven benchmark datasets are considered to compare the methods. The datasets are chosen from the UCI repository \citep{dataset_uci}, Delve dataset repository \citep{dataset_delve} and a well-known article on non-parametric regression \citep{silverman_dataset}. The UCI repository datasets are also documented and available in {\tt R}'s {\tt mlbench} package. Datasets were chosen to cover small and moderate size datasets. The dataset sizes vary from $N$ = 103 (Slump) to $N$ = 8192 (Computer), and the number of regressors vary from $p$ = 1 (Motorcycle) to $p$ = 21 (Parkinson1). Some of the datasets contain only numerical variables and some datasets have numerical and categorical variables. Instances with missing values are omitted.\\

These datasets are listed below (where we can find each dataset name in double quotes and its abbreviation in parentheses, their numbers of predictor and number of instances, respectively denoted by $p$ and $N$). Note that some of these datasets have fewer variables than their source because we systematically removed any instances having null values. The ``Parkinsons Telemonitoring'' dataset \citep{dataset_uci} contains two regression variables named ``motor\_UPDRS'' and ``total\_UPDRS''. We considered it as two distinct datasets named ``Parkinson1'' and ``Parkinson2''. Each dataset has one of the ``motor\_UPDRS'' or ``total\_UPDRS'' variables.

\begin{itemize}

\item ``Computer Activity'' (Computer) \citep{dataset_delve}. We used the small variant of this dataset which contains only 12 of the 32 attributes. $N=8192, p=12$. 

\item ``Bank'' (Bank) \citep{dataset_delve}. We used the 8nm variant of this dataset, which just contains 8
of the 32 attributes, and is highly non-linear with moderate noise. $N=8192, p=8$.

\item ``Parkinsons Telemonitoring'' (Parkinson1) \citep{dataset_uci}. We removed ``motor\_UPDRS'' variable and left ``total\_UPDRS'' as the response variable. $N=5875, p=21$. 

\item ``Parkinsons Telemonitoring'' (Parkinson2) \citep{dataset_uci}. We removed the ``total\_UPDRS'' variable and left ``motor\_UPDRS'' as the response variable. $N=5875, p=21$. 

\item ``Abalone'' (Abalone) \citep{dataset_slump}. $N=4177, p=10$.

\item ``Concrete Compressive Strength'' (Concrete) \citep{dataset_concrete}. $N=1030, p=9$.

\item ``Boston Housing'' (Housing) \citep{dataset_uci}. $N=506, p=14$.

\item ``Auto MPG'' (Auto) \citep{dataset_uci}. $N=392, p=8$. 

\item ``CPU''(CPU) \citep{dataset_uci}. $N=209, p=7$.

\item ``Concrete Slump Test'' (Slump) \citep{dataset_slump}. $N=103, p=10$.

\item ``Motorcycle'' (Motorcycle) \citep{silverman_dataset}. $N=133, p=1$.

\end{itemize}

\subsection{Results on Real Datasets}
\label{ref_exp_bech}
The goal of this section is to compare the above-mentioned interval prediction methods based on their strength while providing $\beta$-content prediction intervals. The models are compared based on \textit{reliability and efficiency of their envelope}. A-BOPI and F-BOPI methods  are used to obtain prediction intervals for Local Linear Regression (LLR). Consequently, we compare those methods with the conventional prediction intervals on the local linear regression (Loess Conv.) and other prediction intervals stated above. For this purpose, we will use Tables \ref{tab_loess_tol_bechmark_80_2} and \ref{tab_loess_tol_bechmark_95_2} which compare Loess Conv., A-BOPI, F-BOPI, LS-SVM Conv. and OLS. For each dataset, we build a unique linear loess models, then we apply on this estimated model the BOPI methods and the conventional method. So the only difference between the results obtained with prediction intervals for linear loess models (A-BOPI, F-BOPI and linear loess) is due to their prediction interval method and not the regression model. Tables \ref{tab_loess_tol_bechmark_80_2} and \ref{tab_loess_tol_bechmark_95_2} provide detailed experimental results. For the sake of clarity and ease of interpretation, different charts are drawn to compare all of the prediction intervals. This comparison measures a method's strength, while providing $\beta$-prediction interval with $\beta= 0.8,0.9,0.95$ and $0.99$. 

\subsubsection{Comparing Methods by Tables}
Outliers, limited number of observations and contrast between  assumptions and the true regression function are among potentials cause of errors in the prediction process. These errors occur in a similar manner when estimating the response variable distribution and they increase with $\beta$. For $\beta=0.9,0.95$, and particularly for $\beta=0.99$, it becomes a critical task to find an efficient interval prediction procedure that is able to find an upper bound of $Y(x)$. However these inter-quantiles are the most used ones in machine learning and statistical hypothesis-testing. Hence, we will compare the methods based on their strength, while providing $\beta$-prediction intervals with $\beta= 0.8,0.9,0.95$ and $0.99$. 

Tables \ref{tab_loess_tol_bechmark_80_2} and \ref{tab_loess_tol_bechmark_95_2} display the direct dataset measures explained in Section \ref{ref_comp_section}, for each dataset. These tables compare five different models: Loess Conv., A-BOPI, F-BOPI, LS-SVM Conv. and OLS. For each dataset of the 11 benchmark datasets described in Section \ref{ref_datasets}, we have to estimate 20 models, (5 methods $\times$ 4 $\beta$'s value). 

\subsubsection{Table description}
\label{ref_table_desc}
In Tables \ref{tab_loess_tol_bechmark_80_2} and \ref{tab_loess_tol_bechmark_95_2}, each cell represents a combination of dataset and $\beta$ which displays $F^{0.05}_{\beta,N}$ for the underlying experiment. The $F^{0.05}_{\beta,N}$ column represents the Wilson Score critical value for a binomial proportion test at a significance level of $0.05$ and the alternate hypothesis as $\beta_{pop} < \beta$, where $\beta_{pop}$ denotes the average proportion of response values that are contained in the tested prediction intervals. So, the null hypothesis claims that the constructed prediction intervals cover on average a proportion $\beta_{pop}$ of response values and $\beta_{pop}$ is greater than or equal to the desired proportion $\beta$. In order to test each model reliability, its $coverage$ value is compared with its corresponding critical value $F^{0.05}_{\beta,N}$, and if $coverage < F^{0.05}_{\beta,N}$, it means that the null hypothesis is rejected, with a significance level of at most $0.05$. In such cases, we consider the model as unreliable.\\

Tables \ref{tab_loess_tol_bechmark_80_2} and \ref{tab_loess_tol_bechmark_95_2} illustrate the coverage probability of the five different models stated before. If the computed coverage is less than $F^{0.05}_{\beta,N}$, the model is considered as non-reliable and this is indicated by legend \small{$\blacktriangleleft$} or \small{$\vartriangleleft$} next to the $coverage$. When there is only one non-reliable model, the legend \small{$\blacktriangleleft$} is used and when there are more than one non-reliable model, the legend \small{$\vartriangleleft$} is used. For each experiment the reliable model having the smallest MIS is written in bold. Two-sided paired t-tests at levels $0.05$, $0.01$ and $0.001$ are used to compare the interval size of the two estimated models which find the smallest MIS and are not rejected for the Wilson Score binomial proportion test at level $0.05$ (reliability test). *, ** and *** signs indicate that the two-sided paired t-tests are respectively statistically significant at levels $0.05$, $0.01$ and $0.001$. 
A-BOPI's value for ${K}^{min}_{lhnpe},{K}^{max}_{lhnpe}$ and $\gamma$ and F-BOPI's value for ${K}^{f}_{lhnpe}$ and $\gamma$ are given Tables \ref{tab_method_80} and \ref{tab_method_95}.


\begin{table}[H]
 \begin{center}
\begin{tabular}{|>{\tiny} p{2cm}  | >{\tiny} p{2.5cm} | >{\tiny} p{2.7cm} | >{\tiny} p{2.5cm}  | }
\hline
Dataset & ``SVM Quantile '' C & ``SVM Quantile CV'' C  &``Loss Conv.'' $K_{loess}$  \\
\hline
\hline

Computer   & 5 & 0.15  &   500  \\ 
\hline
Bank  & 4.2 & 0.2 &   500\\ 
\hline
Parkinson1   & 5 & 0.2  &   80  \\ 
\hline
Parkinson2  & 5 & 0.1 &   70\\ 
\hline
Abalone   & 4 & 0.2  &   700 \\ 
\hline
Concrete   & 4 & 0.1 &  80	\\ 
\hline
Housing   & 4.5 & 1 &    60\\ 
\hline
Auto   & 3.8 & 0.2 &    30 \\ 
\hline
CPU  & 4 & 0.2  &  40	\\ 
\hline
Slump   & 4.5 & 0.05  &  	30\\ 
\hline
Motorcycle  & 4 & 0.1  &   15	\\ 
\hline

\end{tabular}
\caption{Hyper-parameter values for prediction intervals with SVM Quantile, SVM Quantile CV and Loss Conv.} \label{tab_dataset_hyper}
\end{center}
\end{table}


\begin{table}[H]
 \begin{center}
\begin{tabular}{| >{\tiny} p{3.2cm}  |  >{\tiny} p{3cm} | >{\tiny} p{4.2cm} | }
\hline
Dataset &   F-BOPI  &  A-BOPI    \\
\hline
\hline

Computer    &   ${K}^{f}_{lhnpe}= 40$, $\gamma=0.8$  & $({K}^{min}_{lhnpe},{K}^{max}_{lhnpe},\gamma)$ = (30, 50, 0.8)   \\ 
\hline

Bank  &   ${K}^{f}_{lhnpe}= 40$, $\gamma=0.8$  & $({K}^{min}_{lhnpe},{K}^{max}_{lhnpe},\gamma)$ = (30, 50, 0.8)   \\ 
\hline

Parkinson1 &   ${K}^{f}_{lhnpe}= 40$, $\gamma=0.9$  & $({K}^{min}_{lhnpe},{K}^{max}_{lhnpe},\gamma)$ = (20, 60, 0.9)  \\ 
\hline

Parkinson2    &  ${K}^{f}_{lhnpe}$= 50, $\gamma=0.9$ & $({K}^{min}_{lhnpe},{K}^{max}_{lhnpe},\gamma)$ = (30, 60, 0.9)  \\ 
\hline

Abalone     & ${K}^{f}_{lhnpe}$= 40, $\gamma=0.7$  & $({K}^{min}_{lhnpe},{K}^{max}_{lhnpe},\gamma)$ = (30, 50, 0.7)    \\ 
\hline

Concrete  &  ${K}^{f}_{lhnpe}$= 35, $\gamma=0.5$  & $({K}^{min}_{lhnpe},{K}^{max}_{lhnpe},\gamma)$ = (20, 60, 0.9)   \\ 
\hline
Housing  &   ${K}^{f}_{lhnpe}$= 40, $\gamma=0.9$   &  $({K}^{min}_{lhnpe},{K}^{max}_{lhnpe},\gamma)$ = (30, 55, 0.9)  \\ 
\hline
Auto &   ${K}^{f}_{lhnpe}$= 50, $\gamma=0.9$  & $({K}^{min}_{lhnpe},{K}^{max}_{lhnpe},\gamma)$ = (30, 60, 0.9)     \\ 
\hline
CPU   &    ${K}^{f}_{lhnpe}$= 40, $\gamma=0.9$  &  $({K}^{min}_{lhnpe},{K}^{max}_{lhnpe},\gamma)$ = (20, 50, 0.9)   \\ 
\hline
Slump   &    ${K}^{f}_{lhnpe}$= 20, $\gamma=0.5$   &  $({K}^{min}_{lhnpe},{K}^{max}_{lhnpe},\gamma)$ = (15, 30, 0.5)   \\ 
\hline
Motorcycle  & ${K}^{f}_{lhnpe}$= 35, $\gamma=0.55$ &  $({K}^{min}_{lhnpe},{K}^{max}_{lhnpe},\gamma)$ = (20, 35, 0.55)    \\ 

\hline

\end{tabular}
\caption{Hyper-parameter values for BOPI methods with $\beta= 0.8,0.9$ .} \label{tab_method_80}
\end{center}
\end{table}


\begin{table}[H]
 \begin{center}
\begin{tabular}{| >{\tiny} p{3.2cm} |   >{\tiny} p{3cm} | >{\tiny} p{4.2cm} | }
\hline
Dataset &  F-BOPI  &  A-BOPI    \\
\hline
\hline

Computer &   ${K}^{f}_{lhnpe}= 40$, $\gamma=0.9$  & $({K}^{min}_{lhnpe},{K}^{max}_{lhnpe},\gamma)$ = (30, 50, 0.9)   \\ 
\hline

Bank &   ${K}^{f}_{lhnpe}= 40$, $\gamma=0.9$  & $({K}^{min}_{lhnpe},{K}^{max}_{lhnpe},\gamma)$ = (30, 50, 0.9)   \\ 
\hline

Parkinson1 &   ${K}^{f}_{lhnpe}= 40$, $\gamma=0.99$  & $({K}^{min}_{lhnpe},{K}^{max}_{lhnpe},\gamma)$ = (20, 60, 0.99) \\ 
\hline

Parkinson2  &  ${K}^{f}_{lhnpe}$= 50, $\gamma=0.99$ & $({K}^{min}_{lhnpe},{K}^{max}_{lhnpe},\gamma)$ = (30, 60, 0.99)  \\ 
\hline

Abalone & ${K}^{f}_{lhnpe}$= 40, $\gamma=0.9$  & $({K}^{min}_{lhnpe},{K}^{max}_{lhnpe},\gamma)$ = (30, 50, 0.9)    \\ 
\hline
Concrete & ${K}^{f}_{lhnpe}$= 35, $\gamma=0.55$  & $({K}^{min}_{lhnpe},{K}^{max}_{lhnpe},\gamma)$ = (20, 60, 0.9)   \\ 
\hline
Housing& ${K}^{f}_{lhnpe}$= 40, $\gamma=0.9$   &  $({K}^{min}_{lhnpe},{K}^{max}_{lhnpe},\gamma)$ = (30, 50, 0.99)  \\ 
\hline
Auto & ${K}^{f}_{lhnpe}$= 50, $\gamma=0.99$  & $({K}^{min}_{lhnpe},{K}^{max}_{lhnpe},\gamma)$ = (30, 60, 0.99)     \\ 
\hline
CPU &  ${K}^{f}_{lhnpe}$= 40, $\gamma=0.99$  &  $({K}^{min}_{lhnpe},{K}^{max}_{lhnpe},\gamma)$ = (20, 50, 0.99)   \\ 
\hline
Slump     &  ${K}^{f}_{lhnpe}$= 20, $\gamma=0.9$   &  $({K}^{min}_{lhnpe},{K}^{max}_{lhnpe},\gamma)$ = (15, 30, 0.9)   \\ 
\hline
Motorcycle &        ${K}^{f}_{lhnpe}$= 35, $\gamma=0.7$ &  $({K}^{min}_{lhnpe},{K}^{max}_{lhnpe},\gamma)$ = (20, 35, 0.7)    \\ 
\hline

\end{tabular}
\caption{Hyper-parameter values for BOPI methods with $\beta= 0.95,0.99$.} \label{tab_method_95}
\end{center}
\end{table}


\begin{center}
\begin{table}[htbp]
\begin{tabular}{| >{\small} p{1.42cm} | >{\tiny} p{1.3cm} | >{\tiny} c | >{\tiny} c|   >{\tiny} p{0.68cm} || >{\tiny} c |  >{\tiny} c | >{\tiny} p{0.68cm}|}
\hline

\multirow{3}{*}{ Dataset}  & \multirow{3}{*}{  \small{Method}}  & \multicolumn{3}{  c||}{ $\beta=0.8$ }    & \multicolumn{3}{ c|}{$\beta=0.9$}    \\
\cline{3-8}
& & \multirow{2}{*}{ $Cover$ }  & \multirow{2}{*}{$MIS$ ($\sigma_{is}$)} & \multirow{2}{*}{ $F^{0.05}_{0.8,N}$} &   \multirow{2}{*}{$Cover$} &  \multirow{2}{*}{$MIS$ ($\sigma_{is}$)} & \multirow{2}{*}{$F^{0.05}_{0.9,N}$} \\
 &  &  &   &  &   &  &  \\
 \hline \hline


\hline 
\multirow{2}{2pt}{Computer}  & Loess Conv.  &   82.99  &  6.64 (0.03) & \multirow{4}{*}{79.27} &  89.36 \small{$\blacktriangleleft$} & 8.52 (0.04) & \multirow{4}{*}{89.45}\\ 

\cline{2-4}\cline{6-7}
& LS-SVM Conv.  &    93.31  & 25.19 (0.16) &   &  96.56  & 32.34 (0.28) &    \\

\cline{2-4}\cline{6-7}
& OLS  &    94  & 12.77 (0.16) &   &  96.56  & 16.39 (0.21) &    \\

\cline{2-4}\cline{6-7}
& F-BOPI  &     85.22 & 6.8  (2.9)  &   &  92.13  & 8.73 (3.72) &    \\
\cline{2-4}\cline{6-7}
& A-BOPI  & 82.74   &  \textbf{6.34***} (2.72) &  & 90.31 & \textbf{8.13*** }(3.49) &    \\
\cline{2-4}\cline{6-7}


\hline \hline
\multirow{2}{2pt}{Bank}  & Loess Conv.  &   83.05  &  0.05 (0.001) & \multirow{4}{*}{79.27} &  89.94  & 0.06 (0.001) & \multirow{4}{*}{89.45}\\ 

\cline{2-4}\cline{6-7}
& LS-SVM Conv.  &      87.12  & 0.04 (0.001) &   &  91.66  & 0.06 (0.001) &    \\

\cline{2-4}\cline{6-7}
& OLS  &    86.85 & 0.08 (0.0001) &   &  93.24  & 0.1 (0.0001) &    \\

\cline{2-4}\cline{6-7}
& F-BOPI  &     84.88 & 0.04  (0.01)  &   &  92.29  & 0.06 (0.02) &    \\
\cline{2-4}\cline{6-7}
& A-BOPI  & 82.38   &  \textbf{0.04} (0.01) &  & 90.48 & \textbf{0.05***}(0.02) &    \\


\hline \hline
\multirow{2}{2pt}{Parkinson1 }  & Loess Conv.  &   90.93  &  6.81 (0.16) & \multirow{4}{*}{79.14	} &  90.93  & 6.81 (0.16) & \multirow{4}{*}{89.35}\\

\cline{2-4}\cline{6-7}
& LS-SVM Conv.  &   83.33  & 13.89 (0.012) &   &  89.18 \small{$\blacktriangleleft$} & 17.83 (0.15) &    \\


\cline{2-4}\cline{6-7}
& OLS  &    80.42 & 23.79 (0.14) &   &  90.93  & 30.54 (0.0001) &    \\

\cline{2-4}\cline{6-7}
& F-BOPI &     91.55 & 5.48  (4.4)  &   &  94.88  & 7.04 (5.64) &    \\

\cline{2-4}\cline{6-7}
& A-BOPI  & 88.55   &  \textbf{4.39***} (3.78) &  & 92.81 & \textbf{5.64*** }(4.85) &    \\


\hline  \hline
\multirow{2}{2pt}{Parkinson2 }  & Loess Conv.  & 91.48  &  5.21 (0.14) & \multirow{4}{*}{79.14} &  93.86  & 6.69 (0.19) & \multirow{4}{*}{89.35}\\

\cline{2-4}\cline{6-7}
& LS-SVM Conv.  &   83.14  & 9.96 (0.1) &   &  89.49  & 12.79 (0.13) &    \\


\cline{2-4}\cline{6-7}
& OLS  &    77.64 \small{$\blacktriangleleft$} & 18.49 (0.1) &   &  91.15  & 23.74 (0.13) &    \\

\cline{2-4}\cline{6-7}
& F-BOPI  &    91.46   & 4.2  (3.22) &   &  94.64  & 5.4 (4.14) &    \\

\cline{2-4}\cline{6-7}
& A-BOPI  &  89.08   & \textbf{3.52*** }(2.95) & & 93.03  & \textbf{4.52***} (3.79)   & \\


\hline  \hline
\multirow{2}{2pt}{Abalone }  & Loess Conv.  & 83.69 &  5.14 (0.02) & \multirow{4}{*}{78.98} & 90.03   & 6.6 (0.02) & \multirow{4}{*}{89.23} \\

\cline{2-4}\cline{6-7}
& LS-SVM Conv.  &      86.54  & 5.53 (0.02) &   &  91.59  & 7.1 (0.03) &    \\


\cline{2-4}\cline{6-7}
& OLS  &    85.46 & 5.63 (0.04) &   &  91.18  & 7.22 (0.05) &    \\

\cline{2-4}\cline{6-7}
& F-BOPI&    84.4   &  5.19 (1.76) &   &  91.75  & 6.67 (2.26) &    \\
\cline{2-4}\cline{6-7}
& A-BOPI  &  82.1  & \textbf{4.81***}  (1.65)  &  & 89.91   & \textbf{6.18***} (2.12) &  \\



\hline  
\multirow{2}{2pt}{Concrete}  & Loess Conv.  & 81.06  & \textbf{17.1} (0.23)  & \multirow{4}{*}{77.94}&  88.73   & 21.95(0.3) & \multirow{4}{*}{88.46} \\

\cline{2-4}\cline{6-7}
& LS-SVM Conv.  &      82.22 & 17.08 (0.26) &   &  90.19  & 21.93 (0.33) &    \\


\cline{2-4}\cline{6-7}
& OLS  &    80.28 & 26.8 (0.16) &   &   89.7  & 34.41 (0.21) &    \\

\cline{2-4}\cline{6-7}
& F-BOPI &  82.61  & 16.76  (5.73) &   &  91.45  &  \textbf{21.52*} (7.36)  &    \\


\cline{2-4}\cline{6-7}
& A-BOPI & 83.68  & 17.03 (5.91) &   & 93  & 22.2 (7.59) &   \\


\hline   \hline 
\multirow{2}{2pt}{Housing }  & Loess Conv.  & 86.18   &  8.13 (0.32) & \multirow{4}{*}{76.67}  & 91.31   & 10.43 (0.41) & \multirow{4}{*}{87.5}\\

\cline{2-4}\cline{6-7}
& LS-SVM Conv.  &      92.08 & 10.17 (0.41) &   &  94.46   & 13.05 (0.53) &    \\


\cline{2-4}\cline{6-7}
& OLS  &    86.16  & 12.36 (0.25) &   &  92.67  & 15.87 (0.32) &    \\

\cline{2-4}\cline{6-7}
& F-BOPI &    87.97   &  8.67 (3.31) &   &  92.7   & 11.14 (4.25) &    \\

\cline{2-4}\cline{6-7}
& A-BOPI  & 84.59  &  \textbf{7.8**} (2.8)  &  &  91.72  & \textbf{10.01**} (3.6)  & \\

\hline   \hline
\multirow{2}{2pt}{Auto }  & Loess Conv.  & 84.96  &  7.33 (0.3) & \multirow{4}{*}{77.07} & 90.57   & 9.41 (0.38) & \multirow{4}{*}{87.8} \\

\cline{2-4}\cline{6-7}
& LS-SVM Conv.  &      85.72  & 7.04 (0.16) &   &  93.37  & 9.03 (0.21) &    \\


\cline{2-4}\cline{6-7}
& OLS  &   83.16   & 8.64 ( 0.1) &   &  91.82   & 11.1 (0.13) &    \\

\cline{2-4}\cline{6-7}
& F-BOPI  &    87.77  &  7.92 (3.21) &   &  94.15  & 10.17 (4.12) &    \\

\cline{2-4}\cline{6-7}
& A-BOPI  & 83.17  & \textbf{7.03} (2.83)   &  &  90.83  & \textbf{9.02* }(3.64)  & \\
 

\hline  \hline
\multirow{2}{2pt}{CPU  }  & Loess Conv.  & 86.09  &  123.68 (15.13) & \multirow{4}{*}{75.44} & 91.37   & 158.75 (19.42) & \multirow{4}{*}{86.58} \\
\cline{2-4}\cline{6-7}

& LS-SVM Conv.  &      96.16  &  302.38 (17.58) &   &  96.63   & 388.11 (22.56) &    \\


\cline{2-4}\cline{6-7}
& OLS  &   89.47    & 156.15 (9) &   &  93.78   & 200.71 (11.57) &    \\

\cline{2-4}\cline{6-7}
& F-BOPI & 85.16 &  88.07 (64.23) &  &  91.4  & 113.04 (82.44) & \\

\cline{2-4}\cline{6-7}
& A-BOPI   & 80.37  &  \textbf{78.49**} (59.2)  &  &  88.97  & \textbf{100.75**} (20.89) &  \\

\hline

 \end{tabular}

\end{table}
\end{center}



\begin{center}
\begin{table}[htbp]
\begin{tabular}{| >{\small} p{1.42cm} | >{\tiny} p{1.3cm} | >{\tiny} c | >{\tiny} c|   >{\tiny} p{0.68cm} || >{\tiny} c |  >{\tiny} c | >{\tiny} p{0.68cm}|}
\hline

\multirow{3}{*}{ Dataset}  & \multirow{3}{*}{  \normalsize{Method}}  & \multicolumn{3}{  c||}{$\beta=0.8$ }    & \multicolumn{3}{ c|}{$\beta=0.9$}    \\
\cline{3-8}
& & \multirow{2}{*}{ $Cover$ }  & \multirow{2}{*}{$MIS$ ($\sigma_{is}$)} & \multirow{2}{*}{ $F^{0.05}_{0.8,N}$} &   \multirow{2}{*}{$Cover$} &  \multirow{2}{*}{$MIS$ ($\sigma_{is}$)} & \multirow{2}{*}{$F^{0.05}_{0.9,N}$} \\
 &  &  &   &  &   &  &  \\
 \hline \hline


\hline \hline
\multirow{2}{2pt}{Slump }  & Loess Conv.  & 85.72 &  4.7 (0.4) & \multirow{4}{*}{73.51} & 89.54  & 6.03 (0.52) & \multirow{4}{*}{85.13}\\

\cline{2-4}\cline{6-7}
& LS-SVM Conv.  &      87.36  & 7.55 (0.68) &   &  92.18    & 9.69 (0.88) &    \\


\cline{2-4}\cline{6-7}
& OLS  & 84.63     &  6.72 (  0.23) &   &   89.45   & 8.66 (0.3) &    \\

\cline{2-4}\cline{6-7}
& F-BOPI &   85.72   &  4.85 (1.41) &   &  88.54  & 6.23 (1.81) &    \\

\cline{2-4}\cline{6-7}
& A-BOPI  & 83.81  &  \textbf{4.32**} (1.24) &  & 87.63   & \textbf{5.55**} (1.6)  & \\


\hline  
\multirow{2}{2pt}{Motorcycle}  & Loess Conv.  & 78.84 &  57.82 (1.22) & \multirow{4}{*}{74.29} & 89.5  & 74.21 (1.57) & \multirow{4}{*}{85.72}\\

\cline{2-4}\cline{6-7}
& LS-SVM Conv.  &      84  & 64.61 (4.13) &   &  90.16    & 82.92 (5.3) &    \\


\cline{2-4}\cline{6-7}
& OLS  &  78.95    & 120.44 (2.86) &   & 88.67   &  154.93  (2.39) &    \\

\cline{2-4}\cline{6-7}
& F-BOPI &  88.67   &  65.7	 (17.35) &   &  94.77  & 73.36 (28.63) &    \\

\cline{2-4} \cline{6-7}
& A-BOPI  & 85.6 &  \textbf{57.16} (22.31)  &  & 94 & \textbf{72.82} (32.44)  & \\
      
 \hline

\end{tabular}
\caption{Prediction intervals for the linear loess regression model with prediction intervals built on benchmark datasets with desired contents $\beta=0.8$ and $\beta=0.9$. Loess Conv. described by Equation (\ref{eq_tol_convetional_interval2}), F-BOPI and A-BOPI described in Section \ref{ref_pred_loess_app} are used to obtain prediction intervals on the same linear loess model. If the computed coverage probability is less than $F^{0.05}_{\beta,N}$ the model is considered as non-reliable, and it is indicated by legend \small{$\blacktriangleleft$} or \small{$\vartriangleleft$} next to the $coverage$. When there is only one non-reliable model, the legend \small{$\blacktriangleleft$} is used and when there are more than one non-reliable model, the legend \small{$\vartriangleleft$} is used. For each experiment the reliable model having the smallest MIS is written in bold. The *, ** and *** signs indicate that the difference between the bold MIS (smallest reliable) and the second smallest reliable MIS were statistically significant at respectively $0.05$, $0.01$ and $0.001$ level with a two-sided paired t-test.}
\label{tab_loess_tol_bechmark_80_2}
\end{table}
\end{center}


\begin{center}
\begin{table}[htbp]
\begin{tabular}{| >{\small} p{1.42cm} | >{\tiny} p{1.3cm} | >{\tiny} c | >{\tiny} c|   >{\tiny} p{0.68cm} || >{\tiny} c |  >{\tiny} c | >{\tiny} p{0.68cm}|}
\hline

\multirow{3}{*}{  Dataset}  & \multirow{3}{*}{  \normalsize{Method}}  & \multicolumn{3}{ c||}{ $\beta=0.95$ }    & \multicolumn{3}{ c|}{$\beta=0.99$}    \\
\cline{3-8}
& & \multirow{2}{*}{ $Cover$ }  & \multirow{2}{*}{$MIS$ ($\sigma_{is}$)} & \multirow{2}{*}{$F^{0.05}_{0.95,N}$ } &   \multirow{2}{*}{$Cover$} &  \multirow{2}{*}{$MIS$ ($\sigma_{is}$)} & \multirow{2}{*}{$F^{0.05}_{0.99,N}$} \\
 &  &  &   &  &   & &  \\
 \hline \hline
 

\hline 
\multirow{2}{2pt}{Computer  }  & Loess Conv.  &   93.1 \small{$\blacktriangleleft$}  &  10.15 (0.05) & \multirow{4}{*}{94.60} &  97.15 \small{$\vartriangleleft$}  & 13.34 (0.07) & \multirow{4}{*}{98.81}\\ 

\cline{2-4}\cline{6-7}
& LS-SVM Conv.  &     97.39 & 19.53  (0.25)  &   &  98.5 \small{$\vartriangleleft$}  & 25.67 (0.33) &    \\

\cline{2-4}\cline{6-7}
& OLS  &    96.72  & 38.54 (0.34) &   &  97.27 \small{$\vartriangleleft$}  & 50.65 (0.45) &    \\

\cline{2-4}\cline{6-7}
& F-BOPI  &     96.41 & 10.99  (4.68)  &   &  98.64 \small{$\vartriangleleft$} & 14.44 (6.15) &    \\
\cline{2-4}\cline{6-7}
& A-BOPI  & 95.46   &  \textbf{10.22} (4.38) &  & 98.09 \small{$\vartriangleleft$}& 13.43 (5.76) &    \\


\hline \hline
\multirow{2}{2pt}{Bank}  & Loess Conv.  &   93.65 \small{$\vartriangleleft$} &  0.08 (0.001) & \multirow{4}{*}{94.60} &  97.39 \small{$\vartriangleleft$} & 0.1 (0.001) & \multirow{4}{*}{98.81}\\ 

\cline{2-4}\cline{6-7}
& LS-SVM Conv.  &   94.13 \small{$\vartriangleleft$} & 0.07 (0.001) &   &  96.99 \small{$\vartriangleleft$} & 0.09 (0.001) &    \\

\cline{2-4}\cline{6-7}
& OLS  &    95.84  & 0.12 (0.0001) &   &  97.99 \small{$\vartriangleleft$} & 0.16 (0.0001) &    \\

\cline{2-4}\cline{6-7}
& F-BOPI  &     96.27 & 0.07  (0.02)  &   &  98.44 \small{$\vartriangleleft$} & 0.1 (0.03) &    \\
\cline{2-4}\cline{6-7}
& A-BOPI  & 95.3   &  \textbf{0.07} (0.02) &  & 97.93 \small{$\vartriangleleft$} & 0.09 (0.03) &    \\


\hline \hline
\multirow{3}{2pt}{Parkinson1}  & Loess Conv.  & 95.26   &  10.41 (0.24) &  \multirow{4}{*}{94.53} &  96.96 \small{$\vartriangleleft$} & 13.68 (0.32) &  \multirow{4}{*}{98.78} \\
\cline{2-4}\cline{6-7}

& LS-SVM Conv.  &   92.62 \small{$\blacktriangleleft$} & 21.25 (0.18) &   &  97.4 \small{$\vartriangleleft$} & 27.93 (0.24) &    \\
\cline{2-4}\cline{6-7}

& OLS  &    95  & 36.39 (0.21) &   &  99.78  & 47.84 (0.28) &    \\

\cline{2-4}\cline{6-7}
& F-BOPI  &    97.61   &  9.63 (7.7) &  &  98.74  \small{$\vartriangleleft$} &  12.66 (10.2) &  \\

\cline{2-4}\cline{6-7}
& A-BOPI  & 96.31  &  \textbf{7.72***} (6.51)  &  &  98.08 \small{$\vartriangleleft$} &  10.15 (8.56) &      \\



\hline  \hline
\multirow{2}{2pt}{Parkinson2  }  & Loess Conv. & 95.46   & 7.97 (0.22) &  \multirow{4}{*}{94.53} & 97.04  \small{$\vartriangleleft$} &  10.48 (0.3) &  \multirow{4}{*}{98.78} \\

\cline{2-4}\cline{6-7}
& LS-SVM Conv.  &   93.38 \small{$\blacktriangleleft$} & 15.24 (0.15) &   &  97.42 \small{$\vartriangleleft$} & 20.03 (0.2) &    \\


\cline{2-4}\cline{6-7}
& OLS  &    97.08  & 28.29 (0.16) &   &  99.91  & 37.18 (0.21) &    \\

\cline{2-4}\cline{6-7}
& F-BOPI  &  97.4  &  7.26 (5.57) &   &  98.64 \small{$\vartriangleleft$} & 9.54 (7.32) &    \\

\cline{2-4}\cline{6-7}
& A-BOPI  &  96.35  & \textbf{6.1*** }(5.07)  & & 98.13  \small{$\vartriangleleft$} & 8.02 (6.76)  &   \\


\hline  \hline
\multirow{2}{2pt}{Abalone }  & Loess Conv.  & 93.17  \small{$\vartriangleleft$}   &  7.86 (0.03) & \multirow{4}{*}{94.45} & 96.95 \small{$\vartriangleleft$}  & 10.33 (0.04) & \multirow{4}{*}{98.74} \\
\cline{2-4}\cline{6-7}
& LS-SVM Conv.  &      93.96 \small{$\vartriangleleft$} & 8.47 (0.04) &   &  96.76 \small{$\vartriangleleft$}  & 11.13 (0.05) &    \\

\cline{2-4}\cline{6-7}
& OLS  &    93.84 \small{$\vartriangleleft$}  & 8.61 (0.06) &   &  97.14 \small{$\vartriangleleft$} & 11.32 (0.08) &    \\

\cline{2-4}\cline{6-7}
& F-BOPI &    96.09   &  8.72 (2.95) &   &  98.15 \small{$\vartriangleleft$} & 11.47 (3.88) &    \\
\cline{2-4}\cline{6-7}
& A-BOPI  &  94.89  & \textbf{8.06***}  (2.74)  &  & 97.64 \small{$\vartriangleleft$}  & 10.59 (3.61) &  \\



\hline 
\multirow{2}{2pt}{Concrete }  & Loess Conv.  & 94.36   &  26.15 (0.35)  & \multirow{4}{*}{93.88} &  98.82 & 34.37 (0.47)  & \multirow{4}{*}{98.49} \\
\cline{2-4}\cline{6-7}
& LS-SVM Conv.  &      94.36  & 26.13 (0.4) &   &  97.46 \small{$\blacktriangleleft$}  & 34.34 (0.52) &    \\


\cline{2-4}\cline{6-7}
& OLS  &    94.65  &  41.01  (0.25) &   &  99.21  & 53.94 (0.33) &    \\

\cline{2-4}\cline{6-7}
& F-BOPI  &    95.62  & \textbf{25.64*}  (8.77)  &   &  99.02 & \textbf{33.7*} (11.53) &    \\

\cline{2-4}\cline{6-7}
& A-BOPI & 95.72 & 26.46 (9.04) &   & 99.02  & 34.77 (11.88) &  \\

 
\hline   \hline
\multirow{2}{2pt}{Housing }  & Loess Conv.  & 94.26 &  \textbf{12.43 ***} (0.48) & \multirow{4}{*}{93.18} & 97.42 \small{$\vartriangleleft$} & 16.34 (0.64) & \multirow{4}{*}{98.17} \\
\cline{2-4}\cline{6-7}
& LS-SVM Conv.  &      95.64  & 15.55 (0.63) &   &  96.83 \small{$\vartriangleleft$}  & 20.44 (0.83) &    \\


\cline{2-4}\cline{6-7}
& OLS  &    95.24  &  18.93  (0.38) &   &  97.03 \small{$\vartriangleleft$} & 24.92 (0.5) &    \\

\cline{2-4}\cline{6-7}
& F-BOPI &   95.45  &  13.27 (5.07) &   & 98.61  & \textbf{17.44*}(6.66) &    \\

\cline{2-4}\cline{6-7}
& A-BOPI  & 96.24 & 13.8 (5.01) &  & 98.61  & 18.14 (6.58) &  \\

\hline  \hline
\multirow{2}{2pt}{Auto }  & Loess Conv.  & 93.88  & 11.21 (0.45) & \multirow{4}{*}{93.4} & 97.2 \small{$\vartriangleleft$} & 14.74 (0.6) & \multirow{4}{*}{98.27} \\
\cline{2-4}\cline{6-7}
& LS-SVM Conv.  &      95.92  & \textbf{10.76***} (0.25) &   &  98.73   & \textbf{14.15***} (0.33) &    \\


\cline{2-4}\cline{6-7}
& OLS  &    94.89  &  13.23  (0.16) &   &  97.44 \small{$\vartriangleleft$} & 17.43 (0.21) &    \\

\cline{2-4}\cline{6-7}
& F-BOPI  &   97.45   & 13.68  (5.54) &   &  98.71 & 17.98 (7.28) &    \\

\cline{2-4}\cline{6-7}
& A-BOPI  &  96.95 & 12.2 (4.87)  &  & 98.71 & 16.03 (6.4) &  \\


\hline  \hline
\multirow{2}{2pt}{CPU  }  & Loess Conv.  & 92.82  &  189.16 (23.15) & \multirow{4}{*}{92.52} &  96.66 \small{$\vartriangleleft$} & 248.6 (30.42) & \multirow{4}{*}{97.86} \\

\cline{2-4}\cline{6-7}
& LS-SVM Conv.  &      97.11  & 462.46 (26.89) &   &  97.59 \small{$\vartriangleleft$}   & 607.77 (35.34) &    \\


\cline{2-4}\cline{6-7}
& OLS  &    94.25  &  239.54  (13.81) &   &  96.16 \small{$\vartriangleleft$} & 316.03 (18.22) &    \\

\cline{2-4}\cline{6-7}
& F-BOPI  & 96.16 & 154.67 (112.8) &  & 98.07  & \textbf{203.27} (148.24) &  \\

\cline{2-4}\cline{6-7}
& A-BOPI  & 94.25 & \textbf{137.68 }(101.75)  & & 96.64  \small{$\vartriangleleft$} & 180.95 (133.72) & \\


\hline

\end{tabular}
\end{table}
\end{center}

\begin{center}
\begin{table}[htbp]
\begin{tabular}{| >{\small} p{1.42cm} | >{\tiny} p{1.3cm} | >{\tiny} c | >{\tiny} c|   >{\tiny} p{0.68cm} || >{\tiny} c |  >{\tiny} c | >{\tiny} p{0.68cm}|}
\hline

\multirow{3}{*}{  Dataset}  & \multirow{3}{*}{  \normalsize{Method}}  & \multicolumn{3}{ c||}{ $\beta=0.95$ }    & \multicolumn{3}{ c|}{$\beta=0.99$}    \\
\cline{3-8}
& & \multirow{2}{*}{ $Cover$ }  & \multirow{2}{*}{$MIS$ ($\sigma_{is}$)} & \multirow{2}{*}{$F^{0.05}_{0.95,N}$ } &   \multirow{2}{*}{$Cover$} &  \multirow{2}{*}{$MIS$ ($\sigma_{is}$)} & \multirow{2}{*}{$F^{0.05}_{0.99,N}$} \\
 &  &  &   &  &   & &  \\
 \hline \hline
 
\multirow{2}{2pt}{Slump}  & Loess Conv.  &  91.45 & \textbf{7.19***} (0.62) & \multirow{4}{*}{91.46} & 94.45 \small{$\blacktriangleleft$} & 9.45 (0.82) & \multirow{4}{*}{97.38}\\

\cline{2-4}\cline{6-7}
& LS-SVM Conv.  &      96  & 11.55 (1.05) &   &  98    & 15.18 (1.38) &    \\


\cline{2-4}\cline{6-7}
& OLS  &    93.36  &  10.36  (0.36) &   &  99  & 13.73 (0.47) &    \\

\cline{2-4}\cline{6-7}
& F-BOPI  &   97.18  & 9.35  (2.72) &   &  98.09 & 12.29 (3.57) &    \\

\cline{2-4}\cline{6-7}
& A-BOPI  & 96.27  & 8.16 (2.25)  & &  98.09  & \textbf{10.73***} (2.96) & \\


\hline  
\multirow{2}{2pt}{Motorcycle }  & Loess Conv.  & 93.23 &  \textbf{88.43} (1.87) & \multirow{4}{*}{91.89} & 98.51 & \textbf{116.22} (2.46) & \multirow{4}{*}{97.58}\\

\cline{2-4}\cline{6-7}
& LS-SVM Conv.  &      92.46  & 98.81 (6.32) &   &  97.74    & 129.86 (8.3) &    \\


\cline{2-4}\cline{6-7}
& OLS  &    93.18  &  185.07  (2.86) &   &  99.23  & 244.68 (3.79) &    \\

\cline{2-4}\cline{6-7}
& F-BOPI&  96.31  &  105.7	 (27.92) &   &  100  & 138.91 (36.7) &    \\

\cline{2-4}\cline{6-7}
& A-BOPI  & 96.31  &  92.59 (35.9)  &  & 100  & 121.69 (47.18)  & \\

\hline
\end{tabular}
\caption{Prediction intervals for the linear loess regression model with prediction intervals built on benchmark datasets with desired content $\beta=0.95$ and $\beta=0.99$. Loess Conv. described by Equation (\ref{eq_tol_convetional_interval2}), F-BOPI and A-BOPI described in Section \ref{ref_pred_loess_app} are used to obtain prediction intervals on the same linear loess model. If the computed coverage probability is less than $F^{0.05}_{\beta,N}$ the model is considered as non-reliable, and it is indicated by legend \small{$\blacktriangleleft$} or \small{$\vartriangleleft$} next to the $coverage$. When there is only one non-reliable model, the legend \small{$\blacktriangleleft$} is used and when there are more than one non-reliable model, the legend \small{$\vartriangleleft$} is used. For each experiment the reliable model having the smallest MIS is written in bold. The *, ** and *** signs indicate that the difference between the bold MIS (smallest reliable) and the second smallest reliable MIS were statistically significant at respectively $0.05$, $0.01$ and $0.001$ level with a two-sided paired t-test.}
 \label{tab_loess_tol_bechmark_95_2}
\end{table}
\end{center}


\subsubsection{Table commentaries}
By looking at Tables \ref{tab_loess_tol_bechmark_80_2} and \ref{tab_loess_tol_bechmark_95_2}, one could see that almost all method's reliability test (except one case for OLS) are not rejected for $\beta=0.8$ on all benchmark datasets. When the desired proportion is $0.8$, A-BOPI is the most efficient method (in EGSD sense) and then F-BOPI and Loess Conv. are approximately equally efficient. LS-SVM Conv. and OLS result respectively the largest and the second largest Mean Interval Size (MIS).

When $\beta$ is equal to 0.9, all methods, except two cases (Loess Conv. for Computer dataset and LS-SVM Conv. for Parkinson1 dataset), result in reliable prediction intervals. In these two cases, the obtained coverage is smaller than the $F^{0.05}_{0.9,N}$ column which represents the Wilson Score critical value for a binomial proportion test at a significance level of $0.05$ described in Section \ref{ref_table_desc}. This means that the null hypothesis of the binomial proportion test of $\beta_{pop} \geq 0.9$, where $\beta_{pop}$ denotes the average proportion of response value that are contained in the tested prediction intervals, are rejected for two cases (Loess Conv. for Computer dataset and LS-SVM Conv. for Parkinson1 dataset), at a significance level of $0.05$.

 When the desired proportion increases to $0.95$, results are the same for all of the estimated models by A-BOPI and F-BOPI; their reliability tests are not rejected (they are reliable). However Loess Conv. and LS-SVM Conv. have a lower reliability when $\beta= {0.8,0.9}$, and they respectively have $coverage < F^{0.05}_{0.95,N}$ in three and four cases. The reliability test on the conventional methods, Loess Conv. and LS-SVM Conv., are rejected (lead to unreliable estimated models) on larger datasets. So the failure of rejection on the smaller datasets may be caused by a lack of sufficient observations rather than reliable prediction intervals (the power of the reliability test increases with sample size). 
 
When it comes to the comparison of MIS, A-BOPI remains the most efficient\footnote{Note that we compare the Mean Interval Size (MIS) of reliable models, because it makes not a lot of sense to compare the MIS of a reliable method with a non-reliable one. It generally obtains a reliable estimated model with the smallest MIS and its difference with the second smallest MIS (which must be a reliable estimated model) is usually statistically significant at $0.05$ level with a two-sided paired t-test. Moreover, the EGSD measure (introduced in Section \ref{ref_egsd_def}) are used in Section \ref{ref_charts} to compare the efficiency of interval prediction methods indecent of their reliability results. } method, Loess Conv. becomes the second efficient method and F-BOPI and LS-SVM Conv. produce similar results. The desired proportion of $0.99$ is the most difficult one to satisfy. In this case, while F-BOPI is the most reliable method and A-BOPI and OLS are the second most reliable methods, F-BOPI and A-BOPI intervals are much tighter in mean than the OLS ones.\\

We conclude this comment by stressing that all the models given by A-BOPI and F-BOPI are reliable for $\beta=\{0.8, 0.9, 0.95\}$. These methods also provide tighter reliable estimated models than others on the inquired datasets. By looking at Tables \ref{tab_loess_tol_bechmark_80_2} and \ref{tab_loess_tol_bechmark_95_2}, one could see that A-BOPI usually provides a reliable model with the smallest MIS and F-BOPI usually gives a larger coverage than A-BOPI. One could also see that when A-BOPI and F-BOPI are reliable (reliability test not rejected), the conventional estimated models could be not reliable and not the other way around.
 

\subsubsection{Comparing Methods by Charts}
\label{ref_charts}
Figures \ref{ip_chart_80}, \ref{ip_chart_90}, \ref{ip_chart_95} and \ref{ip_chart_99} are coverage charts for our experiments on the benchmark datasets and they compare the coverage of the seven prediction intervals (Loess Conv., A-BOPI, F-BOPI, LS-SVM Conv., OLS, Loess Conv., SVM Quantile and SVM Quantile CV) described in Section \ref{ref_method_def}.

Figures \ref{egsd_80}, \ref{egsd_90}, \ref{egsd_95} and \ref{egsd_99} display EGSD charts for our experiments on the benchmark datasets and they compare the efficiency of the seven aforementioned prediction intervals while ignoring their reliability. These EGSD charts display the normalized EGSD measure described by Equation (\ref{eq_egsd_def}) for all benchmark datasets. For a given dataset, the model having the lowest EGSD value has an Equivalent Gaussian distribution with the smallest variance, which means that it is the most efficient model compared to the others.

\subsubsection{Chart commentaries}
Coverage charts (Figure \ref{egsd_80}, \ref{egsd_90}, \ref{egsd_95} and \ref{egsd_99}), show that SVM Quantile and SVM Quantile CV always obtain coverage smaller than the desired one. For $\beta=\{0.8,0.9\}$ the two methods that usually obtain the larger coverage are respectively LS-SVM Conv. and F-BOPI and other methods are on average similar. This order changes for $\beta=\{0.95,0.99\}$ with F-BOPI having usually the larger coverage and no real ordering for other methods. 
 
Figures \ref{egsd_80}, \ref{egsd_90}, \ref{egsd_95} and \ref{egsd_99} show respectively the EGSD chart for $\beta=\{0.8, 0.9,\allowbreak 0.95, 0.99\}$. One can observe that A-BOPI and F-BOPI models are almost always more efficient than the others on the inquired datasets. If we look in more detail, we can see that A-BOPI usually finds the smallest EGSD value and the conventional method Loess Conv. is the next efficient one. It is interesting to note that while A-BOPI provides reliable models, it has the lowest EGSD (more efficient) compared to other methods.


 \subsection{Discussion of Results}
The introduced methods are compared with the conventional prediction intervals. This comparison is performed with simulation studies on two artificial data generating process (DGP) and a $10$-fold cross validation schema on eleven real benchmark regression datasets with sizes and number of independent variables varying respectively from $N$ = 103 to $N$ = 8192 and from $p$ = 1 to $p$ = 21. Some of the real datasets contain only numerical variables and some datasets have numerical and categorical variables. \\

For $\beta\geq 0.8$, it becomes a critical task to find an efficient and reliable prediction interval. However these proportions are the most used ones in machine learning and statistical hypothesis testing. The experimental part compares F-BOPI and A-BOPI prediction intervals for local linear regression while providing $\beta$ prediction intervals for $\beta=\{0.8, 0.9, 0.95,0.99\}$. This comparison is made with five well-known models of prediction intervals: the conventional prediction interval for local linear regression denoted by Loess Conv., the conventional prediction interval for least-squares SVM (LS-SVM Conv.), prediction intervals for classical linear regression (OLS) and two SVM quantile regression model (SVM Quantile and SVM Quantile CV). These models are described in Section \ref{ref_method_def}.\\
 
When comparing the BOPI methods with other prediction intervals, they appear to be the most reliable methods in both simulated and real cases. The simulation studies with the artificial DGP, rate the conventional method very poorly, so that it produce always non-reliable interval prediction models and the average proportion of response values inside the obtained intervals (coverage) where always less than the desired content. The BOPI methods obtain in all cases more reliable intervals with F-BOPI being more reliable than A-BOPI On the other hand A-BOPI yields intervals that are on average tighter than F-BOPI and both obtain intervals being on average wider than those obtained by Loess Conv..

Experiments on real datasets have shown that BOPI methods provide usually the most efficient solution. A Wilson Score test for binomial proportion at a significance level of $0.05$ and the alternate hypothesis as $\beta_{pop}< \beta$ \footnote{$\beta_{pop}$ denotes the average proportion of response value that are contained in the tested prediction intervals (coverage).} is used to test the reliability of the prediction intervals on real datasets. The conventional methods Loess Conv. and LS-SVM Conv. turn out to be unreliable for higher value of $\beta$. Furthermore, they are usually less efficient than A-BOPI and F-BOPI. Comparison of interval size of the two reliable estimated models which find the smallest MIS using two-sided paired t-tests at levels $0.05$, $0.01$ and $0.001$ shows that A-BOPI generally obtains the better model (minimum MIS) and its difference with the second smallest MIS (which is also reliable) is usually statistically significant at most at 0.05 level. On the other hand, ignoring the reliability, the conventional prediction interval methods Loess Conv. and LS-SVM Conv. rank as the most efficient methods after A-BOPI and F-BOPI. According to the results reported in Figures \ref{egsd_80}, \ref{egsd_90}, \ref{egsd_95} and \ref{egsd_99}, the SVM quantile regression model (SVM Quantile and SVM Quantile CV) are not suited for reliable interval prediction.\\
 
 In a regression context, the conditional mean, the conditional variance and/or the conditional quantile may have different functions. The conditional mean is the general trend of the regression function whereas the conditional quantile is more related to the local distribution of the response variable. Least-squares based interval prediction methods (OLS, LS-SVM Conv. Loess Conv., F-BOPI and A-BOPI) try to indirectly estimate the conditional quantile function. They first estimate the conditional mean and then, they estimate the conditional quantile. On the other hand, quantile regression based methods (SVM Quantile and SVM Quantile CV) directly estimate the conditional quantile. The general trend is easier to predict and its estimator, compared to the conditional quantile, has a higher speed of convergence \citep{koenker_book_2005}. This is why all the tested least-squares based interval prediction methods are more efficient than the quantile regression based methods. Another reason for this superiority may be the absence of a global conditional quantile function. It can occur when the conditional variance of the error distribution is not a global function of the predictors. The proposed methods belong to the class of least-squares based interval prediction methods, so they take advantage of this fast convergence. However they are more reliable and efficient than the conventional methods. The LHNPE assumptions permit to take into account the prediction error oscillation so the introduced prediction intervals consider the local conditional distribution for the response variable. Besides the use of tolerance intervals incorporate the effect of the local sample size used to estimate the prediction intervals.

\section{Discussion and Conclusion}
Having the question of reliable prediction intervals for local linear regression in mind, the authors investigated two new methods (BOPI) for estimating prediction intervals. The main assumptions are that the mean regression function is locally linear and the prediction error is locally homoscedastic and normal. The prediction intervals for the input vector is obtained based on a tolerance interval computed on a restricted set of prediction errors (obtained by a cross validation schema) of the local linear regression. This restricted set is composed of the instances inside the LHNPE bandwidth of the input vector. Two different LHNPE bandwidths are considered, a bandwidth having a fixed number of neighbors and a bandwidth having a variable number of neighbors. In order to test the BOPI methods individually and to compare them with five other interval prediction methods, the authors used the following measures for ranking interval predictions methods: coverage probability, Mean Interval Size and Equivalent Gaussian Standard Deviation. The five aforementioned interval prediction methods were: the conventional interval prediction method (described in Section \ref{ref_tol_convetional}) with local linear regression and least-squares SVM, prediction intervals for classical linear regression and two SVM quantile regression methods. The rankings were performed with a cross validation schema on eleven benchmark regression datasets, and the estimated results were generally in favor of the introduced methods. They also reported a simulation study comparing the BOPI methods with the conventional interval prediction method.

\subsection*{General remarks}
 The advantages, drawbacks and limitations of BOPI are listed below:
\subsubsection*{Advantages}

 \begin{itemize}
 \item[$\bullet$] It is a reliable prediction interval for local linear least squares models;
 \item[$\bullet$] It does not ignore the non-parametric regression bias;
 \item[$\bullet$] It can be used with models having heteroscedastic errors;
  
 \item[$\bullet$] It does not suffer from the crossing quantiles effect; 
 \item[$\bullet$] It is based on local linear regression, which is a well-known regression method.
 \end{itemize}

 \subsubsection*{Drawbacks}
 \begin{itemize}
 \item[$\bullet$] It is limited to local linear regression;
 \item[$\bullet$] It has a greater computational complexity than conventional and quantile regression methods.
 \end{itemize}

\subsubsection*{Limit of Applications}
In the following cases, our methods may obtain \textit{similar results to its alternatives}:
 \begin{itemize}
 \item[$\bullet$] For prediction interval with a very high desired proportion ($0.99$ or more) of the distribution of $Y(x)$;
 \item[$\bullet$] The dataset is almost identically distributed in the feature space.
 \end{itemize}
 
\textit{BOPI are not suited} when: 
 \begin{itemize}
 \item[$\bullet$] There exists regression models having significantly better prediction results than non-parametric regression models;
 \item[$\bullet$] The distribution of prediction errors differs significantly from the normal distribution.
 \end{itemize}
 
For future work, the most promising idea is the extension of these prediction intervals to other regression function, e.g. support vector machines. Another horizon may be its generalization to the one-sided interval prediction problem. One can also apply these methods to interval prediction in time series models.





\markboth{Glossary}{}

\vskip 0.2in

\bibliographystyle{elsewier/elsarticle-harv}
\section*{References}
\bibliography{biblio}

\newpage

\appendix
\section{}{}

\begin{table}[htbp!]
\centering

\captionof{table}{Coverage and MIS for different $\gamma$ on Friedman\#1 DGP} 
\scalebox{1}{
\begin{tabular}{ccccccccc}
 \hline
  \multicolumn{1}{c}{ }  & \multicolumn{3}{c}{  \ $Coverage$}  & \multicolumn{3}{c}{  \ $MIS$}      \\
    $\gamma$ &$ {Conv. }$& ${F-BOPI}$ & ${A-BOPI}$ &
    $ {Conv. }$& ${F-BOPI}$ & ${A-BOPI}$ \\ 
  \hline  \hline
    & 88.06 & 91.09 & 89.48     & 1.335 & 1.468 & 1.400 \\ [-1ex] 
   \raisebox{1.5ex} {0.80}  
 & (2.43) & (2.20) & (2.32) & (0.033) & (0.060) & (0.060) \\  [.5ex] 
   & 88.06 & 92.60 & 91.26   & 1.335 & 1.550 & 1.476 \\ [-1ex] 
    \raisebox{1.5ex} {0.90} 
 & (2.43) & (2.01) & (2.20) & (0.033) & (0.062) & (0.061)  \\  [.5ex]
   & 88.06 & 93.94 & 92.50   & 1.335 & 1.625 & 1.543 \\  [-1ex] 
    \raisebox{1.5ex} {0.95}  
  & (2.43) & (1.85) & (2.03)  & (0.033) & (0.064) & (0.062) \\   [.5ex]
   & 88.06 & 95.99 & 94.83   & 1.335 & 1.780 & 1.680 \\  [-1ex] 
    \raisebox{1.5ex} {0.99}  
  & (2.43) & (1.39) & (1.65)  & (0.033) & (0.068) & (0.065)\\   [.5ex]  \\

   \hline
\end{tabular}}
\bigskip
\caption*{Computed coverage for $\gamma=\{0.8, 0.9,0.95,0.99\}$ and $\beta=0.95$. The coverage values are computed using Friedman\#1 data generating process by a $3$-fold cross validation schema where $\frac{2}{3}$ of the generated sample is used for training and the remaining for test. The generated sample sizes are $N=1500$ and the simulation process is iterated for $N_{sim}=500$ times. The method hyper-parameters are as follows: ${K}^{f}_{lhnpe}=40$ for F-BOPI, $({K}^{min}_{lhnpe}=30,{K}^{max}_{lhnpe}=50)$ for A-BOPI and the regression bandwidth $K_{loess}=100$ is constant for the three interval prediction methods. Loess Conv. is shortened to ``Conv.'' and the standard deviation over the $N_{sim}$ coverage values is shown in parentheses.}

\label{tab_simu_f1covTable}
\end{table}

\begin{table}[htbp!]

\centering
\captionof{table}{Coverage and MIS for different $\gamma$ on Friedman\#2 DGP} 
\scalebox{1}{
\begin{tabular}{ccccccccc}
  \hline

  \multicolumn{1}{c}{ }  & \multicolumn{3}{c}{  \ $Coverage$}  & \multicolumn{3}{c}{  \ $MIS$}      \\
    $\gamma$ &$ {Conv. }$& ${F-BOPI}$ & ${A-BOPI}$ &
    $ {Conv. }$& ${F-BOPI}$ & ${A-BOPI}$ \\ 
  \hline  \hline
    & 85.199 & 89.748 & 88.425 & 11.678 & 13.397 & 12.870 \\ [-1ex] 
   \raisebox{1.5ex} {0.80}  
 & (4.013) & (3.287) & (3.596) & (0.181) & (0.516) & (0.522) \\  [.5ex] 
   & 85.199 & 91.387 & 90.124 & 11.678 & 14.154 & 13.560 \\ [-1ex] 
    \raisebox{1.5ex} {0.90} 
 & (4.013) & (3.094) & (3.324) & (0.181) & (0.543) & (0.542)  \\  [.5ex]
   & 85.199 & 92.721 & 91.382 & 11.678 & 14.832 & 14.170 \\  [-1ex] 
    \raisebox{1.5ex} {0.95}  
  & (4.013) & (2.829) & (3.066) & (0.181) & (0.567) & (0.560) \\   [.5ex]
   & 85.199 & 94.789 & 93.685 & 11.678 & 16.253 & 15.414 \\  [-1ex] 
    \raisebox{1.5ex} {0.99}  
  & (4.013) & (2.402) & (2.635) & (0.181) & (0.617) & (0.599)\\   [.5ex]  \\

   \hline
\end{tabular}}
\bigskip
\caption*{Computed coverage for $\gamma=\{0.8, 0.9,0.95,0.99\}$ and $\beta=0.95$. The coverage values are computed using Friedman\#2 data generating process by a $3$-fold cross validation schema where $\frac{2}{3}$ of the generated sample is used for training and the remaining for test. The generated sample sizes are $N=1500$ and the simulation process is iterated for $N_{sim}=500$ times. The method hyper-parameters are as follows: ${K}^{f}_{lhnpe}=40$ for F-BOPI, $({K}^{min}_{lhnpe}=30,{K}^{max}_{lhnpe}=50)$ for A-BOPI and the regression bandwidth $K_{loess}=100$ is constant for the three interval prediction methods. Loess Conv. is shortened to ``Conv.'' and the standard deviation over the $N_{sim}$ coverage values is shown in parentheses.}
\label{tab_simu_f2covTable} 
\end{table}


\begin{table}[htbp!]

\centering
\captionof{table}{Coverage for different $\beta$ on Friedman\#1 DGP} 
\scalebox{1}{
\begin{tabular}{ccccccccc}
\hline  
 
  \multicolumn{3}{c}{ }  & \multicolumn{6}{c}{  \ $Coverage$}    \\
   \cline{4-9}
\multicolumn{3}{c}{ }  & \multicolumn{3}{c}{  \ $N=1500$}  & \multicolumn{3}{c}{  \ $N=3000$}   \\
    & $\gamma$ & $\beta$ &$ {Conv. }$& ${F-BOPI}$ & ${A-BOPI}$ & ${Conv.}$ & ${F-BOPI}$ &$ {A-BOPI} $\\ 
  \hline  \hline
    &  &  & 67.814 & 81.871 & 79.368 & 67.816 & 81.873 & 79.369 \\ [-1ex] 
  & \raisebox{1.5ex} {0.990} &\raisebox{1.5ex} {0.800} &
 (3.083) & (2.797) & (2.953) & (3.081) & (2.795) & (2.952) \\  [.5ex] 
   &  & & 80.370 & 91.600 & 89.928 & 80.374 & 91.604 & 89.932 \\ [-1ex] 
   \ $N_{sim}=500$  & \raisebox{1.5ex} {0.990} &\raisebox{1.5ex} {0.900} &
  ( 2.987) & (2.123) & (2.247) & (2.987) & (2.122) & (2.246) \\  [.5ex]
      &  &  & 88.056 & 95.992 & 94.827 & 88.060 & 95.996 & 94.831 \\  [-1ex] 
    & \raisebox{1.5ex} {0.990} &\raisebox{1.5ex} {0.950} &
   (2.427) & (1.395) & (1.654) & (2.428) & (1.393) & (1.654) \\   [.5ex]
      &  &  &96.003 & 99.197 & 98.812 & 96.007 & 99.198 & 98.814 \\  [-1ex] 
    & \raisebox{1.5ex} {0.990} &\raisebox{1.5ex} {0.990} & 
  (1.488) & (0.558) & (0.659) & (1.487) & (0.556) & (0.657) \\   [.5ex]  \\
      &  &  & 66.923 & 81.264 & 78.606 & 66.925 & 81.266 & 78.607 \\  [-1ex] 
    & \raisebox{1.5ex} {0.990} &\raisebox{1.5ex} {0.800} &
     (1.768) & (2.084) & (2.197) & (1.768) & (2.084) & (2.196) \\   [.5ex]
      &  &  & 79.143 & 90.906 & 88.955 & 79.143 & 90.908 & 88.957 \\  [-1ex] 
     \ $N_{sim}=1000$   & \raisebox{1.5ex} {0.990} &\raisebox{1.5ex} {0.900} & 
  (1.752) & (1.485) & (1.719) & (1.751) & (1.485) & (1.719) \\   [.5ex]
     &  &  &86.936 & 95.543 & 94.282 & 86.936 & 95.543 & 94.284\\  [-1ex] 
    & \raisebox{1.5ex} {0.990} &\raisebox{1.5ex} {0.950} &
     (1.535) & (1.014) & (1.185) & (1.534) & (1.013) & (1.184) \\   [.5ex]
     &  &  &95.199 & 98.986 & 98.614 & 95.199 & 98.986 & 98.614 \\  [-1ex] 
   & \raisebox{1.5ex} {0.990} &\raisebox{1.5ex} {0.990} &
   (0.845) & (0.453) & (0.492) & (0.844) & (0.453) & (0.491) \\   [.5ex]
   \hline
\end{tabular}}
\bigskip
\caption*{Computed coverage for $\beta=\{0.8, 0.9,0.95,0.99\}$ and $\gamma=0.99$. The coverage values are computed using Friedman\#2 data generating process by a $3$-fold cross validation schema where $\frac{2}{3}$ of the generated sample is used for training and the remaining for test. The generated sample sizes are $N=1500$ (left) and $N=3000$ (right). The simulation process is iterated ($N_{sim}$ times) for $500$ times (up) and $1000$ times (down). The method hyper-parameters are as follows: ${K}^{f}_{lhnpe}=40$ for F-BOPI, $({K}^{min}_{lhnpe}=30,{K}^{max}_{lhnpe}=50)$ for A-BOPI and the regression bandwidth $K_{loess}=100$ is constant for the three interval prediction methods. Loess Conv. is shortened to ``Conv.'' and the standard deviation over the $N_{sim}$ coverage values is shown in parentheses.}
\label{tab_simu_f1Table} 
\end{table}


\begin{table}[htbp!]

\centering
\captionof{table}{Coverage for different $\beta$ on Friedman\#2 DGP}  
\scalebox{1}{
\begin{tabular}{ccccccccc}
  \hline
 
  \multicolumn{3}{c}{ }  & \multicolumn{6}{c}{  \ $Coverage$}    \\
   \cline{4-9}
\multicolumn{3}{c}{ }  & \multicolumn{3}{c}{  \ $N=1500$}  & \multicolumn{3}{c}{  \ $N=3000$}   \\
    & $\gamma$ & $\beta$ &$ {Conv. }$& ${F-BOPI}$ & ${A-BOPI}$ & ${Conv.}$ & ${F-BOPI}$ &$ {A-BOPI} $\\ 
  \hline  \hline
    &  &  &64.033 & 80.659 & 78.316 & 64.026 & 80.652 & 78.311 \\ [-1ex] 
  & \raisebox{1.5ex} {0.990} &\raisebox{1.5ex} {0.800} &
 (3.564) & (4.436) & (4.268) & (3.562) & (4.436) & (4.268) \\  [.5ex] 
   &  & & 77.041 & 90.335 & 88.680 & 77.034 & 90.329 & 88.674 \\ [-1ex] 
   \ $N_{sim}=500$  & \raisebox{1.5ex} {0.990} &\raisebox{1.5ex} {0.900} &
 (4.212) & (3.189) & (3.545) & (4.212) & (3.191) & (3.548) \\  [.5ex]
      &  &  & 85.199 & 94.789 & 93.685 & 85.191 & 94.784 & 93.680 \\  [-1ex] 
    & \raisebox{1.5ex} {0.990} &\raisebox{1.5ex} {0.950} &
   (4.013) & (2.402) & (2.635) & (4.014) & (2.402) & (2.635)\\   [.5ex]
      &  &  &93.689 & 98.528 & 98.069 & 93.682 & 98.526 & 98.066  \\  [-1ex] 
    & \raisebox{1.5ex} {0.990} &\raisebox{1.5ex} {0.990} & 
  (2.588) & (1.048) & (1.256) & (2.589) & (1.048) & (1.257) \\   [.5ex]  \\
      &  &  & 59.460 & 76.089 & 73.690 & 59.465 & 76.092 & 73.694 \\  [-1ex] 
    & \raisebox{1.5ex} {0.990} &\raisebox{1.5ex} {0.800} &
    (3.507) & (4.220) & (4.259) & (3.508) & (4.218) & (4.257) \\   [.5ex]
      &  &  & 71.974 & 86.154 & 84.284 & 71.978 & 86.158 & 84.287 \\  [-1ex] 
     \ $N_{sim}=1000$   & \raisebox{1.5ex} {0.990} &\raisebox{1.5ex} {0.900} & 
  (3.749) & (3.752) & (3.888) & (3.749) & (3.752) & (3.887) \\   [.5ex]
     &  &  & 80.061 & 91.324 & 89.986 & 80.065 & 91.327 & 89.989\\  [-1ex] 
    & \raisebox{1.5ex} {0.990} &\raisebox{1.5ex} {0.950} &
     (3.792) & (3.066) & (3.309) & (3.792) & (3.066) & (3.309) \\   [.5ex]
     &  &  &89.721 & 96.457 & 95.656 & 89.725 & 96.457 & 95.657 \\  [-1ex] 
   & \raisebox{1.5ex} {0.990} &\raisebox{1.5ex} {0.990} &
   (3.057) & (1.757) & (2.011) & (3.058) & (1.756) & (2.011) \\   [.5ex]
   \hline
\end{tabular}}
\bigskip
\caption*{Computed coverage for $\beta=\{0.8, 0.9,0.95,0.99\}$ and $\gamma=0.95$. The coverage values are computed using Friedman\#2 data generating process by a $3$-fold cross validation schema where $\frac{2}{3}$ of the generated sample is used for training and the remaining for test. The generated sample sizes are $N=1500$ (left) and $N=3000$ (right). The simulation process is iterated ($N_{sim}$ times) for $500$ times (up) and $1000$ times (down). The method hyper-parameters are as follows: ${K}^{f}_{lhnpe}=40$ for F-BOPI, $({K}^{min}_{lhnpe}=30,{K}^{max}_{lhnpe}=50)$ for A-BOPI and the regression bandwidth $K_{loess}=100$ is constant for the three interval prediction methods. Loess Conv. is shortened to ``Conv.'' and the standard deviation over the $N_{sim}$ coverage values is shown in parentheses.}

\label{tab_simu_f2Table} 
\end{table}


\label{images}
\begin{figure}
        \centering
        \begin{subfigure}[b]{0.5\textwidth}
                \includegraphics[width=9.4cm,height=8.7 cm]{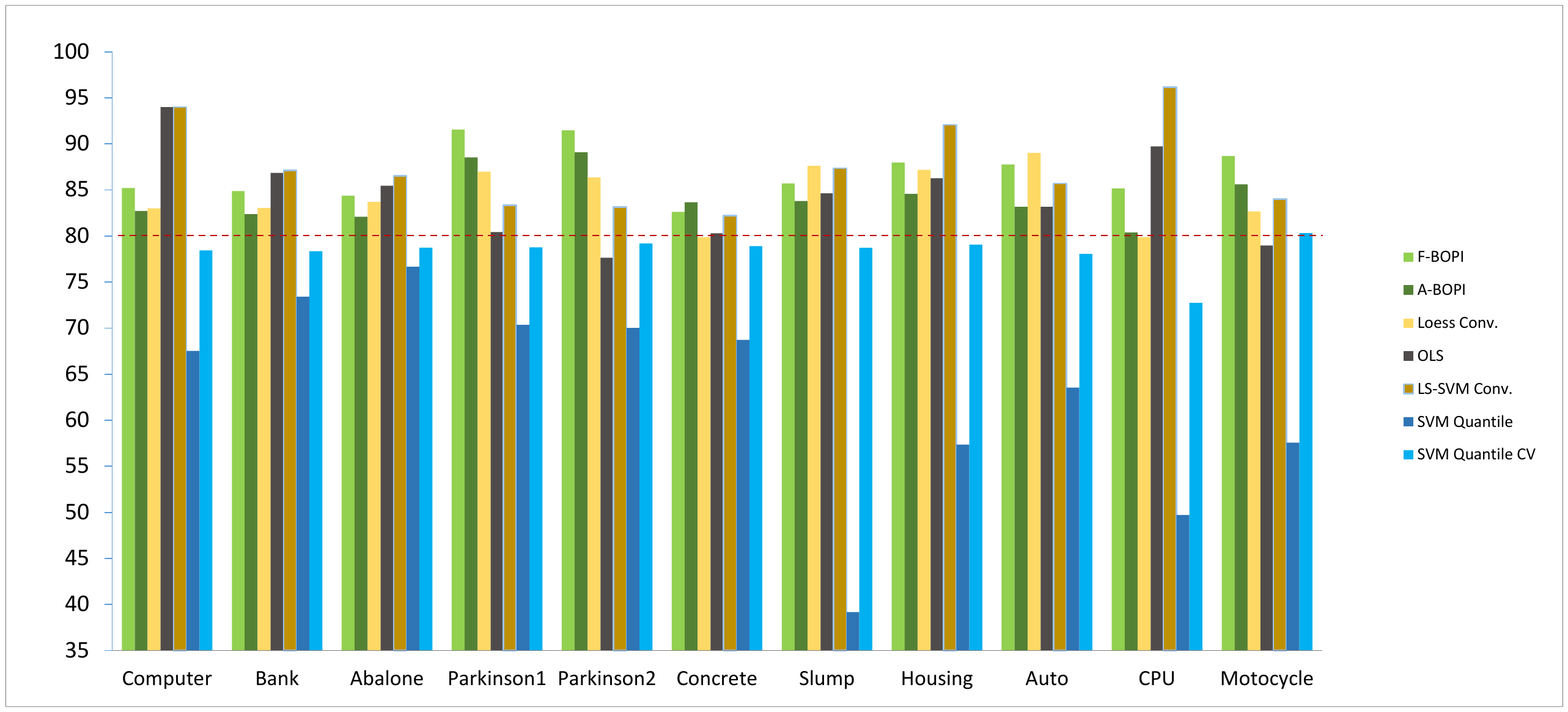}
                \caption{$\beta=0.8$}
                \label{ip_chart_80}
        \end{subfigure}%
        ~ 
        \begin{subfigure}[b]{0.5\textwidth}
                \includegraphics[width=9.4cm,height=9.4cm]{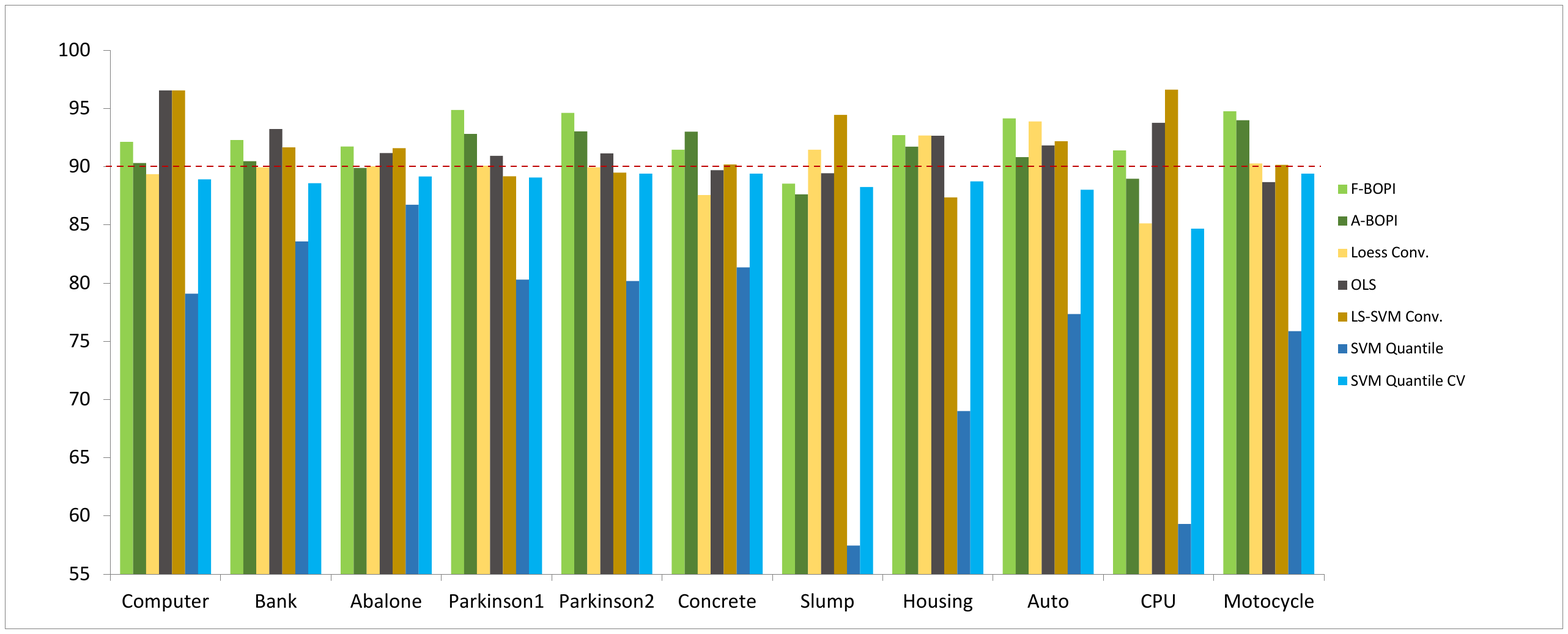}
                \caption{$\beta=0.9$}
                \label{ip_chart_90}
        \end{subfigure}
        
                \begin{subfigure}[b]{0.5\textwidth}
                \includegraphics[width=9.4cm,height=9.4cm]{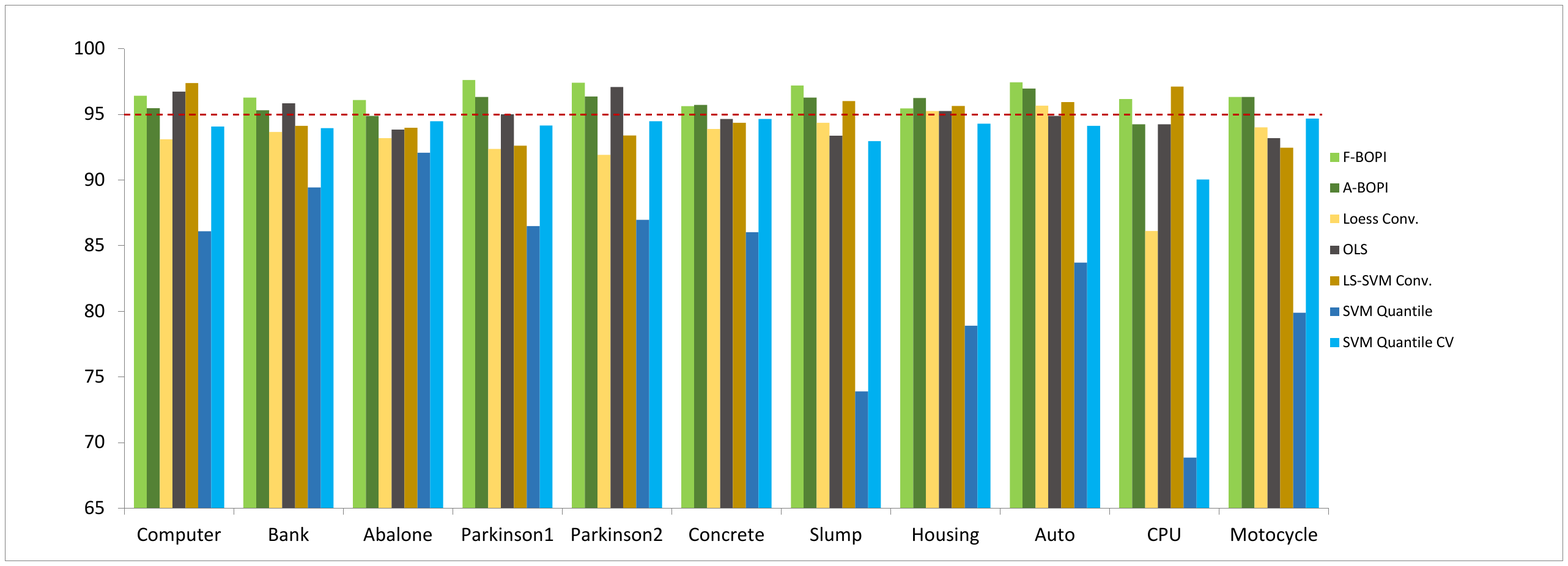}
                \caption{$\beta=0.95$}
                \label{ip_chart_95}
        \end{subfigure}%
        ~ 
        \begin{subfigure}[b]{0.5\textwidth}
                \includegraphics[width=9.4cm,height=9.4cm]{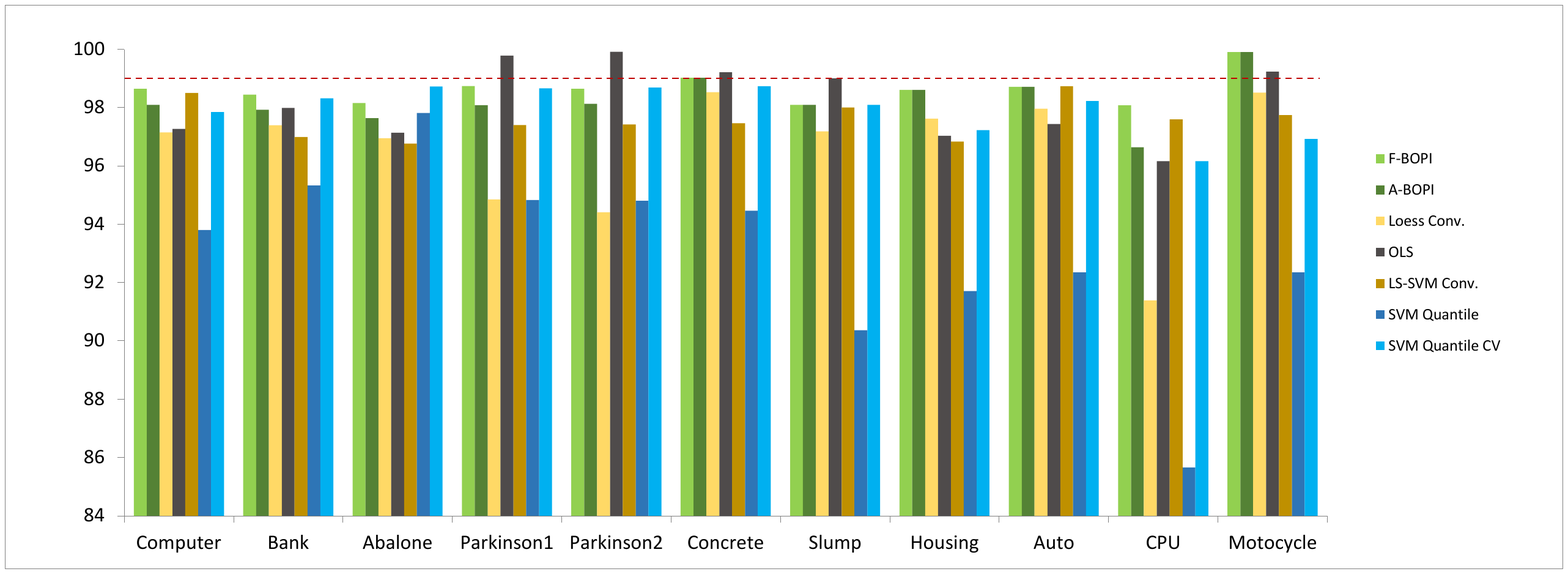}
                \caption{$\beta=0.99$}
                \label{ip_chart_99}
        \end{subfigure}

        ~ 
        \caption{Coverage charts for benchmark datasets with $\beta= \{0.8,0.9,0.95,0.99\}$.}\label{ip_charts}
\end{figure}


 \begin{figure}[htbp!] 
  \vspace{-2.1cm}
 	\includegraphics[width=12cm,height=8cm]{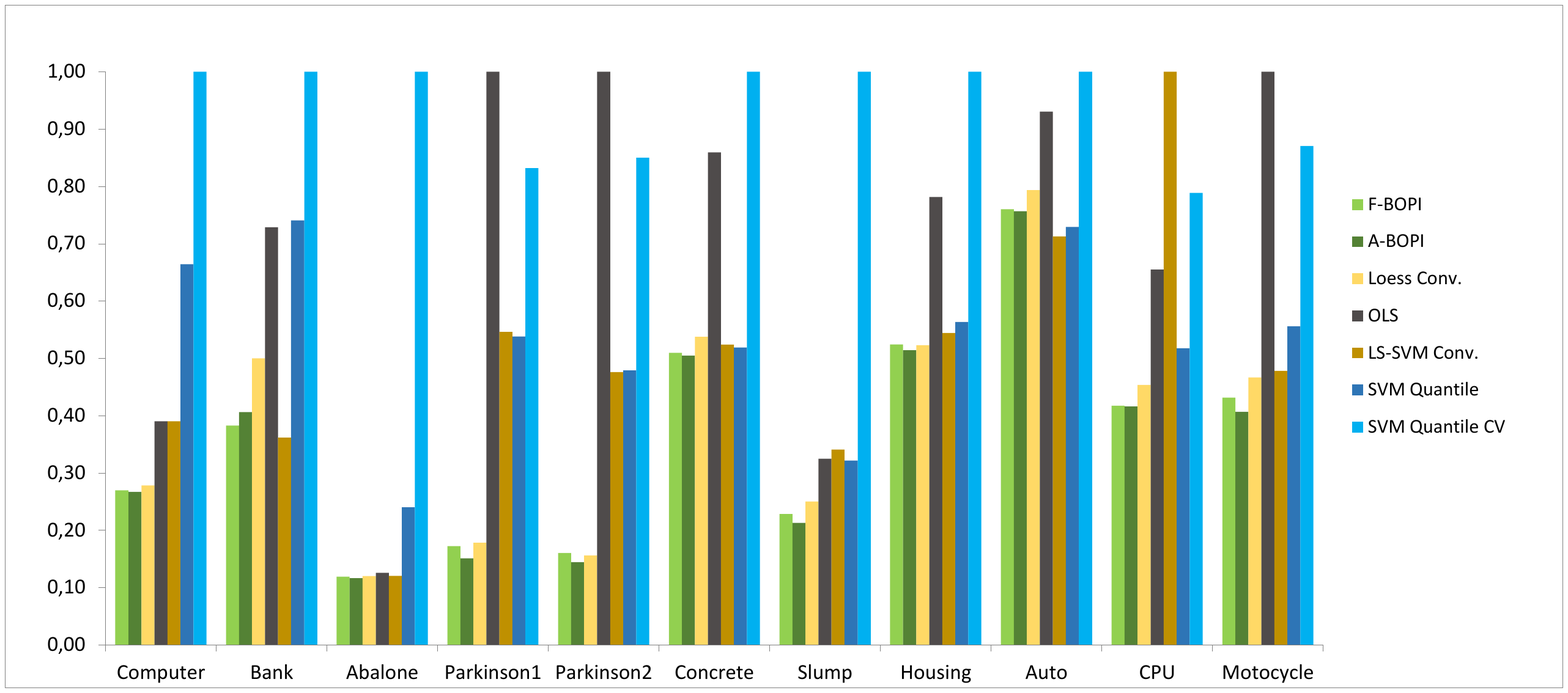}\vspace{-1.7cm}
	\caption{EGSD chart for benchmark datasets with $\beta=0.80$. The smallest value denotes the most efficient band. This measure ignores the reliability.} 
	  \label{egsd_80} 		  
	  \vspace{-0.7cm}
	  	\includegraphics[width=12cm,height=8cm]{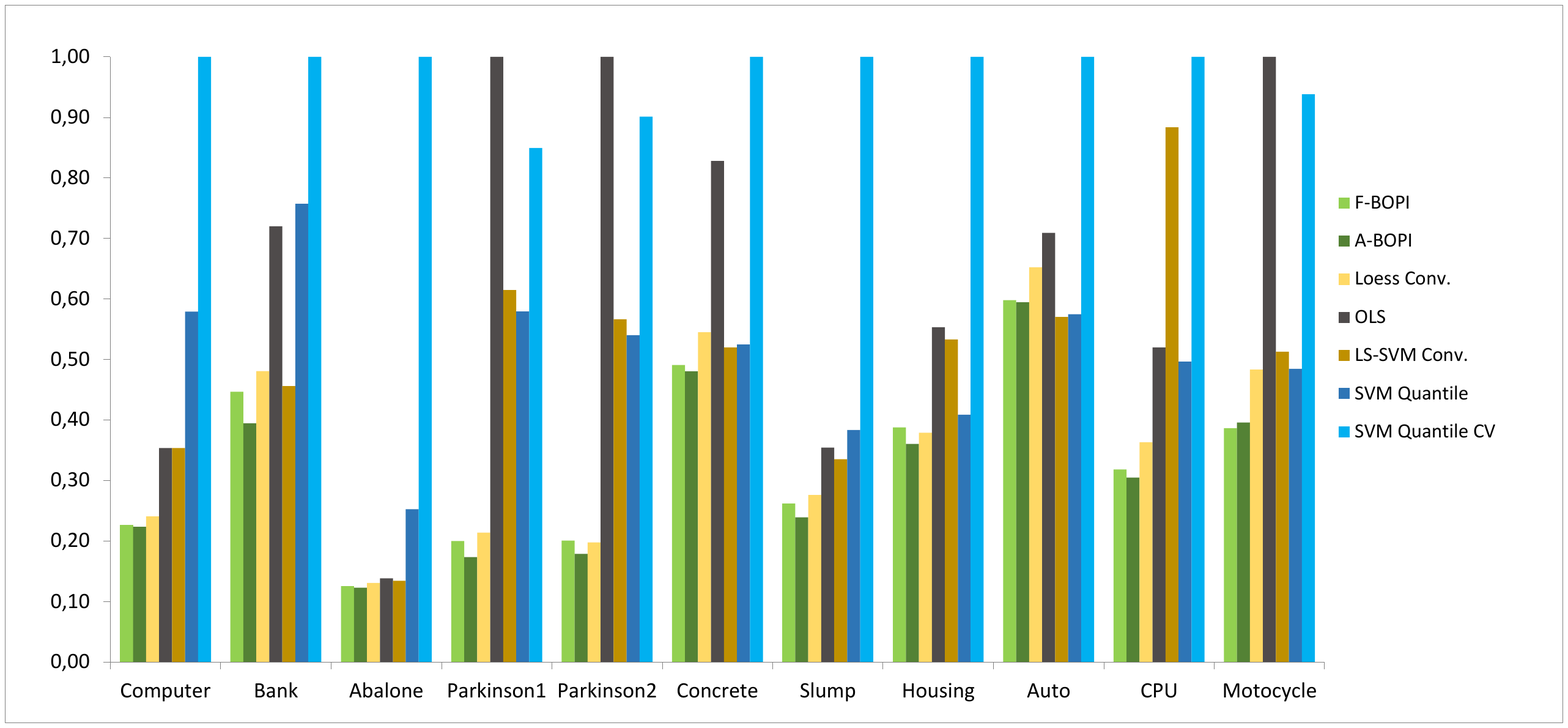}\vspace{-1.7cm}
	\caption{EGSD chart for benchmark datasets with $\beta=0.9$. The smallest value denotes the most efficient band. This measure ignores the reliability.} 
		  \label{egsd_90}

\vspace{-0.7cm}
	\includegraphics[width=12cm,height=8cm]{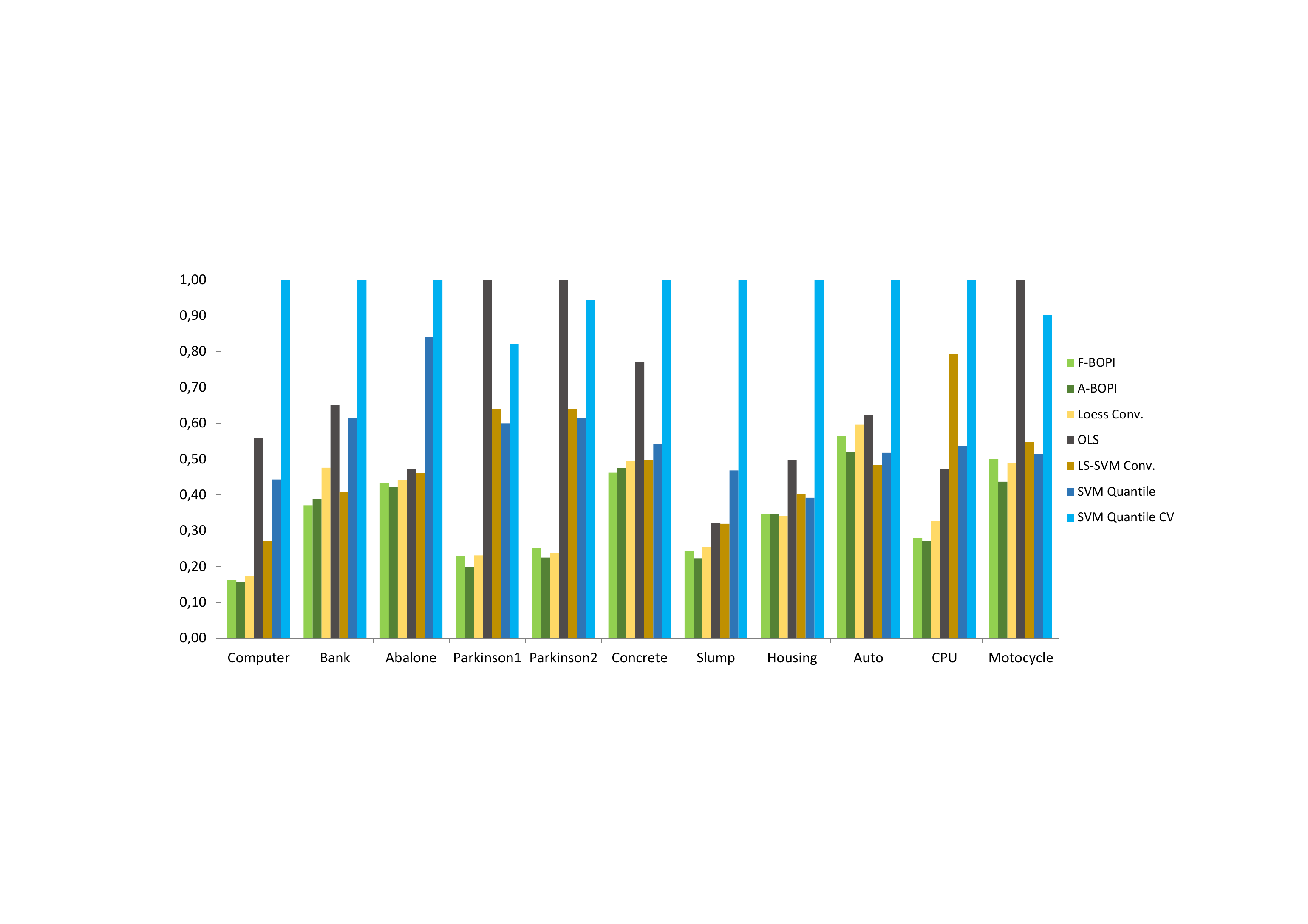}\vspace{-1.7cm}
	\caption{MIS Ratio chart for benchmark datasets with $\beta=0.95$. The smallest value denotes the tightest reliable band.} 
	  \label{egsd_95} 		  	  

\end{figure} 

 \begin{figure}[htbp!] 
	\includegraphics[width=12cm,height=8cm]{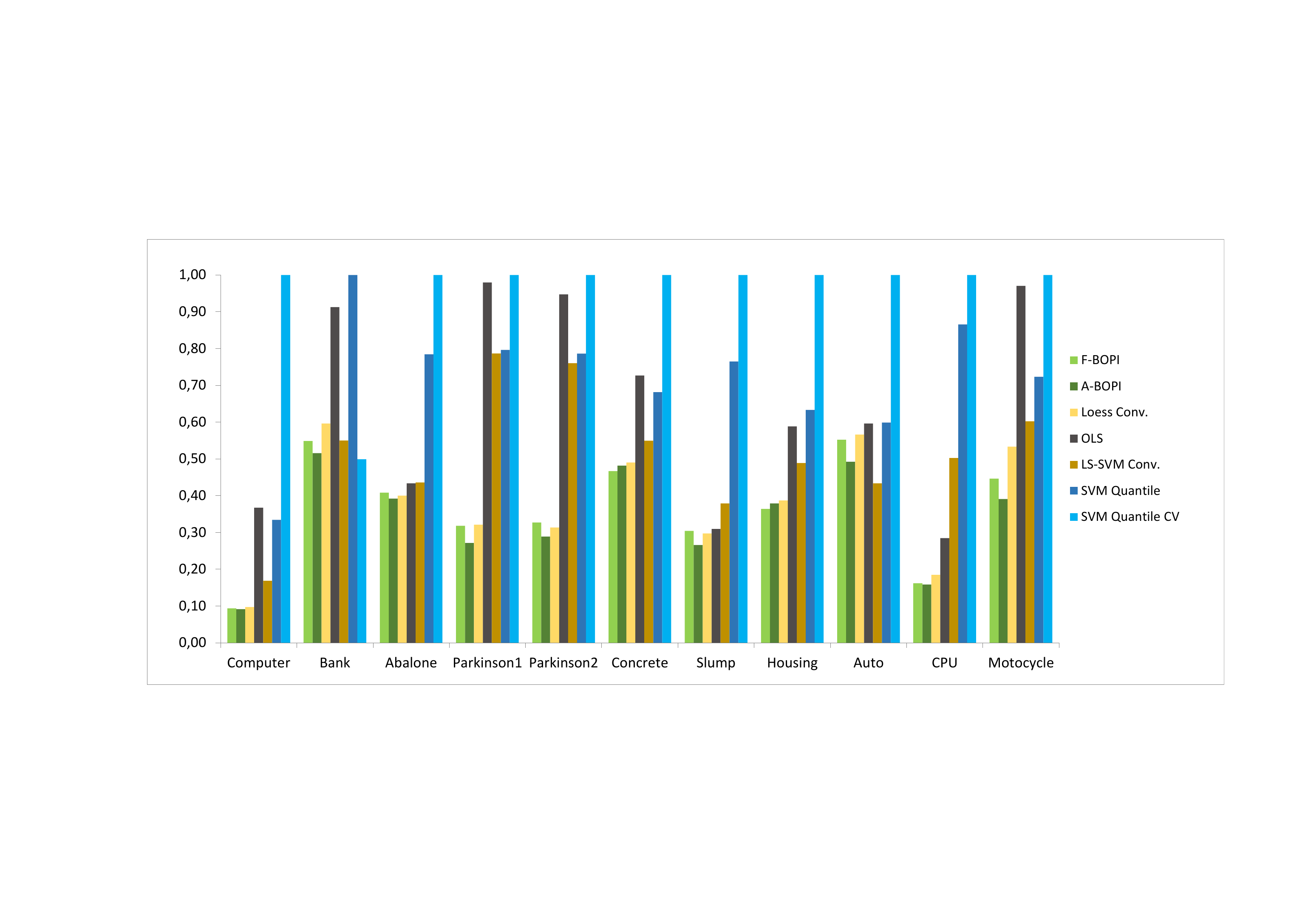}\vspace{-1.7cm}
	\caption{EGSD chart for benchmark datasets with $\beta=0.99$. The smallest value denotes the most efficient band. This measure ignores the reliability.} 
	  \label{egsd_99} 		  
\end{figure} 

\label{proofs}
\subsection{Proof of Proposition \ref{ref_prop_tol}}

Let $Y(x)= f(x)+\varepsilon_x$ and let $\hat{f}(x)$ denote its local linear regression estimator. If this regression estimator satisfies the conditions below:
\begin{itemize}
 \item[$\bullet$] \textit{Normal error distribution:} $\varepsilon_{x} \sim \mathcal{N}(0, \sigma^2_{x})$.
 \item[$\bullet$] \textit{$\hat{f}(x)$ has an almost constant distribution as defined in Definition \ref{def_lhnpe}.} 
\end{itemize}

Then:

\begin{enumerate}[(a)]
\item $\hat{f}(x)$ is an LHNPE regression estimator;
\item The interval $I(x^{*})^{Pred}_{\beta}$ for the input $x^{*}$ obtained by Equation \eqref{eq_tol_def_local_linear_1} is a $\beta$-content prediction interval for $Y(x^{*})$.
\item The sample bias of the prediction error in the LHNPE neighborhood is a consistent estimator of the regression bias:
\begin{equation*}
\underset{ K \rightarrow \infty}{\operatorname{ plim}} \left( \hat{f}(x^{*})- \widehat{bias}_{\hat{f}(x^{*})} \right) = f(x^{*}),
\end{equation*}
where $K/N \rightarrow 0$ as $N \rightarrow \infty$, $K=card(Kset_{x^{*}})$  and $\widehat{bias}_{\hat{f}(x^{*})}$ are respectively the cardinal of $Kset_{x^{*}}$ and the sample bias of $\hat{f}(x^{*})$.

\end{enumerate}

\textit{Proof:} 

Part (a): \citep[pp. 302]{fan_gijbels_1996} have shown that, under certain regularity conditions, the local linear estimator has asymptotically the following normal distribution: $\hat{f}(x) \sim \mathcal{N} \bigg( f(x)+ bias_{\hat{f}(x)},\sigma^2_{\hat{f}(x)} \bigg)$, where $bias_{\hat{f}(x^{*})}= E[\hat{f}(x^{*})- f(x^{*})]$ is the estimator's bias, $\sigma^2_{x^{}}$ is the variance of the error and $\sigma^2_{\hat{f}(x^{*})}$ is the regression estimator variance. The latter result along with the proposition's condition proves (a). \\

Part (b): Let $x^{*}$ denote the input vector and let $\varepsilon^{pred}_{x^{*}}$ denote its prediction error, then by accepting (a), we assume that the prediction errors follow a normal distribution whose variance is approximately the same in the neighborhood of $x^{*}$. So, we have:
\begin{equation*}
\varepsilon^{pred}_{x^{*}}= Y(x^{*})-\hat{f}(x^{*}) = \varepsilon_{x^{*}} + f(x^{*})-\hat{f}(x^{*}),
\end{equation*}
where by definition $\varepsilon_{x^{*}}$ and $\hat{f}(x^{*})$ are independent and $f(x^{*})$ is non-random. Thus we have:
\begin{equation}
\label{eq_pred_error_tol}
\varepsilon^{pred}_{x^{*}}  \sim \mathcal{N}(-bias_{\hat{f}(x^{*})}, \sigma^2_{\varepsilon^{pred}_{x^{*}}} ),
\end{equation}

where $\sigma^2_{\varepsilon^{pred}_{x^{*}}} = \sigma^2_{x^{*}}+ \sigma^2_{\hat{f}(x^{*})}$. Based on the above assumptions, one can use the prediction error of the LHNPE neighbors of $x^{*}$ as an iid sample of $\varepsilon^{pred}_{x^{*}}$ and calculate the prediction interval of the prediction error $I(\varepsilon^{pred}_{x^{*}})^{Pred}_{\beta}$ by using Equation~\eqref{eq_prediction_normal_interval}. In this case, the prediction interval of the prediction error is calculated by replacing $\overline{X}$, $\hat{\sigma}$ and $n$ in Equation~\eqref{eq_prediction_normal_interval} with $-\widehat{bias}_{\hat{f}(x^{*})}$ and $\hat{\sigma}_{\varepsilon^{pred}_{x^{*}}}$ and $K$. These value are estimated as:
\begin{eqnarray*}
\widehat{bias}_{\hat{f}(x^{*})}= (-K)^{-1} \underset{ x_{i} \in Kset_{x^{*}}}{\operatorname{ \sum}} \varepsilon^{pred}_{x_i} \\
\hat{\sigma}_{\varepsilon^{pred}_{x^{*}}}= \left((K-1)^{-1} \sum_{x_{i} \in Kset_{x^{*}}} (\varepsilon^{pred}_{x_i} + \widehat{bias}_{\hat{f}(x^{*})} )^2\right)^{\frac{1}{2}},\\
\end{eqnarray*}

where $K/N \rightarrow 0$ as $N \rightarrow \infty$ and we have $K=card(Kset_{x^{*}})$ and $\widehat{bias}_{\hat{f}(x^{*})}$ are respectively the cardinal of $Kset_{x^{*}}$ and the sample bias of $\hat{f}(x^{*})$. Thus $I(\varepsilon^{pred}_{x^{*}})^{Pred}_{\beta}$ takes into account two kinds of uncertainties: the regression's method uncertainty and the observation error. It results in :

\begin{eqnarray*}
I(\varepsilon^{pred}_{x^{*}})^{Pred}_{\beta}= [L(\varepsilon^{pred}_{x^{*}})^{Pred}_{\beta},U(\varepsilon^{pred}_{x^{*}})^{Pred}_{\beta}]=\\
 - \widehat{bias}_{\hat{f}(x^{*})} \pm \hat{\sigma}_{\varepsilon^{pred}_{x^{*}}} t_{(\frac{1-\beta}{2},K-1)}  \sqrt{1+ \frac{1}{K}}.
\end{eqnarray*}




Equation~\eqref{eq_pred_error_tol} shows that the prediction error $\varepsilon^{pred}_{x^{*}}$ has a normal distribution with the unknown mean $-bias_{\hat{f}(x^{*})}$. The prediction interval for the prediction errors $I(\varepsilon^{pred}_{x^{*}})^{Pred}_{\beta}$ is constructed based on the $Kset_{x^{*}}$, which has a finite sample size, and it is centered on the sample bias $-\widehat{bias}_{\hat{f}(x^{*})}$. However, because of its definition, $I(\varepsilon^{pred}_{x^{*}})^{Pred}_{\beta}$ we have:

\begin{equation*}
P_{\mathcal{T},\varepsilon} \bigg(  L(\varepsilon^{pred}_{x^{*}})^{T}_{\gamma,\beta} \leq  \varepsilon^{pred}_{x^{*}} \leq  U(\varepsilon^{pred}_{x^{*}})^{T}_{\gamma,\beta}  \bigg) \geq \beta,
\end{equation*}
where $\mathcal{T}= (\hat{f}(x^{*}), \hat{\sigma}_{x^{*}})$ is the estimated vector at $x^{*}$. This equation can be rewritten as:

\begin{eqnarray}
\begin{aligned}
P_{\mathcal{T},\varepsilon} \bigg(  L(\varepsilon^{pred}_{x^{*}})^{Pred}_{\beta} \leq  \varepsilon + f(x^{*}) - \hat{f}(x^{*})  \leq  U(\varepsilon^{pred}_{x^{*}})^{Pred}_{\beta} \bigg) \geq \beta   &\\
  = P_{\mathcal{T},\varepsilon} \bigg(   \hat{f}(x^{*}) + L(\varepsilon^{pred}_{x^{*}})^{Pred}_{\beta} \leq  Y(x^{*})  \leq  \hat{f}(x^{*}) + U(   \varepsilon^{pred}_{x^{*}})^{Pred}_{\beta}  \bigg) \geq \beta  &\\
  \label{eq_pred_loess_def}
      = P_{\mathcal{T},\varepsilon} \bigg(  Y(x^{*}) \in  \left( \hat{f}(x^{*})  + I(\varepsilon^{pred}_{x^{*}})^{Pred}_{\beta} \right) \bigg)  \geq \beta .
\end{aligned}
\end{eqnarray}

Having in mind the assumptions, Equation~\eqref{eq_pred_loess_def} could be interpreted as follows: the prediction interval for the response variable is computed by adding the local linear regression estimate to the prediction interval on the prediction error:\\
\begin{equation*}
I(x^{*})^{Pred}_{\gamma,\beta}= \hat{f}(x^{*}) + I(\varepsilon^{pred}_{x^{*}})^{Pred}_{\beta}.
\end{equation*}

Note that, since $\varepsilon^{pred}_{x^{*}}$ has a normal distribution, then $I(\varepsilon^{pred}_{x^{*}})^{Pred}_{\beta}$ is obtained by a prediction interval for normal distribution calculated using Equation~\eqref{eq_prediction_normal_interval}.


Part (c): even though the prediction is biased, the prediction interval contains on average a desired proportion $\beta$ of the conditional distribution of the response variable: the prediction intervals are computed on the prediction error and the prediction error is centered on the sample estimate of negative bias.\\



In order to show that the sample bias converges in mean squared to the local linear regression bias, we will show that the expectation of the sample bias is the local linear regression bias and its asymptotic variance is zero.

\begin{eqnarray*}
E \left[\widehat{bias}_{\hat{f}(x^{*})} \right]= E \left[K^{-1} \underset{ x_{i} \in Kset_{x^{*}}}{\operatorname{ \sum}} \varepsilon^{pred}_{x_i}  \right]\\
=E \left[K^{-1} \underset{ x_{i} \in Kset_{x^{*}}}{\operatorname{ \sum}} \left( \varepsilon_{x^{*}} + f(x^{*})-\hat{f}(x^{*}) \right)  \right]\\
=E \left[K^{-1} \underset{ x_{i} \in Kset_{x^{*}}}{\operatorname{ \sum}} \left( \varepsilon_{x^{*}} \right)  \right]+ E \left[K^{-1} \underset{ x_{i} \in Kset_{x^{*}}}{\operatorname{ \sum}} \left(f(x^{*})-\hat{f}(x^{*}) \right)  \right]\\
=K^{-1} \underset{ x_{i} \in Kset_{x^{*}}}{\operatorname{ \sum}} E \left[ f(x^{*})-\hat{f}(x^{*})  \right]\\
=bias_{\hat{f}(x^{*})}.
\end{eqnarray*}

\begin{eqnarray*}
Var \left[\widehat{bias}_{\hat{f}(x^{*})} \right]= &Var \left[K^{-1} \underset{ x_{i} \in Kset_{x^{*}}}{\operatorname{ \sum}} \left( \varepsilon_{x^{*}} \right)  \right]\\
+ &Var \left[K^{-1} \underset{ x_{i} \in Kset_{x^{*}}}{\operatorname{ \sum}} \left(f(x^{*})-\hat{f}(x^{*}) \right)  \right]\\
=K^{-1} \sigma^2_{\varepsilon^{pred}_{x^{*}}},
\end{eqnarray*}

where $K/N \rightarrow 0$ as $N \rightarrow \infty$ and $K=card(Kset_{x^{*}})$. By definition $\varepsilon_{x^{*}}$ and $\hat{f}(x^{*})$ are independent and $f(x^{*})$ is non-random, thus:

 $$ \underset{ K \rightarrow \infty}{\operatorname{ lim}} E \left[ \left( \widehat{bias}_{\hat{f}(x^{*})} - bias_{\hat{f}(x^{*})}  \right)^2 \right] = \underset{ K \rightarrow \infty}{\operatorname{ lim}} K^{-1} \sigma^2_{\varepsilon^{pred}_{x^{*}}}=0 .$$

which implies a convergence in probability.\\

Under the mentioned conditions the sample $-\widehat{bias}_{\hat{f}(x^{*})}$ is a consistent estimator of $bias_{\hat{f}(x^{*})}$, and it is evident that:

\begin{eqnarray*}
\underset{ K \rightarrow \infty}{\operatorname{ plim}} \left( \hat{f}(x^{*})- \widehat{bias}_{\hat{f}(x^{*})} \right) =\underset{ K \rightarrow \infty}{\operatorname{ plim}} \left( \hat{f}(x^{*})\right) - \underset{ K \rightarrow \infty}{\operatorname{ plim}} \left( \widehat{bias}_{\hat{f}(x^{*})} \right) \\
= f(x^{*}) + bias_{\hat{f}(x^{*})}   - \underset{ K \rightarrow \infty}{\operatorname{ plim}} \left( \widehat{bias}_{\hat{f}(x^{*})} \right)=  f(x^{*})
\end{eqnarray*} \QEDA


\subsection{Proof of Proposition \ref{eq_interv_compare}}

\textit{For any random sample $20 \leq n\leq 10000$, if we set $\gamma$ and $\beta$, then the $\gamma$-coverage $\beta$-content tolerance interval of the standard normal distribution is greater than or equal to its $\beta$-prediction intervals. This is stated formally below:}

\begin{equation*}
\forall n\geq 20 , \gamma \geq 0.7 , \beta \in [0.01,0.99] , \ size(I^{Tol}_{\gamma,\beta}) \geq size(I^{Prev}_{\beta}) .
\end{equation*}

\textit{where $size(I)= U -L$, $I=[L,U]$ and the terms $I^{Tol}_{\gamma,\beta}$ and $I^{Prev}_{\beta}$ refer to $\gamma$-coverage $\beta$-content tolerance interval and $\beta$-prediction interval of the standard normal distribution.}

\textit{Proof:} 

\begin{figure}[htbp!]
	\includegraphics[scale=0.4]{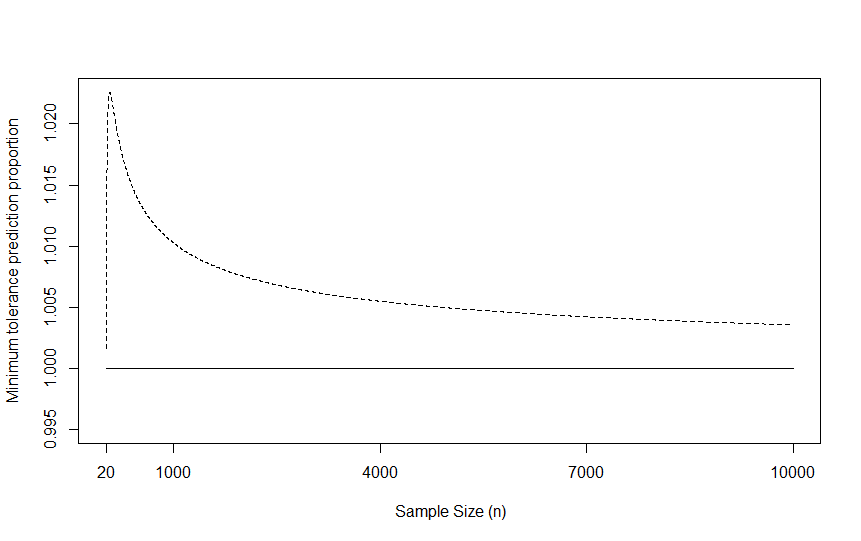} 		
 	\caption{This figure plots the inequality described by Equation~\eqref{eq_tolpred_prop_fig}. The vertical axis is the minimum tolerance prediction proportion obtained within $\beta \in [0.01,0.99]$ and the horizontal axis is the sample size $n$.}   
  \label{fig_tol_pred_prop}
\end{figure}

In order to verify this property, one must show that the proportion of tolerance factor on prediction factor for a normal distribution, when $ 20 \leq n\leq 10000$, is always greater than or equal to 1. The tolerance and prediction factor are the coefficient of $\hat{\sigma}$ in Equation~\eqref{eq_normal_tolrance} and ~\eqref{eq_prediction_normal_interval}. So, the proportion of tolerance factor on prediction factor is the proportion of the coefficient of $\hat{\sigma}$ in Equation~\eqref{eq_normal_tolrance} on the coefficient of $\hat{\sigma}$ in Equation~\eqref{eq_prediction_normal_interval}. We call this proportion the tolerance prediction proportion and it can be simplified as:

\begin{equation}
\label{eq_tolpred_prop}
 \frac{Z_{\frac{1-\beta}{2}} \sqrt{n-1}}{t_{(\frac{1-\beta}{2},n-1)}\sqrt{\chi^{2}_{1-\gamma ,n-1}} }  
\end{equation}
 This property is verified numerically by Figure \ref{fig_tol_pred_prop}. In order to verify this property for sample size $ 20 \leq n\leq 10000$, one has to ensure that for fixed $n$ and $\gamma=0.7$, the minimum tolerance prediction proportion obtained over $\beta \in [0.01,0.99]$ must be greater than or equal to 1 and this inequality must hold for all $ 20 \leq n\leq 10000$. This is described formally as:

\begin{equation}
\label{eq_tolpred_prop_fig}
\forall n \in [20, \ldots,10000], \beta \in [0.01,0.99],  \min_{\beta} \frac{Z_{\frac{1-\beta}{2}} \sqrt{n-1}}{t_{(\frac{1-\beta}{2},n-1)}\sqrt{\chi^{2}_{1-\gamma ,n-1}} } \geq 1,
\end{equation}

where $\gamma=0.7$. The inequality described by Equation~\eqref{eq_tolpred_prop_fig} is proven for $\gamma=0.7$ by Figure (\ref{fig_tol_pred_prop}). We have seen above that the left part of Equation~\eqref{eq_tolpred_prop_fig} is an increasing function of $\gamma$, so this property holds also for small samples with $\gamma \geq 0.7$ \QEDA

\end{document}